\documentclass{amsbook}
\usepackage{amssymb}

\usepackage{graphicx}
\usepackage{amscd}

\theoremstyle{plain}

\newtheorem{corollary}{Corollary}

\newtheorem{definition}{Definition}
\newtheorem{example}{Example}

\newtheorem{lemma}{Lemma}

\newtheorem{theorem}{Theorem}
\numberwithin{equation}{section}

\input{tcilatex}

\begin{document}
\frontmatter
\title[Short Title]{Quantum Statistical Field Theory and Combinatorics}
\author{John Gough}
\address{Department of Computing \& Mathematics\\
Nottingham-Trent University\\
Burton Street\\
Nottingham NG1\ 4BU\\
United Kingdom.}
\email{\\
john.gough@ntu.ac.uk}
\urladdr{}
\maketitle
\dedicatory{To Margarita.}
\tableofcontents

\section{Preface}

The purpose of these notes is to gather together several facts concerning
the combinatorial aspects of diagram notation in field theory, as well as to
say something of the use of combinatorics in probability theory and
stochastic processes. For simplicity we have restricted to Boson systems
only. The motivation comes from the fact that, for the full potential of the
diagram notation to be realized, it must be supported by appropriate
mathematical techniques. In the general physics community, there are those
who never use diagrams and those who never use anything else. In favour of
diagrammatic notation it should be emphasized that it is a powerful and
appealing way of picturing fundamental processes that would otherwise
require unwieldy and uninspired mathematical notation. It is also the
setting in which a great many problems are generated and settled. On the
other, it should be noted the diagrams are artifacts of perturbation theory
and typically describe unphysical situations when treated in isolation. The
main problem is, however, one of mathematical intuition. The author, for
one, has always been troubled by the idea that, in a series expansion of
some Green function or other, some diagrams my have been omitted, or some
have been wrongly included, or some are equivalent but not recognized as
such, or the wrong combinatorial weights are attached, etc. Ones suspects
that a great many physicists habour similar concerns and would like some
independent mechanism to check up on this.

Now, it is a fair comment to say that few people get into theoretical
physics so that they can do combinatorics! Nevertheless, it turns out that
any attempt to study statistical fields through perturbation theory becomes
an exercise in manipulating  expressions (or diagrams representing these
expressions) and sooner or later we find ourselves asking combinatorial
questions. The intention here is to gather together several combinatorial
features arising from statistical field theory, quantum field theory and
quantum probability: all are related in one way or another to the
description of either random variables or observables through their moments
in a given state. The connotation of the word combinatorics for most
physicists is likely to be the tedious duty of counting out all diagrams to
a particular order in a given perturbation problem and attaching the equally
tedious combinatorial weight, if any. On the contrary, there is more
combinatorics than mindless enumeration and the subject develops into a
sophisticated and elegant process which has strong crossover with field
theoretic techniques. It is hoped that the notes delve just enough into
combinatorics, however, so as to cover the essentials we need in field
theory, yet give some appreciation for its power, relevance and subtly.

\ \ \ \ \ \ \ \ \ \ \ \ \ \ \ \ \ \ \ \ \ \ \ \ \ \ \ \ \ \ \ \ \ \ \ \ \ \
\ \ \ \ \ \ \ \ \ \ \ \ \ \ \ \ \ \ \ \ \ \ \ \ \ \ \ \ \ \ \ \ \ \ \ \ \ \
\ \ \ \ \ \ \ \ \ \ \ \ \ \ \ \ John Gough

\textit{Nottingham}

\textit{2005}

\pagebreak

\begin{center}
{\Huge Frequently used symbols}

\bigskip
\end{center}

$B_{n}$ Bell numbers (number of partitions of a set of $n$ items).

$\varepsilon \left( f\right) $ exponential vector for test function $f$.

$\Gamma _{+}\left( \frak{h}\right) $ the Bose Fock space over a Hilbert
space $\frak{h}$.

$h_{n}$ number of hierarchies on a set of $n$ items.

$\frak{H}\left( X\right) $ set of all hierarchies of a set $X$.

$\mu \left( \cdot \right) $ the M\"{o}bius function

$\frak{P}\left( X\right) $ set of all partitions of a set $X$.

$\frak{P}_{n}$ set of all partitions of $\left\{ 1,2,\cdots n\right\} $.

$\frak{P}_{n,m}$ set of all partitions of $\frak{P}_{n}$ consisting of $m$
parts

$\frak{P}^{f}\left( X\right) $ set of all partitions of a set $X$ finer than 
$\left\{ X\right\} $.

$\frak{P}^{c}\left( X\right) $ set of all partitions of a set $X$ coarser
than $\left\{ \left\{ x\right\} :x\in X\right\} $.

$s\left( n,m\right) $ Stirling numbers of the first kind.

$S\left( n,m\right) $ Stirling numbers of the second kind.

$\frak{S}_{n}$ set of permutations on $\left\{ 1,\cdots ,n\right\} $.

$\frak{S}_{n,m}$ set of permutations on $\left\{ 1,\cdots ,n\right\} $
having exactly $m$ cycles.

$W_{t}$ the Wiener process.

\bigskip

$x^{\downarrow }$ falling factorial power.

$x^{\uparrow }$ rising factorial power.

$\oplus $ direct sum.

$\otimes $ tensor product.

$\hat{\otimes}$ symmetrized tensor product.

\bigskip

\mainmatter

\chapter{Combinatorics}

One of the basic problems in combinatorics related to over-counting and
under-counting which occurs when we try to count unlabeled objects as if
they where labeled and labeled objects as if they were unlabeled,
respectively. It is curious that the solution to an early questions in
probability theory - namely, if three dice were rolled, would a 11 be more
likely than 12? - was answered incorrectly by under-counting. The wrong
answer was to say that both were equally likely as the are six ways to get
11 (6+4+1=6+3+2=5+5+1=5+4+2=5+3+3=4+4+3) and six ways to get 12
(6+5+1=6+4+2=6+3+3=5+5+2=5+4+3=4+4+4). The correct solution was provided by
Pascal and takes into account that the dice are distinguishable and so, for
instance, 6+5+1 can occur 3!=6 ways, 5+5+1 can occur 3 ways while 4+4+4 can
occur only one way: so there are 6+6+3+6+3+3=27 ways to get 11 and only
6+6+3+3+6+1=25 ways to get 12. The wrong solution for dice, however, turns
out to be the correct solution for indistinguishable sub-atomic particles
obeying Bose statistics. The same over-counting of microscopic
configurations is at the heart of Gibbs' entropy of mixing paradox in
statistical mechanics.

In this section, we look at some basic enumerations and show how they are of
use in understanding moments in probability theory.

\section{Combinatorics and Moments}

\subsection{Ordinary Moments}

Let $X$\ a random variable and let $\mathbb{F}\left( x\right) =\Pr \left\{
X<x\right\} $. The expected value of some function $f$ of the variable is
then given by the Stieltjes integral $\mathbb{E}\left[ f\left( X\right) %
\right] =\int_{-\infty }^{\infty }f\left( x\right) d\mathbb{F}\left(
x\right) $. We assume that the variable has moments to all orders: that is $%
\mathbb{E}\left[ X^{n}\right] $ exists for each integer $n\geq 0$. The
moment generating function is then defined to be 
\begin{equation}
M\left( t\right) =\sum_{n=0}^{\infty }\frac{1}{n!}\mathbb{E}\left[ X^{n}%
\right] t^{n}=\mathbb{E}\left[ e^{tX}\right]  \label{mgf}
\end{equation}
and so $M\left( t\right) =\int_{-\infty }^{\infty }e^{tx}d\mathbb{F}\left(
x\right) $ is Laplace transform of the probability distribution. The moments
are recovered from $M\left( t\right) $ by differentiating to the appropriate
order and evaluating at $t=0$, vis. 
\begin{equation}
\mathbb{E}\left[ X^{n}\right] =\left. \frac{d^{n}}{dt^{n}}M\left( t\right)
\right| _{t=0}.  \label{mgf diff}
\end{equation}

\subsection{Factorial Moments}

Suppose we now work with the parameter $z=e^{t}-1$, then $M\left( t\right)
\equiv \mathbb{E}\left[ \left( 1+z\right) ^{X}\right] $ and we consider
Taylor expansion about $z=0$. Now note that $\left( 1+z\right)
^{x}=\sum_{n=0}^{\infty }\dbinom{x}{n}z^{n}$ where $\dbinom{x}{n}=\dfrac{%
x\left( x-1\right) \left( x-2\right) \cdots \left( x-n+1\right) }{n!}$. It
is convenient to introduce the notion of falling factorial power 
\begin{equation}
x^{\downarrow n}:=x\left( x-1\right) \left( x-2\right) \cdots \left(
x-n+1\right)  \label{falling factorial powers}
\end{equation}
as well as a rising factorial power 
\begin{equation}
x^{\uparrow n}:=x\left( x+1\right) \left( x+2\right) \dots \left(
x+n-1\right) .  \label{rising factorial powers}
\end{equation}
(We also set $x^{\downarrow 0}=x^{\uparrow 0}=1$.)

It then follows that 
\begin{equation}
\mathbb{E}\left[ z^{X}\right] =\sum_{n=0}^{\infty }\frac{1}{n!}\mathbb{E}%
\left[ X^{\downarrow n}\right] z^{n}  \label{factorial mgf}
\end{equation}
and we refer to $\mathbb{E}\left[ X^{\downarrow n}\right] $ as the $n$-th
falling factorial moment: clearly $\mathbb{E}\left[ z^{X}\right] $ acts as
their generating function.

\subsection{Stirling's Numbers}

We now derive the connection between ordinary and falling factorial moments.
We begin by remarking that the right-hand sides of $\left( \ref{falling
factorial powers}\right) $ and $\left( \ref{rising factorial powers}\right) $%
\ may be expanded to obtain a polynomial in $x$ of degree $n$ with integer
coefficients. The Stirling numbers of the first and second kind are defined
as the coefficients appearing in the relations 
\begin{equation}
x^{\uparrow n}\equiv \sum_{m}s\left( n,m\right) x^{m};\quad x^{n}\equiv
\sum_{m}S\left( n,m\right) x^{\downarrow m}  \label{Stirling expansion}
\end{equation}
Evidently, the Stirling numbers of the first kind are integers satisfying $%
s\left( n,m\right) \geq 0$. It turns out that this is also true of the
second kind numbers. We also have $s\left( n,m\right) =0=S\left( n,m\right) $
for $m\neq 1,\dots ,n$. It is easy to see that $x^{\downarrow n}\equiv
\sum_{m}\left( -1\right) ^{n+m}s\left( n,m\right) x^{m}$ and $x^{n}\equiv
\sum_{m}\left( -1\right) ^{n+m}S\left( n,m\right) x^{\uparrow m}$ and from
this we see that the Stirling numbers are dual in the sense that 
\begin{equation}
\sum_{k}\left( -1\right) ^{n+k}s\left( n,k\right) S\left( k,m\right) =\delta
_{nm}.  \label{Strirling duality}
\end{equation}
The Stirling's numbers satisfy the recurrence relations (Stirling's
Identities) 
\begin{eqnarray}
s\left( n+1,m\right) &=&s\left( n,m-1\right) +ns\left( n,m\right) ;
\label{s recurrence} \\
S\left( n+1,m\right) &=&S\left( n,m-1\right) +mS\left( n,m\right)
\label{S recurrence}
\end{eqnarray}
with $s\left( 1,1\right) =1=S\left( 1,1\right) $. [Evidently $s\left(
n,n\right) =S\left( n,n\right) =S\left( n,1\right) =1$. The relations $%
x^{\uparrow \left( n+1\right) }=x^{\uparrow n}\times \left( x+n\right) $ and 
$x\times x^{\downarrow m}=\left( x-m+m\right) x^{\downarrow m}=x^{\downarrow
\left( m+1\right) }+mx^{\downarrow m}$ lead to the identities when we equate
coefficients.]

This means that the Stirling numbers may then be generated recursively using
a construction similar to Pascal's triangle, vis.

\begin{equation*}
\begin{tabular}{ll}
$s\left( n,m\right) $ & $S\left( n,m\right) $ \\ 
$
\begin{tabular}{r|rrrrrr}
$n\diagdown m$ & $1$ & $2$ & $3$ & $4$ & $5$ & $6$ \\ \hline
$1$ & $1$ &  &  &  &  &  \\ 
$2$ & $1$ & $1$ &  &  &  &  \\ 
$3$ & $2$ & $3$ & $1$ &  &  &  \\ 
$4$ & $6$ & $11$ & $6$ & $1$ &  &  \\ 
$5$ & $24$ & $50$ & $35$ & $10$ & $1$ &  \\ 
$6$ & $120$ & $274$ & $225$ & $85$ & $15$ & $1$%
\end{tabular}
$ & 
\begin{tabular}{r|rrrrrr}
$n\diagdown m$ & $1$ & $2$ & $3$ & $4$ & $5$ & $6$ \\ \hline
$1$ & $1$ &  &  &  &  &  \\ 
$2$ & $1$ & $1$ &  &  &  &  \\ 
$3$ & $1$ & $3$ & $1$ &  &  &  \\ 
$4$ & $1$ & $7$ & $6$ & $1$ &  &  \\ 
$5$ & $1$ & $15$ & $25$ & $10$ & $1$ &  \\ 
$6$ & $1$ & $31$ & $90$ & $65$ & $15$ & $1$%
\end{tabular}
\end{tabular}
\end{equation*}

From relation $\left( \ref{Stirling expansion}\right) $ we see that the
ordinary and falling factorial moments are related by 
\begin{eqnarray}
\mathbb{E}\left[ X^{n}\right] &\equiv &\sum_{m}S\left( n,m\right) \mathbb{E}%
\left[ X^{\downarrow m}\right] ,  \notag \\
\mathbb{E}\left[ X^{\downarrow n}\right] &\equiv &\sum_{m}\left( -1\right)
^{n+m}s\left( n,m\right) \mathbb{E}\left[ X^{m}\right] .
\label{ordinary-factorial moments}
\end{eqnarray}

\subsection{Cumulant Moments}

Cumulant moments $\kappa _{n}$ are defined through the relation $%
\sum_{n=1}^{\infty }\dfrac{1}{n!}\kappa _{n}t^{n}=\ln M\left( t\right) $ or 
\begin{equation*}
\sum_{n=0}^{\infty }\frac{1}{n!}\mathbb{E}\left[ X^{n}\right] t^{n}=\exp
\left\{ \sum_{n=1}^{\infty }\dfrac{1}{n!}\kappa _{n}t^{n}\right\}
\end{equation*}
and one sees from expanding and comparing coefficients of $t$ that 
\begin{eqnarray*}
\mathbb{E}\left[ X\right] &=&\kappa _{1,} \\
\mathbb{E}\left[ X^{2}\right] &=&\kappa _{2}+\kappa _{1}^{2}, \\
\mathbb{E}\left[ X^{3}\right] &=&\kappa _{3}+3\kappa _{2}\kappa _{1}+\kappa
_{1}^{3}, \\
\mathbb{E}\left[ X^{4}\right] &=&\kappa _{4}+4\kappa _{3}\kappa _{1}+3\kappa
_{2}^{2}+6\kappa _{2}\kappa _{1}^{2}+\kappa _{1}^{4}, \\
&&\text{etc.,}
\end{eqnarray*}
or inversely 
\begin{eqnarray*}
\kappa _{1} &=&\mathbb{E}\left[ X\right] , \\
\kappa _{2} &=&\mathbb{E}\left[ X^{2}\right] -\mathbb{E}\left[ X\right] ^{2}
\\
\kappa _{3} &=&\mathbb{E}\left[ X^{3}\right] -3\mathbb{E}\left[ X^{2}\right] 
\mathbb{E}\left[ X\right] +2\mathbb{E}\left[ X\right] ^{3} \\
\kappa _{4} &=&\mathbb{E}\left[ X^{4}\right] -4\mathbb{E}\left[ X^{3}\right] 
\mathbb{E}\left[ X\right] -3\mathbb{E}\left[ X^{2}\right] ^{2}+12\mathbb{E}%
\left[ X^{2}\right] \mathbb{E}\left[ X\right] ^{2}-6\mathbb{E}\left[ X\right]
^{4}, \\
&&\text{etc.}
\end{eqnarray*}

\subsection{Examples}

\begin{example}[Standard Gaussian]
We take $\mathbb{F}\left( x\right) =\left( 2\pi \right) ^{1/2}\int_{-\infty
}^{x}e^{-y^{2}/2}dy$ leading to the moment generating function 
\begin{equation*}
M\left( t\right) =e^{t^{2}/2}.
\end{equation*}
We see that all cumulant moments vanish except $\kappa _{2}=1$. Expanding
the moment generating function yields 
\begin{equation*}
\mathbb{E}\left[ X^{n}\right] =\left\{ 
\begin{array}{cc}
\dfrac{\left( 2k\right) !}{2^{k}k!}, & n=2k; \\ 
0, & n=2k+1.
\end{array}
\right.
\end{equation*}
\end{example}

\begin{example}[Poisson]
We take $X$ to be discrete with $\Pr \left\{ X=n\right\} =\frac{1}{n!}%
\lambda ^{n}e^{-\lambda }$ for $n=0,1,2,\cdots $. The parameter $\lambda $
must be positive. The moment generating function is readily computed and we
obtain 
\begin{equation*}
M\left( t\right) =\exp \left\{ \lambda \left( e^{t}-1\right) \right\} .
\end{equation*}
Taking the logarithm shows that all cumulant moments of the Poisson
distribution are equal to $\lambda $: 
\begin{equation*}
\kappa _{n}=\lambda \text{, for all }n=0,1,2,\cdots .
\end{equation*}
Likewise, $M\left( t\right) \equiv \exp \left\{ \lambda z\right\} $ where $%
z=e^{t}-1$ and so we see that the falling factorial moments are 
\begin{equation*}
\mathbb{E}\left[ X^{\downarrow n}\right] =\lambda ^{n}.
\end{equation*}
The ordinary moments are more involved. From $\left( \ref{ordinary-factorial
moments}\right) $ we see that they are polynomials of degree $n$ in $\lambda
:$%
\begin{equation}
\mathbb{E}\left[ X^{n}\right] =\sum_{m}S\left( n,m\right) \lambda ^{m}
\label{Poisson moments}
\end{equation}
and in particular $\lambda $ is the mean. We may expand the moment
generating function through the following series of steps 
\begin{eqnarray*}
M\left( t\right) &=&\sum_{m=0}^{\infty }\frac{1}{m!}\lambda ^{m}\left(
e^{t}-1\right) ^{m} \\
&=&\sum_{m=0}^{\infty }\frac{1}{m!}\lambda ^{m}\sum_{k=0}^{m}\binom{m}{k}%
\left( -1\right) ^{m-k}e^{kt} \\
&=&\sum_{m=0}^{\infty }\frac{1}{m!}\lambda ^{m}\sum_{k=0}^{m}\binom{m}{k}%
\left( -1\right) ^{m-k}\sum_{n=0}^{\infty }\frac{1}{n!}\left( kt\right) ^{n}
\end{eqnarray*}
and comparison with $\left( \ref{Poisson moments}\right) $ yields the
Stirling identity 
\begin{equation*}
S\left( n,m\right) =\frac{1}{m!}\sum_{k=0}^{m}\binom{m}{k}\left( -1\right)
^{m-k}k^{n}.
\end{equation*}
\end{example}

\begin{example}[Gamma]
Let $X$ be a positive continuous variable with $\mathbb{F}\left( x\right)
=\Gamma \left( \lambda \right) ^{-1}\int_{0}^{x}y^{\lambda -1}e^{-y}dy$,
(the Gamma function is $\Gamma \left( \lambda \right) =\int_{0}^{\infty
}y^{\lambda -1}e^{-y}dy$), the its moment generating function is 
\begin{equation*}
M\left( t\right) =\left( 1-t\right) ^{-\lambda }.
\end{equation*}
Now $\left( 1-t\right) ^{-\lambda }=\sum_{n=0}^{\infty }\dbinom{-\lambda }{n}%
\left( -t\right) ^{n}\equiv \sum_{n=0}^{\infty }\frac{1}{n!}\lambda
^{\uparrow n}t^{n}$ and so its moments are 
\begin{equation}
\mathbb{E}\left[ X^{n}\right] =\lambda ^{\uparrow n}=\sum_{m}s\left(
n,m\right) \lambda ^{m}.  \label{Gamma moments}
\end{equation}
\end{example}

\section{Fundamental Enumerations}

We now want to draw attention to the fact that the various families of
numbers, appearing in the formulas for the moments of our main examples
above, have an importance in combinatorics \cite{Riodain}.

For instance, the factor $\dfrac{\left( 2k\right) !}{2^{k}k!}$ occurring in
the moments of the Gaussian are well-known as the count of the number of
ways to partition a set of $2k$ items into pairs. The Stirling numbers also
arise as counts of classes of permutations and partitions, as we shall see
below.

\subsection{Pair Partitions}

A pair partition of the set $\left\{ 1,2,\dots ,2k\right\} $ consists of $k$
pairs taken from the set so that every element of the set is paired with
another. We shall denote the collection of all such pair partitions by $%
\mathcal{P}_{2k}$. Evidently, $\left| \mathcal{P}_{2k}\right| =\dfrac{\left(
2k\right) !}{2^{k}k!}$: we have $2k\times \left( 2k-1\right) $ choices for
the first pair, then $\left( 2k-2\right) \times \left( 2k-3\right) $ for the
second, etc. This gives a total of $\left( 2k\right) !$ however we have
over-counted by a factor of $k!$, as we do not label the pairs, and by $%
2^{k} $, as we do not label the elements within each pair either. It is
convenient to set $\left| \mathcal{P}_{2k+1}\right| =0$ since clearly we
cannot partition up an odd number of elements into pairs.

\subsection{Permutations}

The set of permutations, $\frak{S}_{n}$, over $\left\{ 1,\dots ,n\right\} $
forms a non-abelian group under composition. We shall use the notation $%
\sigma ^{0}=id,\sigma ^{1}=\sigma ,\sigma ^{2}=\sigma \circ \sigma $, etc.

Given a permutation $\sigma \in \frak{S}_{n}$, the \emph{orbit} of a number $%
i\in \left\{ 1,\dots ,n\right\} $ under $\sigma $ is the sequence $\left\{
i,\sigma \left( i\right) ,\sigma ^{2}\left( i\right) ,\dots \right\} $. As
the orbit must lie within $\left\{ 1,\dots ,n\right\} $ it is clear that $%
\sigma ^{k}\left( i\right) =i$ for some $k>0$: the smallest such value is
called the \emph{period }of the orbit and clearly orbit repeats itself
beyond this point $\left( \sigma ^{n+k}\left( i\right) =\sigma ^{n}\left(
i\right) \right) $. The ordered collection $\left[ i,\sigma \left( i\right)
;\sigma ^{2}\left( i\right) ;\dots ;\sigma ^{k-1}\left( i\right) \right] $
is referred to as a \emph{cycle} or more explicitly a $k$\emph{-cycle}.
Cycles will be considered to be equivalent under cyclic permutation in the
sense that $\left[ x_{1};x_{2};\dots ;x_{k}\right] $ is not distinguished
from $\left[ x_{2};x_{3};\dots ;x_{k};x_{1}\right] $, etc. Thus each $k$%
-cycle is equivalent to $k$ sequences depending on where on the orbit we
choose to start. Clearly orbits arising from the same permutation $\sigma $
either coincide or are completely disjoint; this simple observation leads to
the cyclic factorization theorem for permutations: each permutation $\sigma $%
\ can be uniquely written as a collection of disjoint cycles\emph{.}

\bigskip

\begin{lemma}
Let $\frak{S}_{n,m}$\ be the set of permutations in $\frak{S}_{n}$\ having
exactly $m$\ cycles. Then the number of permutations in $\frak{S}_{n,m}$\ is
given by the Stirling numbers of the first kind 
\begin{equation}
\left| \frak{S}_{n,m}\right| =s\left( n,m\right) .
\end{equation}
\end{lemma}

\begin{proof}
This is proved by showing that $\left| \frak{S}_{n,m}\right| $ satisfies the
same recurrence relation $\left( \ref{s recurrence}\right) $\ as the
Stirling numbers of the first kind, that is $\left| \frak{S}_{n+1,m}\right|
=\left| \frak{S}_{n,m-1}\right| +n\left| \frak{S}_{n,m}\right| .$Now $\left| 
\frak{S}_{n+1,m}\right| $ counts the number of permutations of $\left\{
1,2,\dots ,n+1\right\} $ having $m$ cycles. Of these, some will have $n+1$
as a fixed point which here means that $\left[ n+1\right] $ is unit-cycle:
as we have $m-1$ cycles left to be made up from the remaining labels $%
\left\{ 1,2,\dots ,n\right\} $, we see that there are $\left| \frak{S}%
_{n,m-1}\right| $ such permutations. Otherwise, the label $n+1$ lies within
a cycle of period two or more: now if we take any permutation in $\frak{S}%
_{n,m}$ then we could insert the label $n+1$ before any one of the labels $%
i\in \left\{ 1,\dots ,n\right\} $ in the cyclic decomposition - there are $%
n\left| \frak{S}_{n,m}\right| $ such possibilities and the second situation.
Clearly $\left| \frak{S}_{1,1}\right| =1=\left| \frak{S}_{n,n}\right| $
while $\left| \frak{S}_{n,m}\right| =0$ if $m>n$. Therefore $\left| \frak{S}%
_{n,m}\right| \equiv s\left( n,m\right) $.
\end{proof}

\subsection{Partitions}

Let $\mathcal{X}$ be a set. We denote by $\frak{P}\left( \mathcal{X}\right) $
the collection of all partitions of $\mathcal{X}$, that is, $\mathcal{A}%
=\left\{ A_{1},\dots ,A_{m}\right\} \in $ $\frak{P}\left( \mathcal{X}\right) 
$ if the $A_{j}$ are mutually-disjoint non-empty subsets of $\mathcal{X}$
having $\mathcal{X}$ as their union. The subsets $A_{j}$ making up a
partition are called parts. The set $\mathcal{X}$ is trivially a partition -
the partition of $\mathcal{X}$ consisting of just one part, namely $\mathcal{%
X}$ itself. All other partitions are called proper partitions.

If $\mathcal{X}=\left\{ 1,\dots ,n\right\} $ then the collection of
partitions of $\mathcal{X}$\ will be denoted as $\frak{P}_{n}$.while the
collection of partitions of $\mathcal{X}$\ having exactly $m$ parts will be
denoted as $\frak{P}_{n,m}$.

\bigskip

\begin{lemma}
The number of partitions of $\left\{ 1,\dots ,n\right\} $\ having exactly $m$%
\ parts is the Stirling number of the second kind $S\left( n,m\right) $: 
\begin{equation}
\left| \frak{P}_{n,m}\right| =S\left( n,m\right) .
\end{equation}
\end{lemma}

\begin{proof}
To prove this, we first of all show that we have the formula $\left| \frak{P}%
_{n+1,m}\right| =\left| \frak{P}_{n,m-1}\right| +m\left| \frak{P}%
_{n,m}\right| $. This is relatively straightforward. We see that $\left| 
\frak{P}_{n+1,m}\right| $ counts the number of partitions of a set $\mathcal{%
X}=\left\{ 1,\dots ,n,n+1\right\} $ having $m$ parts. Some of these will
have the singleton $\left\{ n+1\right\} $ as a part: there will be $\left| 
\frak{P}_{n,m-1}\right| $ of these as we have to partition the remaining
elements $\left\{ 1,\dots ,n\right\} $ into $m-1$ parts. The others will
have $n+1$ appearing with at least some other elements in a part: we have $%
\left| \frak{P}_{n,m}\right| $ partitions of $\left\{ 1,\dots ,n\right\} $
into $m$ parts and we then may place $n+1$ into any one of these $m$ parts
yielding $m\left| \frak{P}_{n,m}\right| $\ possibilities. Clearly $\left| 
\frak{P}_{1,1}\right| =1$ and $\left| \frak{P}_{n,n}\right| =1$ while $%
\left| \frak{P}_{n,m}\right| =0$ if $m>n$. The numbers $\left| \frak{P}%
_{n,m}\right| $ therefore satisfy the same generating relations $\left( \ref
{S recurrence}\right) $ as the $S\left( n,m\right) $ and so are one and the
same.
\end{proof}

As a corollary, we get the following result.

The total number of partitions that can be made from $n$ symbols, termed the 
$n$-th Bell number and denoted by $B_{n}$, is given by 
\begin{equation}
B_{n}=\left| \frak{P}_{n}\right| \equiv \sum_{m=1}^{n}S\left( n,m\right)
\end{equation}
The first few Bell numbers are 
\begin{equation*}
\begin{tabular}{l||l|l|l|l|l|l|l|l}
$n$ & $1$ & $2$ & $3$ & $4$ & $5$ & $6$ & $7$ & $8$ \\ \hline
$B_{n}$ & $1$ & $2$ & $5$ & $15$ & $52$ & $203$ & $877$ & $4140$%
\end{tabular}
\end{equation*}

For instance, the set $\left\{ 1,2,3,4\right\} $ can be partitioned into 2
parts in $S\left( 4,2\right) =7$ ways, vis.

\begin{equation*}
\begin{array}{cc}
\begin{array}{c}
\left\{ \left\{ 1,2\right\} ,\left\{ 3,4\right\} \right\} , \\ 
\left\{ \left\{ 1,3\right\} ,\left\{ 2,4\right\} \right\} , \\ 
\left\{ \left\{ 1,4\right\} ,\left\{ 2,3\right\} \right\} ,
\end{array}
& 
\begin{array}{c}
\left\{ \left\{ 1\right\} ,\left\{ 2,3,4\right\} \right\} , \\ 
\left\{ \left\{ 2\right\} ,\left\{ 1,3,4\right\} \right\} , \\ 
\left\{ \left\{ 3\right\} ,\left\{ 1,2,4\right\} \right\} , \\ 
\left\{ \left\{ 4\right\} ,\left\{ 1,2,3\right\} \right\} ,
\end{array}
\end{array}
\end{equation*}
and into 3 parts in $S\left( 4,3\right) =6$, vis. 
\begin{eqnarray*}
&&\left\{ \left\{ 1\right\} ,\left\{ 2\right\} ,\left\{ 3,4\right\} \right\}
,\text{ }\left\{ \left\{ 1\right\} ,\left\{ 3\right\} ,\left\{ 2,4\right\}
\right\} , \\
&&\left\{ \left\{ 1\right\} ,\left\{ 4\right\} ,\left\{ 2,3\right\} \right\}
,\text{ }\left\{ \left\{ 2\right\} ,\left\{ 3\right\} ,\left\{ 1,4\right\}
\right\} , \\
&&\left\{ \left\{ 2\right\} ,\left\{ 4\right\} ,\left\{ 2,4\right\} \right\}
,\text{ }\left\{ \left\{ 3\right\} ,\left\{ 4\right\} ,\left\{ 1,2\right\}
\right\} .
\end{eqnarray*}

\subsubsection{Occupation Numbers for Partitions}

Given $\mathcal{A}\in $\ $\frak{P}\left( \mathcal{X}\right) $, we let $%
n_{j}\left( \mathcal{A}\right) $ denote the number of parts in $\mathcal{A}$
having size $j$. We shall refer to the $n_{j}$ as \emph{occupation numbers}
and we introduce the functions 
\begin{equation*}
N\left( \mathcal{A}\right) =\sum_{j\geq 1}n_{j}\left( \mathcal{A}\right)
,\;E\left( \mathcal{A}\right) =\sum_{j\geq 1}jn_{j}\left( \mathcal{A}\right)
.
\end{equation*}

If $\mathcal{X}=\left\{ 1,\dots ,n\right\} $\ then the collection of
partitions of $\mathcal{X}$\ will be denoted as $\frak{P}_{n}$.while the
collection of partitions having exactly $m$\ parts will be denoted as $\frak{%
P}_{n,m}$.

Note that if $\mathcal{A}\in \frak{P}_{n,m}$ then $N\left( \mathcal{A}%
\right) =m$ and $E\left( \mathcal{A}\right) =n$.

\bigskip

It is sometimes convenient to replace sums over partitions with sums over
occupation numbers. Recall that a partition $\mathcal{A}$ will have
occupation numbers $\mathbf{n}=\left( n_{1},n_{2},n_{3},\dots \right) $ and
we have $N\left( \mathcal{A}\right) =n_{1}+n_{2}+n_{3}+\cdots $ and $E\left( 
\mathcal{A}\right) =n_{1}+2n_{2}+3n_{3}+\cdots $.

We will need to count the number of partitions with $E\left( \mathcal{A}%
\right) =n$ leading to the same set of occupation numbers $\mathbf{n}$; this
is given by 
\begin{equation}
\rho \left( \mathbf{n}\right) =\frac{1}{\left( 1!\right) ^{n_{1}}\left(
2!\right) ^{n_{2}}\left( 3!\right) ^{n_{3}}\cdots }\;\frac{n!}{%
n_{1}!n_{2}!n_{3}!\cdots }.  \label{no. partitions}
\end{equation}
The argument is that there are $n!$ ways to distribute the $n$ objects
however we do not distinguish the $n_{j}$ parts of size $j$ nor their
contents. We remark that the multinomials defined by 
\begin{equation*}
B_{n,m}\left( z_{1},z_{2},\cdots \right) =\sum_{\mathbf{n}}^{E=n,N=m}\rho
\left( \mathbf{n}\right) \,z_{1}^{n_{1}}z_{2}^{n_{2}}\cdots
\end{equation*}
are known as the Bell polynomials.

\subsubsection{Coarse Graining and M\"{o}bius Inversion}

A partial ordering of $\frak{P}\left( \mathcal{X}\right) $ is given by
saying that $\mathcal{A}\preccurlyeq \mathcal{B}$ if every part of $\mathcal{%
A}$ is a union of one or more parts of $\mathcal{B}$. In such situations we
say that $\mathcal{A}$ is coarser than $\mathcal{B}$, or equivalently that $%
\mathcal{B}$ is finer than $\mathcal{A}$. The partition consisting of just
the set $\mathcal{X}$ itself is coarser then every other partition of $%
\mathcal{X}$. Likewise, the partition consisting of only singletons is the
finest possible.

Whenever $\mathcal{A}\preccurlyeq \mathcal{B}$\ we denote by $n_{j}\left( 
\mathcal{A}|\mathcal{B}\right) $ the count of the number of parts of $%
\mathcal{A}$ that occur as the union of exactly $j$ parts of $\mathcal{B}$.
We also introduce the factor 
\begin{equation*}
\mu \left( \mathcal{A}|\mathcal{B}\right) =-\prod_{j\geq 1}\left\{ -\left(
j-1\right) !\right\} ^{n_{j}\left( \mathcal{A}|\mathcal{B}\right) }.
\end{equation*}

\bigskip

\begin{theorem}
Let $\Psi :\frak{P}\left( \mathcal{X}\right) \mapsto \mathbb{C}$\ be given
and let a function be defined by $\Phi \left( \mathcal{B}\right) =\sum_{%
\mathcal{A}\preccurlyeq \mathcal{B}}\Psi \left( \mathcal{A}\right) $.\ The
relation may be inverted to give 
\begin{equation*}
\Psi \left( \mathcal{B}\right) =\sum_{\mathcal{A}\preccurlyeq \frak{B}}\mu
\left( \mathcal{A}|\mathcal{B}\right) \,\Phi \left( \mathcal{A}\right)
\end{equation*}
\end{theorem}

\begin{proof}
Essentially we must show that 
\begin{equation*}
\sum_{\mathcal{A}\preccurlyeq \mathcal{B}\preccurlyeq \mathcal{C}}\mu \left( 
\mathcal{A}|\mathcal{B}\right) =\delta _{\mathcal{A},\mathcal{C}}.
\end{equation*}
Suppose that $A_{j}\in \mathcal{A}$, then we may write $A_{j}$ as the union
of $k_{j}$, say, parts of $\mathcal{B}$ and $r_{j}$, say, parts of $\mathcal{%
C}$. Evidently we will have $1\leq r_{j}\leq k_{j}$. By considering all the
possible partitions of these $k_{j}$ parts of $\mathcal{C}$ (for each $%
j=1,...,l$) we end up with all the partitions $\mathcal{B}$ finer than $%
\mathcal{C}$ but coarser than $\mathcal{A}$. The sum above then becomes 
\begin{equation*}
\prod_{j=1}^{N\left( \mathcal{A}\right) }\left\{
\sum_{r_{j}=1}^{k_{j}}\left( -1\right) ^{r_{j}}\left( r_{j}-1\right)
!S\left( k_{j},r_{j}\right) \right\} ,
\end{equation*}
however, observing that $\left( r-1\right) !=s\left( r,1\right) $ and using
the duality of the first and second kind Stirling numbers $\left( \ref
{Strirling duality}\right) $, we see that this is proportional to $%
\prod_{j=1}^{l}\delta _{1,k_{j}}$. This gives the result.
\end{proof}

\bigskip

Note that if $\psi $ is a function of the subsets of $\mathcal{X}$ and if $%
\phi \left( A\right) =\sum_{B\subseteq A}\psi \left( B\right) $ then we have
the relation $\psi \left( A\right) =\sum_{B\subseteq A}\left( -1\right)
^{\left| A\right| -\left| B\right| }\phi \left( B\right) $ which is the
so-called M\"{o}bius inversion formula \cite{Meyer}. The above result is
therefore the corresponding M\"{o}bius inversion formula for functions of
partitions.

\subsection{Hierarchies}

Let $X$ be a set and let 
\begin{eqnarray*}
\frak{P}^{f}\left( X\right) &:&=\text{ ``set of all finer partitions of }X%
\text{'',} \\
\frak{P}^{c}\left( X\right) &:&=\text{ ``set of all coarser partitions of }X%
\text{''}
\end{eqnarray*}
where we mean $\frak{P}\left( \mathcal{X}\right) $\ excluding the coarsest,
finest partition respectively.

A hierarchy on $X$is a directed tree having subsets of $X$ as nodes, where $%
A $ is further along a branch from $B$ if and only if $A\subset B$, and
where $X$ is the root of the tree and\ $\left\{ x\right\} ,$ $x\in X$, are
the leaves (terminal nodes).

For instance, let $X=\left\{ 1,\cdots ,6\right\} $ then a hierarchy is given
by taking three nodes $A=\left\{ 2,3\right\} $, $B=\left\{ 1,2,3\right\} $
and $C=\left\{ 4,5,6\right\} $.

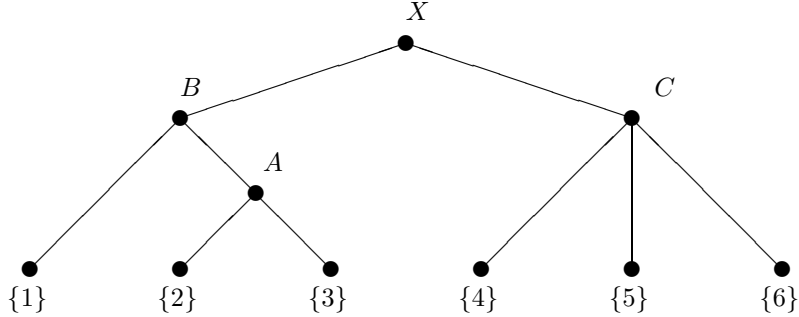
\begin{center}
%TCIMACRO{
%\TeXButton{figure 1}{\setlength{\unitlength}{.1cm}
%\begin{picture}(120,45)
%\label{pic1} 
%
%
%\put(10,10){\circle*{2}}
%\put(30,10){\circle*{2}}
%\put(50,10){\circle*{2}}
%\put(70,10){\circle*{2}}
%\put(90,10){\circle*{2}}
%\put(110,10){\circle*{2}}
%\put(40,20){\circle*{2}}
%\put(30,30){\circle*{2}}
%\put(90,30){\circle*{2}}
%\put(60,40){\circle*{2}}
%
%\put(10,10){\line(1,1){20}} 
%\put(30,10){\line(1,1){10}} 
%\put(50,10){\line(-1,1){20}} 
%\put(30,30){\line(3,1){30}} 
%\put(70,10){\line(1,1){20}} 
%\put(90,10){\line(0,1){20}} 
%\put(110,10){\line(-1,1){20}} 
%\put(90,30){\line(-3,1){30}} 
%
%\put(60,43){$X$}
%\put(30,33){$B$}
%\put(93,33){$C$}
%\put(41,23){$A$}
%\put(7,5){$\{ 1 \} $}
%\put(27,5){$\{ 2 \} $}
%\put(47,5){$\{ 3 \} $}
%\put(67,5){$\{ 4 \} $}
%\put(87,5){$\{ 5 \} $}
%\put(107,5){$\{ 6 \} $}
%
%
%\end{picture}
%}}%
%BeginExpansion
\setlength{\unitlength}{.1cm}
\begin{picture}(120,45)
\label{pic1}

\put(10,10){\circle*{2}}
\put(30,10){\circle*{2}}
\put(50,10){\circle*{2}}
\put(70,10){\circle*{2}}
\put(90,10){\circle*{2}}
\put(110,10){\circle*{2}}
\put(40,20){\circle*{2}}
\put(30,30){\circle*{2}}
\put(90,30){\circle*{2}}
\put(60,40){\circle*{2}}

\put(10,10){\line(1,1){20}} 
\put(30,10){\line(1,1){10}} 
\put(50,10){\line(-1,1){20}} 
\put(30,30){\line(3,1){30}} 
\put(70,10){\line(1,1){20}} 
\put(90,10){\line(0,1){20}} 
\put(110,10){\line(-1,1){20}} 
\put(90,30){\line(-3,1){30}} 

\put(60,43){$X$}
\put(30,33){$B$}
\put(93,33){$C$}
\put(41,23){$A$}
\put(7,5){$\{ 1 \} $}
\put(27,5){$\{ 2 \} $}
\put(47,5){$\{ 3 \} $}
\put(67,5){$\{ 4 \} $}
\put(87,5){$\{ 5 \} $}
\put(107,5){$\{ 6 \} $}

\end{picture}
%
%EndExpansion
\end{center}

There are two equivalent ways to describes hierarchies, both of which are
useful.

\subsubsection{Bottom-up description}

Let us consider the following sequence of partitions: 
\begin{eqnarray*}
\mathcal{A}^{\left( 1\right) } &=&\left\{ A_{1}^{\left( 1\right)
},A_{2}^{\left( 1\right) },A_{3}^{\left( 1\right) }\right\} \in \frak{P}%
^{c}\left( \mathcal{X}\right) :\text{ }A_{1}^{\left( 1\right) }=\left\{
1\right\} ,\,A_{2}^{\left( 1\right) }=\left\{ 2,3\right\} ,\,A_{3}^{\left(
1\right) }=\left\{ 4,5,6\right\} ; \\
\mathcal{A}^{\left( 2\right) } &=&\left\{ A_{1}^{\left( 2\right)
},A_{2}^{\left( 2\right) }\right\} \in \frak{P}^{c}\left( \mathcal{A}%
^{\left( 1\right) }\right) :\text{ }A_{1}^{\left( 2\right) }=\left\{
A_{1}^{\left( 1\right) },A_{2}^{\left( 1\right) }\right\} ,\,A_{2}^{\left(
2\right) }=\left\{ A_{3}^{\left( 1\right) }\right\} ; \\
\mathcal{A}^{\left( 3\right) } &=&\left\{ A_{1}^{\left( 3\right) }\right\}
\in \frak{P}^{c}\left( \mathcal{A}^{\left( 2\right) }\right) :\text{ }%
A_{1}^{\left( 3\right) }=\left\{ A_{1}^{\left( 2\right) },A_{2}^{\left(
2\right) }\right\} .
\end{eqnarray*}
This equivalently describes or example above.

In general, every hierarchy is equivalent to such a sequence 
\begin{equation*}
\mathcal{A}^{\left( 1\right) }\in \frak{P}^{c}\left( \mathcal{X}\right) ,%
\mathcal{A}^{\left( 2\right) }\in \frak{P}^{c}\left( \mathcal{A}^{\left(
1\right) }\right) ,\mathcal{A}^{\left( 3\right) }\in \frak{P}^{c}\left( 
\mathcal{A}^{\left( 2\right) }\right) ,\cdots
\end{equation*}
and as the partitions are increasing in coarseness (by combining parts of
predecessors) the sequence must terminate. That is, there will be an $m$,
which we refer to as the order of the hierarchy, such that $\mathcal{A}%
^{\left( m\right) }=\left\{ \mathcal{A}^{\left( m-1\right) }\right\} $.
Evidently, $m$ measures the number of edges along the longest branch of the
tree.

In our example, we can represent the hierarchy as 
\begin{eqnarray*}
\mathcal{H} &=&A_{1}^{\left( 3\right) } \\
&=&\left\{ A_{1}^{\left( 2\right) },A_{2}^{\left( 2\right) }\right\} \\
&=&\left\{ \left\{ A_{1}^{\left( 1\right) },A_{2}^{\left( 1\right) }\right\}
,\left\{ A_{3}^{\left( 1\right) }\right\} \right\} \\
&=&\left\{ \left\{ \left\{ 1\right\} ,\left\{ 2,3\right\} \right\} ,\left\{
\left\{ 4,5,6\right\} \right\} \right\} .
\end{eqnarray*}
This is an order three partition - each of the original elements of $X$ sits
inside three braces.

\subsubsection{Top-down description}

Alternatively, we obtain the same hierarchy by first partitioning $\left\{
1,\cdots ,6\right\} $ as $\frak{B}^{\left( 1\right) }=\left\{ B_{1}^{\left(
1\right) },B_{2}^{\left( 1\right) }\right\} $ where $B_{1}^{\left( 1\right)
}=\left\{ 1,2,3\right\} $ and $B_{2}^{\left( 1\right) }=\left\{
4,5,6\right\} $, then partitioning $B_{1}^{\left( 1\right) }$ as $\left\{
1\right\} $ and $\left\{ 2,3\right\} $, and finally partitioning all the
parts at this stage into singletons.

In general, every hierarchy can be viewed as a progression: 
\begin{equation*}
\frak{B}^{\left( 1\right) }\in \frak{P}^{f}\left( X\right) ,\left\{ \frak{B}%
_{B}^{\left( 2\right) }\in \frak{P}^{f}\left( B\right) :B\in \frak{B}%
^{\left( 1\right) }\right\} ,\left\{ \frak{B}_{B}^{\left( 3\right) }\in 
\frak{P}^{f}\left( B\right) :B\in \frak{B}^{\left( 2\right) }\right\}
,\cdots .
\end{equation*}
Eventually, this progression must bottom out as we can only subdivide $X$
into finer partitions so many times. Again the maximal number of
subdivisions is again given by the order of the hierarchy.

\bigskip

We shall now introduce some notation. Let $\frak{H}\left( X\right) $ denote
the collection of all hierarchies on a set $X$. We would like to know the
values of $h_{n}$, the number of hierarchies on a set of $n$ elements. We
may work out the lowest enumerations:

\bigskip

When $n=2$ we have the one tree 
%TCIMACRO{
%\TeXButton{figure 2}{\setlength{\unitlength}{.1cm}
%\begin{picture}(4,4)
%\label{pic1} 
%
%
%\put(2,4){\circle*{1}}
%\put(0,0){\circle*{1}}
%\put(4,0){\circle*{1}}
%
%\put(0,0){\line(1,2){2}}
%\put(4,0){\line(-1,2){2}}
%
%\end{picture}
%} }%
%BeginExpansion
\setlength{\unitlength}{.1cm}
\begin{picture}(4,4)
\label{pic1}

\put(2,4){\circle*{1}}
\put(0,0){\circle*{1}}
\put(4,0){\circle*{1}}

\put(0,0){\line(1,2){2}}
\put(4,0){\line(-1,2){2}}

\end{picture}
%
%EndExpansion
and so $h_{2}=1$. When $n=3$ we have the topologically distinct trees

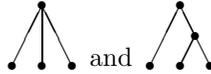
\begin{center}
%TCIMACRO{
%\TeXButton{figure 3}{\setlength{\unitlength}{.1cm}
%\begin{picture}(8,8)
%\label{pic1} 
%
%
%\put(4,8){\circle*{1}}
%\put(0,0){\circle*{1}}
%\put(4,0){\circle*{1}}
%\put(8,0){\circle*{1}}
%
%\put(0,0){\line(1,2){4}}
%\put(8,0){\line(-1,2){4}}
%\put(4,0){\line(0,1){8}}
%\end{picture}
%} }%
%BeginExpansion
\setlength{\unitlength}{.1cm}
\begin{picture}(8,8)
\label{pic1}

\put(4,8){\circle*{1}}
\put(0,0){\circle*{1}}
\put(4,0){\circle*{1}}
\put(8,0){\circle*{1}}

\put(0,0){\line(1,2){4}}
\put(8,0){\line(-1,2){4}}
\put(4,0){\line(0,1){8}}
\end{picture}
%
%EndExpansion
\ and 
%TCIMACRO{
%\TeXButton{figure 4}{\setlength{\unitlength}{.1cm}
%\begin{picture}(8,8)
%\label{pic1} 
%
%
%\put(4,8){\circle*{1}}
%\put(0,0){\circle*{1}}
%\put(4,0){\circle*{1}}
%\put(8,0){\circle*{1}}
%\put(6,4){\circle*{1}}
%
%\put(0,0){\line(1,2){4}}
%\put(8,0){\line(-1,2){4}}
%\put(4,0){\line(1,2){2}}
%
%\end{picture}
%}}%
%BeginExpansion
\setlength{\unitlength}{.1cm}
\begin{picture}(8,8)
\label{pic1}

\put(4,8){\circle*{1}}
\put(0,0){\circle*{1}}
\put(4,0){\circle*{1}}
\put(8,0){\circle*{1}}
\put(6,4){\circle*{1}}

\put(0,0){\line(1,2){4}}
\put(8,0){\line(-1,2){4}}
\put(4,0){\line(1,2){2}}

\end{picture}
%
%EndExpansion
\end{center}

\noindent and, when we count the number of ways to attach the leaves, we
have $h_{3}=1+3=4$ possibilities.

When $n=4$ when have the topologically distinct trees

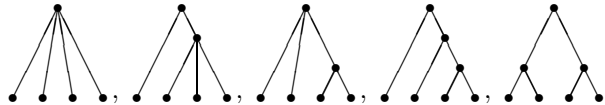
\begin{center}
%TCIMACRO{
%\TeXButton{figure 5}{\setlength{\unitlength}{.1cm}
%\begin{picture}(12,12)
%\label{pic1} 
%
%
%\put(6,12){\circle*{1}}
%\put(0,0){\circle*{1}}
%\put(4,0){\circle*{1}}
%\put(8,0){\circle*{1}}
%\put(12,0){\circle*{1}}
%
%
%\put(0,0){\line(1,2){6}}
%\put(4,0){\line(1,6){2}}
%\put(8,0){\line(-1,6){2}}
%\put(12,0){\line(-1,2){6}}
%
%\end{picture}
%}}%
%BeginExpansion
\setlength{\unitlength}{.1cm}
\begin{picture}(12,12)
\label{pic1}

\put(6,12){\circle*{1}}
\put(0,0){\circle*{1}}
\put(4,0){\circle*{1}}
\put(8,0){\circle*{1}}
\put(12,0){\circle*{1}}

\put(0,0){\line(1,2){6}}
\put(4,0){\line(1,6){2}}
\put(8,0){\line(-1,6){2}}
\put(12,0){\line(-1,2){6}}

\end{picture}
%
%EndExpansion
, 
%TCIMACRO{
%\TeXButton{figure 6}{\setlength{\unitlength}{.1cm}
%\begin{picture}(12,12)
%\label{pic1} 
%
%
%\put(6,12){\circle*{1}}
%\put(0,0){\circle*{1}}
%\put(4,0){\circle*{1}}
%\put(8,0){\circle*{1}}
%\put(12,0){\circle*{1}}
%\put(8,8){\circle*{1}}
%
%
%\put(0,0){\line(1,2){6}}
%\put(4,0){\line(1,2){4}}
%\put(8,0){\line(0,1){8}}
%\put(12,0){\line(-1,2){6}}
%
%\end{picture}
%}}%
%BeginExpansion
\setlength{\unitlength}{.1cm}
\begin{picture}(12,12)
\label{pic1}

\put(6,12){\circle*{1}}
\put(0,0){\circle*{1}}
\put(4,0){\circle*{1}}
\put(8,0){\circle*{1}}
\put(12,0){\circle*{1}}
\put(8,8){\circle*{1}}

\put(0,0){\line(1,2){6}}
\put(4,0){\line(1,2){4}}
\put(8,0){\line(0,1){8}}
\put(12,0){\line(-1,2){6}}

\end{picture}
%
%EndExpansion
, 
%TCIMACRO{
%\TeXButton{figure 7}{\setlength{\unitlength}{.1cm}
%\begin{picture}(12,12)
%\label{pic1} 
%
%
%\put(6,12){\circle*{1}}
%\put(0,0){\circle*{1}}
%\put(4,0){\circle*{1}}
%\put(8,0){\circle*{1}}
%\put(12,0){\circle*{1}}
%\put(10,4){\circle*{1}}
%
%
%\put(0,0){\line(1,2){6}}
%\put(4,0){\line(1,6){2}}
%\put(8,0){\line(1,2){2}}
%\put(12,0){\line(-1,2){6}}
%
%\end{picture}
%}}%
%BeginExpansion
\setlength{\unitlength}{.1cm}
\begin{picture}(12,12)
\label{pic1}

\put(6,12){\circle*{1}}
\put(0,0){\circle*{1}}
\put(4,0){\circle*{1}}
\put(8,0){\circle*{1}}
\put(12,0){\circle*{1}}
\put(10,4){\circle*{1}}

\put(0,0){\line(1,2){6}}
\put(4,0){\line(1,6){2}}
\put(8,0){\line(1,2){2}}
\put(12,0){\line(-1,2){6}}

\end{picture}
%
%EndExpansion
, 
%TCIMACRO{
%\TeXButton{figure 8}{\setlength{\unitlength}{.1cm}
%\begin{picture}(12,12)
%\label{pic1} 
%
%
%\put(6,12){\circle*{1}}
%\put(0,0){\circle*{1}}
%\put(4,0){\circle*{1}}
%\put(8,0){\circle*{1}}
%\put(12,0){\circle*{1}}
%
%\put(8,8){\circle*{1}}
%\put(10,4){\circle*{1}}
%
%\put(0,0){\line(1,2){6}}
%\put(4,0){\line(1,2){4}}
%\put(8,0){\line(1,2){2}}
%\put(12,0){\line(-1,2){6}}
%
%\end{picture}
%}}%
%BeginExpansion
\setlength{\unitlength}{.1cm}
\begin{picture}(12,12)
\label{pic1}

\put(6,12){\circle*{1}}
\put(0,0){\circle*{1}}
\put(4,0){\circle*{1}}
\put(8,0){\circle*{1}}
\put(12,0){\circle*{1}}

\put(8,8){\circle*{1}}
\put(10,4){\circle*{1}}

\put(0,0){\line(1,2){6}}
\put(4,0){\line(1,2){4}}
\put(8,0){\line(1,2){2}}
\put(12,0){\line(-1,2){6}}

\end{picture}
%
%EndExpansion
, 
%TCIMACRO{
%\TeXButton{figure 9}{\setlength{\unitlength}{.1cm}
%\begin{picture}(12,12)
%\label{pic1} 
%
%
%\put(6,12){\circle*{1}}
%\put(0,0){\circle*{1}}
%\put(4,0){\circle*{1}}
%\put(8,0){\circle*{1}}
%\put(12,0){\circle*{1}}
%
%\put(2,4){\circle*{1}}
%\put(10,4){\circle*{1}}
%
%\put(4,0){\line(-1,2){2}}
%\put(8,0){\line(1,2){2}}
%\put(0,0){\line(1,2){6}}
%\put(12,0){\line(-1,2){6}}
%
%\end{picture}
%}}%
%BeginExpansion
\setlength{\unitlength}{.1cm}
\begin{picture}(12,12)
\label{pic1}

\put(6,12){\circle*{1}}
\put(0,0){\circle*{1}}
\put(4,0){\circle*{1}}
\put(8,0){\circle*{1}}
\put(12,0){\circle*{1}}

\put(2,4){\circle*{1}}
\put(10,4){\circle*{1}}

\put(4,0){\line(-1,2){2}}
\put(8,0){\line(1,2){2}}
\put(0,0){\line(1,2){6}}
\put(12,0){\line(-1,2){6}}

\end{picture}
%
%EndExpansion
\end{center}

\noindent which implies that $h_{4}=1+4+6+12+3=26$. We find that 
\begin{equation*}
\begin{tabular}{l||l|l|l|l|l|l|l|l}
$n$ & $1$ & $2$ & $3$ & $4$ & $5$ & $6$ & $7$ & $8$ \\ \hline
$h_{n}$ & $1$ & $1$ & $4$ & $26$ & $236$ & $2752$ & $39208$ & $660032$%
\end{tabular}
\end{equation*}
This sequence is well known in combinatorics and appears as sequence A000311
on the ATT classification. It is known that the exponential generating
series $h\left( x\right) =\sum_{n}\frac{1}{n!}h_{n}x^{n}$ converges and
satisfies $\exp h\left( x\right) =2h\left( x\right) -x+1$.

\section{Prime Decompositions}

There results of this section are really necessary for our discussion on
field theory, but have been included for completeness. As is well known, the
primes are the indivisible natural numbers and every natural number can be
decomposed uniquely into a product of primes. Each natural number determines
an unique sequence of occupation numbers consisting of the number of time as
particular prime divides into that number. In our discussion below, we meet
a calculation of functions that can be defined by their action on primes.
Here we encounter an argument for replacing a sum (over occupation
sequences) of products with a product of sums (over individual occupation
numbers). This argument will recur elsewhere.

It turns out that the cumulant moments play a role similar to primes insofar
as they are indivisible, in a sense to be made explicit, and every ordinary
moment can be uniquely decomposed into cumulant moments.

\subsection{The Prime Numbers}

The \textit{natural numbers} are positive integers $\mathbb{N}=\left\{
1,2,3,\cdots \right\} $. If a natural number $m$ goes into another natural
number $n$ with no remainder, then we say that $m$ \textit{divides} $n$ and
write this as $m|n$. Given a natural number $n$, we shall denote the \textit{%
number of divisors} of $n$ by $d\left( n\right) $, that is 
\begin{equation*}
d\left( n\right) :=\#\left\{ m:m|n\right\} .
\end{equation*}
Likewise, the \textit{sum of the divisors} of $n$ is denoted as 
\begin{equation*}
s\left( n\right) :=\dsum_{k}^{k|n}k.
\end{equation*}
A natural number, $p$, is \textit{prime} if is has no divisors other than
itself and $1$. That is, $d\left( p\right) =2$. The collection of primes
will be denoted as $\mathcal{P}$ and we list them, in increasing order, as 
\begin{equation*}
p_{1}=2,\,p_{2}=3,\,p_{3}=5,\,p_{4}=7,\,p_{5}=11\text{, etc.}
\end{equation*}

\bigskip

\begin{theorem}[Prime Decomposition Theorem]
\textit{Any natural number }$m$\textit{\ can be uniquely decomposed as a
product of primes:} 
\begin{equation}
m=\prod_{j=1}^{\infty }\left( p_{j}\right) ^{n_{j}}.
\label{Prime decomposition}
\end{equation}
\end{theorem}

The numbers $n_{j}=n_{j}\left( m\right) $ give the number of times the $j^{%
\text{th}}$prime, $p_{j},$ divides an integer $m$. In this way, we see that
there is a one-to-one correspondence between the natural numbers and the
collection of ``occupation numbers'' $\mathbf{n}=\left( n_{j}\right)
_{j=1}^{\infty }$ where have $0<\sum_{j=1}^{\infty }n_{j}<\infty $.

\bigskip

\begin{theorem}[Euclid]
\textit{There are infinitely many primes.}
\end{theorem}

\begin{proof}
Suppose that we new the first $N$ primes where $N<\infty $, we then
construct the number $q=\prod_{j=1}^{N}p_{j}+1$. If we try to divide $q$ by
any of the known primes $p_{1},\cdots p_{N}$, we get a remainder of one each
time. Since any potential divisor of $q$ must be factorizable as a product
of the known primes, we conclude that $q$ has no divisors other that itself
and one and is therefore prime itself. Therefore the list of prime numbers
is endless.
\end{proof}

\bigskip

If $m$ and $n$ have no common factors then we say that they are \textit{%
relatively prime} and write this as $m\perp n$. The Euler phi function, $%
\varphi \left( n\right) ,$ counts the number of natural numbers less than
and relatively prime to $n$: 
\begin{equation*}
\varphi \left( n\right) :=\#\left\{ m:m<n,m\perp n\right\} .
\end{equation*}

\subsection{Dirichlet Generating Functions}

Let $a=\left( a_{n}\right) _{n=1}^{\infty }$\ be a sequence of real numbers.
Its\textit{\ Dirichlet generating function }is defined to be 
\begin{equation*}
\frak{D}_{a}\left( s\right) :=\sum_{n=1}^{\infty }\frac{a_{n}}{n^{s}}.
\end{equation*}
Let $a$\ and $b$\ be sequences, then their Dirichlet convolution is the
sequence $c=a\ast _{d}c$\ defined by 
\begin{equation*}
c_{n}=\sum_{m}^{m|n}a_{m}b_{n/m}.
\end{equation*}

\begin{lemma}
\textit{Let }$a$\textit{\ and }$b$\textit{\ be sequences, then }$\frak{D}_{a}%
\frak{D}_{b}=\frak{D}_{c}$\textit{\ where the sequence }$c$\textit{\ is the
Dirichlet convolution of }$a$\textit{\ and }$b$\textit{.}
\end{lemma}

\begin{proof}
\begin{equation*}
\frak{D}_{a}\frak{D}_{b}=\sum_{n=1}^{\infty }\sum_{m=1}^{\infty }\frac{a_{n}%
}{n^{s}}.\frac{b_{m}}{m^{s}}=\sum_{n=1}^{\infty }\sum_{m=1}^{\infty }\frac{%
a_{n}b_{m}}{\left( nm\right) ^{s}}\equiv \sum_{k=1}^{\infty }\frac{c_{k}}{%
k^{s}}.
\end{equation*}
\end{proof}

Let $f$\ be a function defined on the natural numbers. The function is said
to be\textit{\ multiplicative} if 
\begin{equation*}
f\left( nm\right) =f\left( n\right) f\left( m\right) ,\qquad \text{whenever
\ }m\perp n\text{.}
\end{equation*}
If we furthermore have $f\left( nm\right) =f\left( n\right) f\left( m\right) 
$, for every pair of natural numbers, then the function is said to be 
\textit{strongly multiplicative}. For instance, $f\left( n\right) =n^{s}$ is
strongly multiplicative. However, the Euler phi function is multiplicative,
but not strongly so.

\begin{lemma}
\textit{Let }$f$\textit{\ be a multiplicative function of the natural
numbers. Then} 
\begin{equation*}
\frak{D}_{f}\left( s\right) =\prod_{p\in \mathcal{P}}\sum_{m\geq 0}\frac{%
f\left( p^{m}\right) }{p^{ns}}.
\end{equation*}
\end{lemma}

\begin{proof}
This is a consequence of the unique prime decomposition (\ref{Prime
decomposition})\ for any natural number. If $f$ is multiplicative and $%
m=\prod_{j=1}^{\infty }\left( p_{j}\right) ^{n_{j}}$, then $f\left( m\right)
=\prod_{j=1}^{\infty }f\left( p_{j}^{n_{j}}\right) $. The Dirichlet
generating function for the sequence $f\left( m\right) $ is then 
\begin{eqnarray*}
\frak{D}_{f}\left( s\right) &=&\sum_{m\geq 1}\frac{f\left( m\right) }{m^{s}}
\\
&=&\sum_{\mathbf{n}}\prod_{j=1}^{\infty }f\left( p_{j}^{n_{j}}\right)
p_{j}^{-n_{j}s} \\
&=&\prod_{j=1}^{\infty }\sum_{n=0}^{\infty }f\left( p_{j}^{n}\right)
p_{j}^{-ns} \\
&=&\prod_{p\in \mathcal{P}}\sum_{n\geq 0}\frac{f\left( p^{n}\right) }{p^{ns}}%
.
\end{eqnarray*}
\end{proof}

\bigskip

The replacement $\sum_{\mathbf{n}}\prod_{j=1}^{\infty }g\left( n_{j}\right)
\leftrightarrow \prod_{j=1}^{\infty }\sum_{n=0}^{\infty }g\left( n\right) $
used above is an elementary trick which we shall call the $\sum \prod
\leftrightarrow \prod \sum $ trick. (It's the one that is used to compute
the grand canonical partition function for the free Bose gas!)

\bigskip

\begin{lemma}
\textit{Let }$f$\textit{\ be a strongly multiplicative function of the
natural numbers. Then} 
\begin{equation*}
\frak{D}_{f}\left( s\right) =\prod_{p\in \mathcal{P}}\left( 1-\frac{f\left(
p\right) }{p^{s}}\right) ^{-1}.
\end{equation*}
\end{lemma}

\begin{proof}
If $f$ is strongly multiplicative, then $f\left( p^{n}\right) =f\left(
p\right) ^{n}$. We then encounter the geometric sequence 
\begin{equation*}
\sum_{n\geq 0}\frac{f\left( p^{n}\right) }{p^{ns}}=\sum_{n\geq 0}\frac{%
f\left( p\right) ^{n}}{p^{ns}}=\left( 1-\frac{f\left( p\right) }{p^{s}}%
\right) ^{-1}.
\end{equation*}
\end{proof}

\subsection{The Riemann Zeta Function}

The Riemann zeta function, $\zeta \left( s\right) $, is the Dirichlet
generating function for the constant sequence $1=\left( 1\right)
_{n=1}^{\infty }$. That is, 
\begin{equation}
\zeta \left( s\right) :=\sum_{n=1}^{\infty }\frac{1}{n^{s}}.
\label{Riemann Zeta}
\end{equation}

We note that $\zeta \left( 0\right) $ and $\zeta \left( 1\right) $ are
clearly divergent. However, $\zeta \left( 2\right) =\sum_{n=1}^{\infty }%
\frac{1}{n^{2}}=\dfrac{\pi ^{2}}{6}$. It is clear that $\zeta \left(
s\right) $ is in fact an analytic function of $s$ for $\func{Re}\left(
s\right) >1$.

\bigskip

\begin{lemma}[Euler]
\textit{For }$\func{Re}\left( s\right) >1$\textit{,} 
\begin{equation}
\zeta \left( s\right) =\prod_{p\in \mathcal{P}}\left( 1-\frac{1}{p^{s}}%
\right) ^{-1}.  \label{Riemann Zeta (Euler Form)}
\end{equation}
\end{lemma}

\begin{proof}
This follows from the observation that $\zeta \left( s\right) =\frak{D}%
_{1}\left( s\right) $ and that the constant sequence $1$ corresponds to the
trivially strongly multiplicative function $f\left( n\right) =1$.
\end{proof}

\bigskip

An immediate corollary to the previous lemma is that $\zeta ^{2}\left(
s\right) =\frak{D}_{d}\left( s\right) =\sum_{n=1}^{\infty }\dfrac{d\left(
n\right) }{n^{s}}$.

\subsection{The M\"{o}bius Function}

The M\"{o}bius function\ is the multiplicative function $\mu $\ determined
by 
\begin{equation*}
\mu \left( p^{n}\right) :=\left\{ 
\begin{array}{cc}
1, & n=0; \\ 
-1, & n=1; \\ 
0, & n\geq 2.
\end{array}
\right.
\end{equation*}
for each prime $p$.

\bigskip

\begin{lemma}
$\frak{D}_{\mu }=\dfrac{1}{\zeta }.$\textit{\ }
\end{lemma}

\begin{proof}
$\frak{D}_{\mu }\left( s\right) =\sum_{n\geq 1}\dfrac{\mu \left( n\right) }{%
n^{s}}=\prod_{p\in \mathcal{P}}\sum_{n\geq 0}\dfrac{\mu \left( p^{n}\right) 
}{p^{ns}}=\prod_{p\in \mathcal{P}}\left( 1-p^{-s}\right) =\dfrac{1}{\zeta
\left( s\right) }.$
\end{proof}

\begin{lemma}
\textit{Let }$\left( b_{n}\right) _{n=1}^{\infty }$\textit{\ be a given
sequence and set }$a_{n}=\sum_{k}^{k|n}b_{k}$\textit{, then} 
\begin{equation*}
b_{n}=\sum_{k}^{k|n}\mu \left( \frac{n}{k}\right) a_{k}.
\end{equation*}
\end{lemma}

\begin{proof}
We evidently have that $\frak{D}_{a}=\frak{D}_{b}\zeta $ and so $\frak{D}%
_{b}=\zeta ^{-1}\frak{D}_{a}=\frak{D}_{\mu }\frak{D}_{a}$.
\end{proof}

\bigskip

As an application, we take $b_{n}=n$. Then $a_{n}=\sum_{k}^{k|n}k$ which is
just the sum, $s\left( n\right) $, of the divisors of $n$. We deduce that $%
n=\sum_{k}^{k|n}\mu \left( \frac{n}{k}\right) s\left( k\right) $.

\chapter{Boson Fock Space}

Why do the Gaussian and Poissonian distributions emerge from basic
enumerations? We shall try and answer this by looking at Bosonic field and
their quantum expectations.

\section{Boson Statistics}

Identical sub-atomic particles are indistinguishable to the extent that
there states must be invariant under arbitrary exchange (or, more, generally
permutation) of their labels we might attach to them. Suppose we have a
system of $n$ identical particles, then it happens in Nature that one of two
possibilities can arise: either the total wave function $\psi \left(
x_{1},x_{2},\cdots ,x_{n}\right) $ is completely symmetric, or completely
anti-symmetric under interchange of particle labels. The former type are
called bosons and we have 
\begin{equation*}
\psi \left( x_{\sigma \left( 1\right) },x_{\sigma \left( 2\right) },\cdots
,x_{\sigma \left( n\right) }\right) =\psi \left( x_{1},x_{2},\cdots
,x_{n}\right) \text{, for all }\sigma \in \frak{S}_{n}\text{,}
\end{equation*}
while the latter species are called fermions and we have 
\begin{equation*}
\psi \left( x_{\sigma \left( 1\right) },x_{\sigma \left( 2\right) },\cdots
,x_{\sigma \left( n\right) }\right) =\left( -1\right) ^{\sigma }\psi \left(
x_{1},x_{2},\cdots ,x_{n}\right) \text{, for all }\sigma \in \frak{S}_{n}%
\text{,}
\end{equation*}
where $\left( -1\right) ^{\sigma }$ denotes the sign of the permutation.
(Here, the coordinates $x_{j}$ give all necessary labels we might attach to
a particle: position, spin, color, etc.)

\subsection{Fock Space}

Let $\frak{h}$ be a fixed Hilbert space and denote by $\frak{h}^{\hat{\otimes%
}n}$ the closed linear span of symmetrized vectors of the type 
\begin{equation}
\psi _{1}\hat{\otimes}\cdots \hat{\otimes}\psi _{n}:=\frac{1}{n!}%
\sum_{\sigma \in \frak{S}_{n}}\,\psi _{\sigma (1)}\otimes \cdots \otimes
\psi _{\sigma (n)}.
\end{equation}
where $\psi _{j}\in \frak{h}$. To understand better the form of $\frak{h}^{%
\hat{\otimes}n}$, let $\left\{ e_{j}:j=1,\cdots \right\} $ be an orthonormal
basis of $\frak{h}$. Suppose that we have $n$ particles with $n_{1}$
particles are in state $e_{1}$, $n_{2}$ in state $e_{2}$, etc., then there
is accordingly only one total state describing this: setting $\mathbf{n}%
=\left\{ n_{j}:j=1,2,\cdots \right\} $, the appropriately normalized state
is described by the vector -known as a number vector - given by 
\begin{equation}
\left| \mathbf{n}\right\rangle =\binom{n}{n_{1},n_{2},\cdots }^{1/2}\;%
\underset{n_{1}\text{ factors}}{\left( \underbrace{e_{1}\hat{\otimes}\cdots 
\hat{\otimes}e_{1}}\right) }\hat{\otimes}\underset{n_{2}\text{ factors}}{%
\left( \underbrace{e_{2}\hat{\otimes}\cdots \hat{\otimes}e_{2}}\right) }\hat{%
\otimes}\cdots .
\end{equation}
The span of such states, with the obvious restriction that $\sum_{j}n_{j}=n$%
, yields $\frak{h}^{\hat{\otimes}n}$.

The Fock space over $\frak{h}$ is then the space spanned by all vectors $%
\left| \mathbf{n}\right\rangle $ when only the restriction $%
\sum_{j}n_{j}<\infty $ applies. We also include a no-particle state called
the Fock vacuum and which we denote by $\Omega $. The Fock space is then the
direct sum 
\begin{equation}
\Gamma _{+}\left( \frak{h}\right) =\bigoplus_{n=0}^{\infty }\frak{h}^{\hat{%
\otimes}n}.
\end{equation}
The vacuum space $\frak{h}^{\hat{\otimes}0}$ is taken to be spanned by the
single vector $\Omega $.

Often, the number state vectors are not the most convenient to use and we
now give an alternative class. The exponential vector map $\varepsilon :%
\frak{h}\mapsto \Gamma _{+}\left( \frak{h}\right) $ is defined by 
\begin{equation}
\varepsilon \left( f\right) =\oplus _{n=0}^{\infty }\left( \frac{1}{\sqrt{n!}%
}f^{\otimes n}\right)  \label{Expvectors}
\end{equation}
with $f^{\otimes n}$ the $n$-fold tensor product of $f$ with itself. The
Fock vacuum is, in particular, given by $\Omega =\varepsilon \left( 0\right) 
$. We note that $\left\langle \varepsilon \left( f\right) |\varepsilon
\left( g\right) \right\rangle =\sum_{n\geq 0}\frac{1}{n!}\left\langle
f|g\right\rangle ^{n}=\exp \left\langle f|g\right\rangle $, whence the name
exponential vectors. (These vectors are called Bargmann vector states in the
physics literature, while their normalized versions are known as coherent
state vectors.) The set $\varepsilon \left( \frak{h}\right) $, consisting of
all exponential vectors generated by the test functions in $\frak{h}$, is
linearly independent in $\Gamma _{+}\left( \frak{h}\right) $. Moreover, they
have the property that $\varepsilon \left( S\right) $ will be dense in the
Boson Fock space whenever $S$ is dense in $\frak{h}$.

\subsection{Creation, Annihilation and Conservation}

Boson creation and annihilation fields are then defined on $\Gamma
_{+}\left( \frak{h}\right) $ by the actions:

\begin{eqnarray}
B^{+}\left( \phi \right) \,f_{1}\hat{\otimes}\cdots \hat{\otimes}f_{n} &:&=%
\sqrt{n+1}\;\phi \hat{\otimes}f_{1}\hat{\otimes}\cdots \hat{\otimes}f_{n}; 
\notag \\
B^{-}\left( \phi \right) \,f_{1}\hat{\otimes}\cdots \hat{\otimes}f_{n} &:&=%
\frac{1}{\sqrt{n}}\,\sum_{j}\left\langle \phi |f_{j}\right\rangle \,f_{1}%
\hat{\otimes}\cdots \hat{\otimes}\widehat{f_{j}}\hat{\otimes}\cdots \otimes
f_{n}.  \notag
\end{eqnarray}
They satisfy the canonical commutation relations

\begin{eqnarray}
\left[ B^{-}\left( \phi \right) ,B^{+}\left( \psi \right) \right]
&=&\left\langle \phi |\psi \right\rangle ,  \notag \\
\left[ B^{-}\left( \phi \right) ,B^{-}\left( \psi \right) \right] &=&0=\left[
B^{+}\left( \phi \right) ,B^{+}\left( \psi \right) \right] .  \label{CCR}
\end{eqnarray}

Likewise, let $M$ be an operator acting on $\frak{h}$. Its differential
second quantization is the operator $d\Gamma _{+}\left( M\right) $ acting on 
$\Gamma _{+}\left( \frak{h}\right) $ and defined by

\begin{equation*}
d\Gamma _{+}\left( M\right) \left( f_{1}\hat{\otimes}\cdots \hat{\otimes}%
f_{n}\right) :=\left( Mf_{1}\right) \hat{\otimes}\cdots \hat{\otimes}%
f_{n}+\cdots +f_{1}\hat{\otimes}\cdots \hat{\otimes}\left( Mf_{n}\right) .
\end{equation*}

\section{States}

\subsection{Gaussian Fields}

Consider $\left\langle \Omega |\,B^{\varepsilon \left( k\right) }\left(
f_{n}\right) \cdots B^{\varepsilon \left( 2\right) }\left( f_{2}\right)
B^{\varepsilon \left( 1\right) }\left( f_{1}\right) \,\Omega \right\rangle $%
, where the $\varepsilon $'s stand for $+$ or $-$, this is a vacuum
expectation and we may compute by the following scheme: every time we
encounter an expression $B^{-}\left( f_{i}\right) B^{+}\left( f_{j}\right) $
we replace it with $B^{+}\left( f_{j}\right) B^{-}\left( f_{i}\right)
+\left\langle f_{i}|f_{j}\right\rangle $. The term $\left\langle
f_{i}|f_{j}\right\rangle $ is scalar and can be brought outside the
expectation leaving a product of two less fields to average. Ultimately we
must pair up every creator with an annihilator otherwise we get a zero.
Therefore only the even moments are non-zero and we obtain 
\begin{equation}
\sum_{\varepsilon \in \left\{ +,-\right\} ^{n}}\left\langle \Omega
|\,B^{\varepsilon \left( 2n\right) }\left( f_{2n}\right) \cdots
B^{\varepsilon \left( 2\right) }\left( f_{2}\right) B^{\varepsilon \left(
1\right) }\left( f_{1}\right) \,\Omega \right\rangle =\sum_{\left(
p_{j},q_{j}\right) _{j=1}^{n}\in \mathcal{P}_{2n}}\prod_{j=1}^{n}\left%
\langle f_{p_{j}}|f_{q_{j}}\right\rangle .  \label{Gaussian}
\end{equation}
Here $\left( p_{j},q_{j}\right) _{j=1}^{n}\in \mathcal{P}_{2n}$ is a pair
partition: the $p_{j}$ correspond to annihilators and the $q_{j}$ to
creators so we must have $p_{j}>q_{j}$ for each $j$; the ordering of the
pairs is unimportant so for definiteness we take $q_{n}>\cdots >q_{2}>q_{1}$%
. We may picture this as follows: for each $i\in \left\{ 1,2,\dots
,2n\right\} $ we have a vertex; with $B^{+}\left( f_{i}\right) $ we
associate a creator vertex with weight $f_{i}$ and with $B^{-}\left(
f_{i}\right) $ we associate an annihilator vertex with weight $f_{i}$. A
matched creation-annihilation pair $\left( p_{j},q_{j}\right) $\ is called a 
\emph{contraction} over creator vertex $q_{j}$ and annihilator vertex $p_{j}$
and corresponds to a multiplicative factor $\left\langle
f_{p_{j}}|f_{q_{j}}\right\rangle $ and is shown pictorially as a single line,

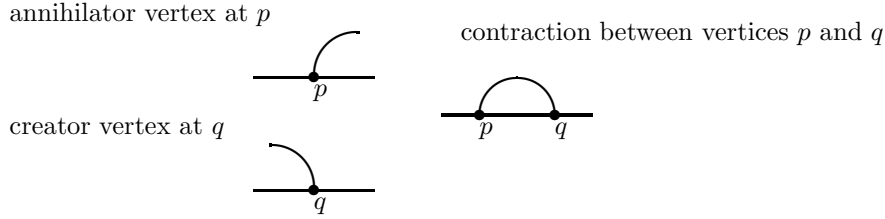
\begin{center}
%TCIMACRO{
%\TeXButton{picture1}{\setlength{\unitlength}{.05cm}
%\begin{picture}(120,60)
%\label{pic1} 
%\thicklines
%\put(10,5){\line(1,0){32}} 
%\put(26,5){\circle*{3}}
%\put(14,5){\oval(24,24)[tr]}
%\put(26,0){$q$}
%\put(-55,20){creator vertex at $q$}
%\put(10,35){\line(1,0){32}} 
%\put(26,35){\circle*{3}}
%\put(38,35){\oval(24,24)[tl]}
%\put(26,30){$p$}
%
%\put(-55,50){annihilator vertex at $p$}
%\put(60,25){\line(1,0){40}}
%\put(70,25){\circle*{3}}
%\put(90,25){\circle*{3}}
%\put(80,25){\oval(20,20)[t]}
%\put(65,45){contraction between vertices $p$ and $q$}
%\put(70,20){$p$}
%\put(90,20){$q$}
%\end{picture}
%}}%
%BeginExpansion
\setlength{\unitlength}{.05cm}
\begin{picture}(120,60)
\label{pic1} 
\thicklines
\put(10,5){\line(1,0){32}} 
\put(26,5){\circle*{3}}
\put(14,5){\oval(24,24)[tr]}
\put(26,0){$q$}
\put(-55,20){creator vertex at $q$}
\put(10,35){\line(1,0){32}} 
\put(26,35){\circle*{3}}
\put(38,35){\oval(24,24)[tl]}
\put(26,30){$p$}

\put(-55,50){annihilator vertex at $p$}
\put(60,25){\line(1,0){40}}
\put(70,25){\circle*{3}}
\put(90,25){\circle*{3}}
\put(80,25){\oval(20,20)[t]}
\put(65,45){contraction between vertices $p$ and $q$}
\put(70,20){$p$}
\put(90,20){$q$}
\end{picture}
%
%EndExpansion
\end{center}

We then consider a sum over all possible diagrams describing.

Setting all the $f_{j}$ equal to a fixed test function $f$ and let $Q\left(
f\right) =B^{+}\left( f\right) +B^{-}\left( f\right) $ then we obtain 
\begin{equation*}
\left\langle \Omega |\,Q\left( f\right) ^{k}\,\Omega \right\rangle
=\left\langle \Omega |\,\left[ B^{+}\left( f\right) +B^{-}\left( f\right) %
\right] ^{k}\,\Omega \right\rangle =\left\| f\right\| ^{k}\left| \mathcal{P}%
_{k}\right| \text{.}
\end{equation*}
The observable $Q\left( f\right) $ therefore has a mean-zero Gaussian
distribution in the Fock vacuum state: 
\begin{equation*}
\left\langle \Omega |\,e^{i\left[ B^{+}\left( f\right) +B^{-}\left( f\right) %
\right] }\,\Omega \right\rangle =e^{-\frac{1}{2}\left\| f\right\| ^{2}}.
\end{equation*}
For instance, we have $\left| \mathcal{P}_{4}\right| =\frac{4!}{2^{2}2!}=3$
and the three pair partitions $\left\{ \left( 4,3\right) \left( 2,1\right)
\right\} $, $\left\{ \left( 4,2\right) (3,1)\right\} $ and $\left\{ \left(
4,1\right) \left( 3,2\right) \right\} $ are pictured below.

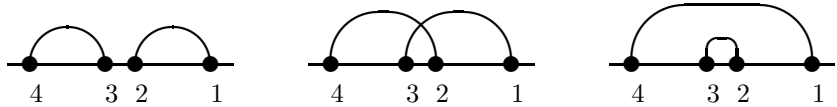
\begin{center}
%TCIMACRO{
%\TeXButton{picture2}{\setlength{\unitlength}{.1cm}
%\begin{picture}(120,20)
%\label{pic1} 
%\thicklines
%\put(5,10){\line(1,0){30}} 
%\put(8,10){\circle*{2}}
%\put(18,10){\circle*{2}}
%\put(22,10){\circle*{2}}
%\put(32,10){\circle*{2}}
%\put(8,5){4}
%\put(18,5){3}
%\put(22,5){2}
%\put(32,5){1}
%\put(13,10){\oval(10,10)[t]}
%\put(27,10){\oval(10,10)[t]}
%
%\put(45,10){\line(1,0){30}} 
%\put(48,10){\circle*{2}}
%\put(58,10){\circle*{2}}
%\put(62,10){\circle*{2}}
%\put(72,10){\circle*{2}}
%\put(48,5){4}
%\put(58,5){3}
%\put(62,5){2}
%\put(72,5){1}
%\put(55,10){\oval(14,14)[t]}
%\put(65,10){\oval(14,14)[t]}
%
%\put(85,10){\line(1,0){30}} 
%\put(88,10){\circle*{2}}
%\put(98,10){\circle*{2}}
%\put(102,10){\circle*{2}}
%\put(112,10){\circle*{2}}
%\put(88,5){4}
%\put(98,5){3}
%\put(102,5){2}
%\put(112,5){1}
%\put(100,10){\oval(4,7)[t]}
%\put(100,10){\oval(24,16)[t]}
%
%\end{picture}
%}}%
%BeginExpansion
\setlength{\unitlength}{.1cm}
\begin{picture}(120,20)
\label{pic1} 
\thicklines
\put(5,10){\line(1,0){30}} 
\put(8,10){\circle*{2}}
\put(18,10){\circle*{2}}
\put(22,10){\circle*{2}}
\put(32,10){\circle*{2}}
\put(8,5){4}
\put(18,5){3}
\put(22,5){2}
\put(32,5){1}
\put(13,10){\oval(10,10)[t]}
\put(27,10){\oval(10,10)[t]}

\put(45,10){\line(1,0){30}} 
\put(48,10){\circle*{2}}
\put(58,10){\circle*{2}}
\put(62,10){\circle*{2}}
\put(72,10){\circle*{2}}
\put(48,5){4}
\put(58,5){3}
\put(62,5){2}
\put(72,5){1}
\put(55,10){\oval(14,14)[t]}
\put(65,10){\oval(14,14)[t]}

\put(85,10){\line(1,0){30}} 
\put(88,10){\circle*{2}}
\put(98,10){\circle*{2}}
\put(102,10){\circle*{2}}
\put(112,10){\circle*{2}}
\put(88,5){4}
\put(98,5){3}
\put(102,5){2}
\put(112,5){1}
\put(100,10){\oval(4,7)[t]}
\put(100,10){\oval(24,16)[t]}

\end{picture}
%
%EndExpansion
\end{center}

We remark that (\ref{Gaussian}) is the basic Wick's theorem in quantum field
theory \cite{Glimm}, and can be realized in terms of Hafnians \cite
{Caianeillo}. However, the result as it applies to multinomial moments of
Gaussian variables goes back to Isserlis \cite{Isserlis}\ in 1918.

\subsection{Poissonian Fields}

More generally, we shall consider fields $B^{\pm }\left( .\right) $\ on some
Fock space on $\Gamma _{+}\left( \frak{h}\right) $. Consider the vacuum
average of an expression of the type 
\begin{equation*}
\left\langle \Omega |\,B^{+}\left( f_{n}\right) ^{\alpha \left( n\right)
}B^{-}\left( g_{n}\right) ^{\beta \left( n\right) }\cdots B^{+}\left(
f_{1}\right) ^{\alpha \left( 1\right) }B^{-}\left( g_{1}\right) ^{\beta
\left( 1\right) }\,\Omega \right\rangle
\end{equation*}
where the $\alpha _{j}$'s and $\beta _{j}$'s are powers taking the values$\
0 $ or $1$. This time, in the diagrammatic description, we have $n$ vertices
with each vertex being one of four possible types:

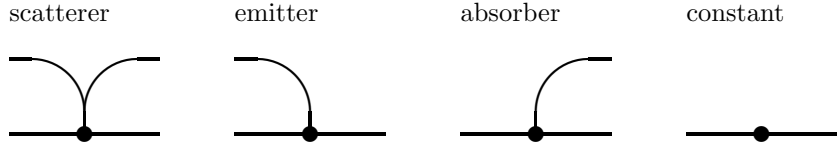
\begin{center}
%TCIMACRO{
%\TeXButton{picture3}{\setlength{\unitlength}{.1cm}
%\begin{picture}(120,30)
%\label{pic1} 
%\thicklines
%
%\put(5,10){\line(1,0){20}} 
%\put(15,10){\circle*{2}}
%\put(5,10){\oval(20,20)[tr]}
%\put(25,10){\oval(20,20)[tl]}
%\put(5,25){scatterer}
%
%\put(35,10){\line(1,0){20}} 
%\put(45,10){\circle*{2}}
%\put(35,10){\oval(20,20)[tr]}
%\put(35,25){emitter}
%
%\put(65,10){\line(1,0){20}} 
%\put(75,10){\circle*{2}}
%\put(85,10){\oval(20,20)[tl]}
%\put(65,25){absorber}
%
%
%\put(95,10){\line(1,0){20}} 
%\put(105,10){\circle*{2}}
%\put(95,25){constant}
%
%\end{picture}
%}}%
%BeginExpansion
\setlength{\unitlength}{.1cm}
\begin{picture}(120,30)
\label{pic1} 
\thicklines

\put(5,10){\line(1,0){20}} 
\put(15,10){\circle*{2}}
\put(5,10){\oval(20,20)[tr]}
\put(25,10){\oval(20,20)[tl]}
\put(5,25){scatterer}

\put(35,10){\line(1,0){20}} 
\put(45,10){\circle*{2}}
\put(35,10){\oval(20,20)[tr]}
\put(35,25){emitter}

\put(65,10){\line(1,0){20}} 
\put(75,10){\circle*{2}}
\put(85,10){\oval(20,20)[tl]}
\put(65,25){absorber}

\put(95,10){\line(1,0){20}} 
\put(105,10){\circle*{2}}
\put(95,25){constant}

\end{picture}
%
%EndExpansion
\end{center}

The typical situation is depicted below:

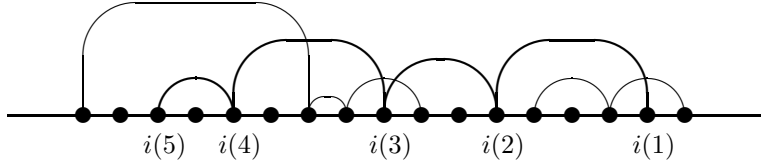
\begin{center}
%TCIMACRO{
%\TeXButton{picture4}{\setlength{\unitlength}{.1cm}
%\begin{picture}(120,30)
%\label{pic1} 
%\thicklines
%
%\put(10,10){\line(1,0){100}} 
%\put(20,10){\circle*{2}}
%\put(25,10){\circle*{2}}
%\put(30,10){\circle*{2}}
%\put(35,10){\circle*{2}}
%\put(40,10){\circle*{2}}
%\put(45,10){\circle*{2}}
%\put(50,10){\circle*{2}}
%\put(55,10){\circle*{2}}
%\put(60,10){\circle*{2}}
%\put(65,10){\circle*{2}}
%\put(70,10){\circle*{2}}
%\put(75,10){\circle*{2}}
%\put(80,10){\circle*{2}}
%\put(85,10){\circle*{2}}
%\put(90,10){\circle*{2}}
%\put(95,10){\circle*{2}}
%\put(100,10){\circle*{2}}
%
%\put(85,10){\oval(20,20)[t]}
%\put(67.5,10){\oval(15,15)[t]}
%\put(50,10){\oval(20,20)[t]}
%\put(35,10){\oval(10,10)[t]}
%
%\put(28,5){$i(5)$}
%\put(38,5){$i(4)$}
%\put(58,5){$i(3)$}
%\put(73,5){$i(2)$}
%\put(93,5){$i(1)$}
%
%
%\thinlines
%\put(35,10){\oval(30,30)[t]}
%\put(60,10){\oval(10,10)[t]}
%\put(95,10){\oval(10,10)[t]}
%\put(85,10){\oval(10,10)[t]}
%\put(52.5,10){\oval(5,5)[t]}
%
%
%\end{picture}
%}}%
%BeginExpansion
\setlength{\unitlength}{.1cm}
\begin{picture}(120,30)
\label{pic1} 
\thicklines

\put(10,10){\line(1,0){100}} 
\put(20,10){\circle*{2}}
\put(25,10){\circle*{2}}
\put(30,10){\circle*{2}}
\put(35,10){\circle*{2}}
\put(40,10){\circle*{2}}
\put(45,10){\circle*{2}}
\put(50,10){\circle*{2}}
\put(55,10){\circle*{2}}
\put(60,10){\circle*{2}}
\put(65,10){\circle*{2}}
\put(70,10){\circle*{2}}
\put(75,10){\circle*{2}}
\put(80,10){\circle*{2}}
\put(85,10){\circle*{2}}
\put(90,10){\circle*{2}}
\put(95,10){\circle*{2}}
\put(100,10){\circle*{2}}

\put(85,10){\oval(20,20)[t]}
\put(67.5,10){\oval(15,15)[t]}
\put(50,10){\oval(20,20)[t]}
\put(35,10){\oval(10,10)[t]}

\put(28,5){$i(5)$}
\put(38,5){$i(4)$}
\put(58,5){$i(3)$}
\put(73,5){$i(2)$}
\put(93,5){$i(1)$}

\thinlines
\put(35,10){\oval(30,30)[t]}
\put(60,10){\oval(10,10)[t]}
\put(95,10){\oval(10,10)[t]}
\put(85,10){\oval(10,10)[t]}
\put(52.5,10){\oval(5,5)[t]}

\end{picture}
%
%EndExpansion
\end{center}

Evidently we must again join up all creation and annihilation operators into
pairs; we however get creation, multiple scattering and annihilation as the
rule; otherwise we have a stand-alone constant term at a vertex. In the
figure, we can think of a particle being created at vertex $i\left( 1\right) 
$ then scattered at $i\left( 2\right) ,i\left( 3\right) ,i\left( 4\right) $
successively before being annihilated at $i\left( 5\right) $. (This
component has been highlighted using thick lines.) Now the argument: each
such component corresponds to a unique part, here $\left\{ i\left( 5\right)
,i\left( 4\right) ,i\left( 3\right) ,i\left( 2\right) ,i\left( 1\right)
\right\} $, having two or more elements; singletons may also occur and these
are just the constant term vertices. Therefore every such diagram
corresponds uniquely to a partition of $\left\{ 1,\dots ,n\right\} $. Once
this link is made, it is easy to see that 
\begin{eqnarray}
&&\sum_{\alpha ,\beta \in \left\{ 0,1\right\} ^{n}}\left\langle \Omega
|\,B^{+}\left( f_{n}\right) ^{\alpha \left( n\right) }B^{-}\left(
g_{n}\right) ^{\beta \left( n\right) }\cdots B^{+}\left( f_{1}\right)
^{\alpha \left( 1\right) }B^{-}\left( g_{1}\right) ^{\beta \left( 1\right)
}\,\Omega \right\rangle  \notag \\
&=&\sum_{\mathcal{A}\in \frak{P}_{n}}\;\prod_{\left\{ i\left( k\right)
>\cdots >i\left( 2\right) >i\left( 1\right) \right\} \in \mathcal{A}%
}\left\langle g_{i\left( k\right) }|f_{i\left( k-1\right) }\right\rangle
\cdots \left\langle g_{i\left( 3\right) }|f_{i\left( 2\right) }\right\rangle
\left\langle g_{i\left( 2\right) }|f_{i\left( 1\right) }\right\rangle .
\label{Poissonian}
\end{eqnarray}

If we now take all the $f_{j}$ and $g_{j}$ equal to a fixed $f$ then we
arrive at 
\begin{eqnarray*}
\left\langle \Omega |\,\left[ \left( B^{+}\left( f\right) +1\right) \left(
B^{-}\left( f\right) +1\right) \right] ^{n}\,\Omega \right\rangle
&=&\sum_{m=0}^{n}\sum_{\Gamma \in \frak{P}_{n,m}}\left\| f\right\| ^{n-m} \\
&=&\sum_{m=0}^{n}S\left( n,m\right) \left\| f\right\| ^{2(n-m)}.
\end{eqnarray*}
Note that a part of size $k$ contributes $\left\| f\right\| ^{2(k-1)}$ so a
partition in $\frak{P}_{n,m}$ with parts of size $k_{1},\dots k_{m}$
contributes $\left\| f\right\| 2^{(k_{1}+\cdots +k_{m}-m)}=\left\| f\right\|
^{2(n-m)}$.

It therefore follows that the observable 
\begin{equation}
N\left( f\right) :=\left( B^{+}\left( \frac{f}{\left\| f\right\| }\right)
+\left\| f\right\| \right) \left( B^{-}\left( \frac{f}{\left\| f\right\| }%
\right) +\left\| f\right\| \right)
\end{equation}
has a Poisson distribution of intensity $\left\| f\right\| ^{2}$ in the Fock
vacuum state.

\subsection{Exponentially Distributed Fields}

Is there a similar interpretation for Stirling numbers of the first kind as
well? Here we should be dealing with cycles within permutations rather than
parts in a partition. Consider the representation of a cycle $\left( i\left(
1\right) ,i\left( 2\right) ,\dots ,i\left( 6\right) \right) $ below:

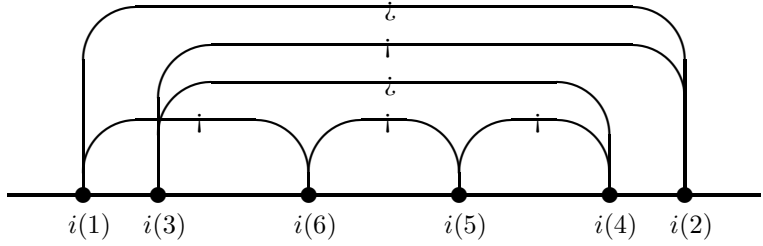
\begin{center}
%TCIMACRO{
%\TeXButton{picture5}{\setlength{\unitlength}{.1cm}
%\begin{picture}(120,40)
%\label{pic1} 
%\thicklines
%
%\put(10,10){\line(1,0){100}} 
%\put(20,10){\circle*{2}}
%\put(30,10){\circle*{2}}
%\put(50,10){\circle*{2}}
%\put(70,10){\circle*{2}}
%\put(90,10){\circle*{2}}
%\put(100,10){\circle*{2}}
%
%\put(18,5){$i(1)$}
%\put(28,5){$i(3)$}
%\put(48,5){$i(6)$}
%\put(68,5){$i(5)$}
%\put(88,5){$i(4)$}
%\put(98,5){$i(2)$}
%
%\put(60,10){\oval(20,20)[t]}
%\put(60,10){\oval(80,50)[t]}
%\put(60,10){\oval(60,30)[t]}
%\put(35,10){\oval(30,20)[t]}
%\put(65,10){\oval(70,40)[t]}
%\put(80,10){\oval(20,20)[t]}
%
%\put(60,19){<}
%\put(60,34){>}
%\put(60,29){<}
%\put(35,19){<}
%\put(60,24){>}
%\put(80,19){<}
%
%\end{picture}
%}}%
%BeginExpansion
\setlength{\unitlength}{.1cm}
\begin{picture}(120,40)
\label{pic1} 
\thicklines

\put(10,10){\line(1,0){100}} 
\put(20,10){\circle*{2}}
\put(30,10){\circle*{2}}
\put(50,10){\circle*{2}}
\put(70,10){\circle*{2}}
\put(90,10){\circle*{2}}
\put(100,10){\circle*{2}}

\put(18,5){$i(1)$}
\put(28,5){$i(3)$}
\put(48,5){$i(6)$}
\put(68,5){$i(5)$}
\put(88,5){$i(4)$}
\put(98,5){$i(2)$}

\put(60,10){\oval(20,20)[t]}
\put(60,10){\oval(80,50)[t]}
\put(60,10){\oval(60,30)[t]}
\put(35,10){\oval(30,20)[t]}
\put(65,10){\oval(70,40)[t]}
\put(80,10){\oval(20,20)[t]}

\put(60,19){<}
\put(60,34){>}
\put(60,29){<}
\put(35,19){<}
\put(60,24){>}
\put(80,19){<}

\end{picture}
%
%EndExpansion
\end{center}

To make the sense of the cycle clear, we are forced to use arrows and
therefore we have two types of lines. In any such diagrammatic
representation of a permutation we will encounter five types of vertex:

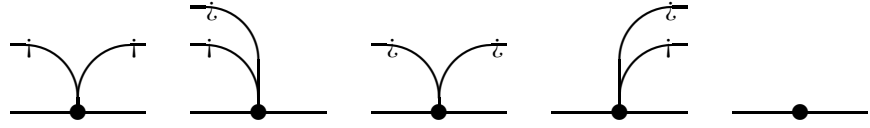
\begin{center}
%TCIMACRO{
%\TeXButton{picture6}{\setlength{\unitlength}{.1cm}
%\begin{picture}(120,30)
%\label{pic1} 
%\thicklines
%
%\put(3,10){\line(1,0){18}} 
%\put(27,10){\line(1,0){18}} 
%\put(51,10){\line(1,0){18}} 
%\put(75,10){\line(1,0){18}} 
%\put(99,10){\line(1,0){18}} 
%
%\put(12,10){\circle*{2}}
%\put(36,10){\circle*{2}}
%\put(60,10){\circle*{2}}
%\put(84,10){\circle*{2}}
%\put(108,10){\circle*{2}}
%
%\put(3,10){\oval(18,18)[tr]} 
%\put(21,10){\oval(18,18)[tl]}
%\put(5,18){<}
%\put(19,18){<}
%
%
%\put(27,10){\oval(18,18)[tr]} 
%\put(29,18){<}
%\put(27,10){\oval(18,28)[tr]} 
%\put(29,23){>}
%
%
%\put(51,10){\oval(18,18)[tr]} 
%\put(69,10){\oval(18,18)[tl]}
%\put(53,18){>}
%\put(67,18){>}
%
%\put(93,10){\oval(18,18)[tl]} 
%\put(90,18){<}
%\put(93,10){\oval(18,28)[tl]} 
%\put(90,23){>}
%
%
%\end{picture}
%}}%
%BeginExpansion
\setlength{\unitlength}{.1cm}
\begin{picture}(120,30)
\label{pic1} 
\thicklines

\put(3,10){\line(1,0){18}} 
\put(27,10){\line(1,0){18}} 
\put(51,10){\line(1,0){18}} 
\put(75,10){\line(1,0){18}} 
\put(99,10){\line(1,0){18}} 

\put(12,10){\circle*{2}}
\put(36,10){\circle*{2}}
\put(60,10){\circle*{2}}
\put(84,10){\circle*{2}}
\put(108,10){\circle*{2}}

\put(3,10){\oval(18,18)[tr]} 
\put(21,10){\oval(18,18)[tl]}
\put(5,18){<}
\put(19,18){<}

\put(27,10){\oval(18,18)[tr]} 
\put(29,18){<}
\put(27,10){\oval(18,28)[tr]} 
\put(29,23){>}

\put(51,10){\oval(18,18)[tr]} 
\put(69,10){\oval(18,18)[tl]}
\put(53,18){>}
\put(67,18){>}

\put(93,10){\oval(18,18)[tl]} 
\put(90,18){<}
\put(93,10){\oval(18,28)[tl]} 
\put(90,23){>}

\end{picture}
%
%EndExpansion
\end{center}

An uncontracted (constant) vertex indicates a fixed point for the
permutation.

Let us consider the one-dimensional case first. The above suggests that we
should use two independent (that is, commuting) Bose variables, say $%
b_{1}^{\pm },b_{2}^{\pm }$. Let us set 
\begin{equation*}
b^{\pm }f:=b_{1}^{\pm }\otimes 1_{2}+1_{1}\otimes b_{2}^{\mp }
\end{equation*}
then we see that $b^{+}$ will commute with $b^{-}$. We note that 
\begin{equation*}
b^{+}b^{-}=b_{1}^{+}b_{1}^{-}\otimes 1_{2}+1_{2}\otimes
b_{2}^{+}b_{2}^{-}+b_{1}^{-}\otimes b_{2}^{-}+b_{1}^{+}\otimes b_{2}^{+}+1
\end{equation*}
and here we see the five vertex terms we need.

\bigskip

Let $\Omega _{1}$ and $\Omega _{2}$ be the vacuum state for $b_{1}^{\pm }$
and $b_{2}^{\pm }$ respectively, then let $\Omega =\Omega _{1}\otimes \Omega
_{2}$ be the joint vacuum state. We wish to show that $N$ has a Gamma
distribution of unit power (an Exponential distribution!) in this state.

First of all, let $Q=b^{+}+b^{-}$ so that $Q=Q_{1}\otimes 1_{2}+1_{1}\otimes
Q_{2}$ where $Q_{j}=b_{j}^{+}+b_{j}^{-}$. Now each $Q_{j}$ has a standard
Gaussian distribution in the corresponding vacuum state $\left( \left\langle
\Omega _{j}|\,e^{tQ_{j}}\,\Omega _{j}\right\rangle =e^{t^{2}/2}\right) $and
so 
\begin{equation*}
\left\langle \Omega |\,e^{tQ}\,\Omega \right\rangle =\left\langle \Omega
_{1}|\,e^{tQ_{1}}\,\Omega _{1}\right\rangle \left\langle \Omega
_{2}|\,e^{tQ_{2}}\,\Omega _{2}\right\rangle =e^{t^{2}}.
\end{equation*}
Therefore $\left\langle \Omega |\,Q^{2n}\,\Omega \right\rangle =2^{n}\left| 
\mathcal{P}_{2n}\right| =\frac{\left( 2n\right) !}{n!}$. However $%
Q^{2n}=\sum_{m}\binom{2n}{m}\left( b^{+}\right) ^{m}\left( b^{-}\right)
^{2n-m}$ (remember that $b^{+}$ and $b^{-}$ commute!) and so $\left\langle
\Omega |\,Q^{2n}\,\Omega \right\rangle \equiv \binom{2n}{n}\left\langle
\Omega |\,\left( b^{+}\right) ^{n}\left( b^{-}\right) ^{n}\,\Omega
\right\rangle $ which should equal $\binom{2n}{n}\left\langle \Omega
|\,\left( b^{+}b^{-}\right) ^{n}\,\Omega \right\rangle $. Therefore 
\begin{equation*}
\left\langle \Omega |\,\left( b^{+}b^{-}\right) ^{n}\,\Omega \right\rangle
=n!
\end{equation*}
and so, for $t<1$, 
\begin{equation*}
\left\langle \Omega |\,\exp \left\{ tb^{+}b^{-}\right\} \,\Omega
\right\rangle =\frac{1}{1-t}.
\end{equation*}
Therefore $N=b^{+}b^{-}$ has an exponential distribution in the joint vacuum
state.

\bigskip

The generalization of this result to Bosonic fields over a Hilbert space $%
\frak{h}$\ is straightforward enough. First we need the notion of a
conjugation map on Hilbert spaces: this is a conjugate-linear map $j:\frak{h}%
\mapsto \frak{h}$. For instance, let $\left\{ e_{n}\right\} _{n}$ be a fixed
complete basis for $\frak{h}$ then an example of a conjugation is given by $%
j\left( \sum_{n}c_{n}e_{n}\right) =\sum_{n}c_{n}^{\ast }e_{n}$.

\bigskip

\begin{theorem}
Let $\frak{h}$ be a separable Hilbert space and $j$\ an conjugation on $%
\frak{h}$. Let $B^{\pm }\left( .\right) $\ be operator fields define on $%
\Gamma _{+}\left( \frak{h}\right) \otimes \Gamma _{+}\left( \frak{h}\right) $%
\ by 
\begin{equation}
B^{\pm }\left( f\right) =B_{1}^{\pm }\left( f\right) \otimes
1_{2}+1_{1}\otimes B_{2}^{\mp }\left( jf\right)
\end{equation}
where the $B_{i}^{\pm }\left( .\right) $\ are usual Bosonic fields on the
factors $\Gamma _{+}\left( \frak{h}\right) $, $i\left( =1,2\right) $. Then $%
\left[ B^{-}\left( f\right) ,B^{+}\left( g\right) \right] =0$\ for all\ $%
f,g\in h$\ and if $N\left( f,g\right) :=B^{+}\left( f\right) B^{-}\left(
g\right) $\ then we have the following expectations in the joint Fock vacuum
state $\Omega =\Omega _{1}\otimes \Omega _{2}:$%
\begin{eqnarray}
&&\left\langle \Omega |\,N\left( f_{n},g_{n}\right) \cdots N\left(
f_{1},g_{1}\right) \,\Omega \right\rangle  \notag \\
&=&\sum_{\sigma \in \frak{S}_{n}}\;\prod_{\left( i\left( 1\right) ,i\left(
2\right) ,\dots ,i\left( k\right) \right) \in \sigma }\left\langle
g_{i\left( k\right) }|f_{i\left( k-1\right) }\right\rangle \cdots
\left\langle g_{i\left( 2\right) }|f_{i\left( 1\right) }\right\rangle \times
\left\langle g_{i\left( 1\right) }|f_{i\left( k\right) }\right\rangle
\label{Exponential}
\end{eqnarray}
\end{theorem}

The proof should be obvious from our discussions above so we omit it. The
sum is over all permutations $\sigma \in \frak{S}_{n}$ and each permutation
is decomposed into its cycles; the product is then over all cycles $\left(
i\left( 1\right) ,i\left( 2\right) ,\dots ,i\left( k\right) \right) $ making
up a particular permutation. We note that the representation corresponds to
a type of infinite dimensional limit to the double-Fock representation for
thermal states \cite{br}.

\subsection{Thermal Fields}

There is a well known trick for representing thermal states of Bose systems.
To begin with, let $\frak{h}=L^{2}\left( \mathbb{R}^{3}\right) $ be the
space of momentum-representation wave functions, then the thermal state $%
\left\langle \,\cdot \,\right\rangle _{\beta ,\mu }$ is characterized as the
Gaussian (sometimes referred to as quasi-free), mean zero state with 
\begin{equation*}
\left\langle \,B^{+}\left( f\right) B^{-}\left( f\right) \,\right\rangle
_{\beta ,\mu }=\int |f\left( k\right) |^{2}\varrho _{\beta ,\mu }\left(
k\right) d^{3}k
\end{equation*}
where 
\begin{equation*}
\varrho _{\beta ,\mu }\left( k\right) =\left( e^{\beta \left[ E\left(
k\right) -\mu \right] }-1\right) ^{-1}.
\end{equation*}
(We have the physical interpretation of $\beta $ as inverse temperature, $%
\mu $ as chemical potential, and $E\left( k\right) $ the energy spectrum
function.) Let us denote by $\varrho _{\beta ,\mu }$ the operation of
pointwise multiplication by the function $\varrho _{\beta ,\mu }\left( \cdot
\right) $ on $\frak{h}$, that is, $\left( \varrho _{\beta ,\mu }f\right)
\left( k\right) =\varrho _{\beta ,\mu }\left( k\right) f\left( k\right) $.
We check that $\left\langle \,\left[ B^{+}\left( f\right) +B^{-}\left(
f\right) \right] ^{2}\,\right\rangle _{\beta ,\mu }=\left\langle f|C_{\beta
,\mu }\,f\right\rangle $ with $C_{\beta ,\mu }=2\varrho _{\beta ,\mu
}+1\equiv \coth \dfrac{\beta \left[ E-\mu \right] }{2}$.

\begin{theorem}
Let $\frak{h}$ be a separable Hilbert space and $j$\ an conjugation on $%
\frak{h}$. Let $B^{\pm }\left( .\right) $\ be operator fields define on $%
\Gamma _{+}\left( \frak{h}\right) \otimes \Gamma _{+}\left( \frak{h}\right) $%
\ by 
\begin{equation}
B^{\pm }\left( f\right) =B_{1}^{\pm }\left( \sqrt{\varrho +1}f\right)
\otimes 1_{2}+1_{1}\otimes B_{2}^{\mp }\left( j\sqrt{\varrho }f\right)
\end{equation}
where the $B_{i}^{\pm }\left( .\right) $\ are usual Bosonic fields on the
factors $\Gamma _{+}\left( \frak{h}\right) $, $i\left( =1,2\right) $, and $%
\varrho $ is a positive operator on $\frak{h}$. Then the fields satisfy the
canonical commutation relations $\left( \ref{CCR}\right) $ and their moments
in the joint Fock vacuum state $\Omega =\Omega _{1}\otimes \Omega _{2}$ are
precisely the same as for the thermal state when we take $\varrho =\varrho
_{\beta ,\mu }$.
\end{theorem}

\bigskip

To see this, note that 
\begin{eqnarray*}
\left\langle \Omega |\,e^{i\left[ B^{+}\left( f\right) +B^{-}\left( f\right) %
\right] }\,\Omega \right\rangle &=&\left\langle \Omega _{1}|\,e^{i\left[
B_{1}^{+}\left( f\right) +B_{1}^{-}\left( f\right) \right] }\,\Omega
_{1}\right\rangle \left\langle \Omega _{2}|\,e^{i\left[ B_{2}^{+}\left(
f\right) +B_{2}^{-}\left( f\right) \right] }\,\Omega _{2}\right\rangle \\
&=&\exp \left\{ -\frac{1}{2}\left\langle f|\left( \varrho +1\right)
f\right\rangle \right\} \exp \left\{ -\frac{1}{2}\left\langle f|\left(
\varrho +1\right) f\right\rangle \right\} \\
&=&\exp \left\{ -\frac{1}{2}\left\langle f|Cf\right\rangle \right\}
\end{eqnarray*}
where $C=2\varrho +1$ is the covariance of the state. The first factor
describes the so-called spontaneous emissions and absorptions while the
second factor describes the so-called stimulated emissions and absorptions.
In the zero temperature limit, $\beta \rightarrow \infty $, we find that $%
\varrho \rightarrow 0$ and so the second factor fields (stimulated emissions
and absorptions) become negligible.

\chapter{Field Theory}

\section{Introduction}

Let us suppose that we have a fixed space $\Phi $ of functions over a set $%
\Lambda $. A function $\varphi =\left\{ \varphi _{x}:x\in \Lambda \right\} $
in $\Phi $ will be called a \textit{field realization}. By a \textit{field} $%
\phi $ we mean an observable taking values in $\Phi $ (that is to say, the
field realization are ``eigen-values'' for the field) and we suppose that we
are supplied with state which we shall denote as $\left\langle \,\cdot
\,\right\rangle $. Formally, we may think of the field as a family of
operators $\phi =\left\{ \phi _{x}:x\in \Lambda \right\} $ and we suppose
that $\left[ \phi _{x},\phi _{y}\right] =0$ for all $x,y\in \Lambda $. Since
we are assuming that the field is commutative, we may in fact revert to
classical probability and think of the field as being a random variable
taking values in $\Phi $. The expectation of a (typically nonlinear)
functional $F=F\left[ \phi \right] $ of the field is to be understood as an
integral over $\Phi $: 
\begin{equation}
\left\langle F\left[ \phi \right] \right\rangle =\int_{\Phi }F\left[ \varphi %
\right] \,d\nu \left( \varphi \right) ,  \label{field expectation}
\end{equation}
where $\nu $ is the probability measure over $\Phi $ corresponding to the
state. We shall frequently use the notation $\left\langle \,\cdot
\,\right\rangle =\left\langle \,\cdot \,\right\rangle _{\nu }$ when we want
to emphasize the correspondence.

The label $x\in \Lambda $ is assumed to give all relevant information such
as position, spin, etc. When $\Lambda $ is a finite set, then the
mathematical treatment is straightforward. Otherwise, we find ourselves
having to resort to infinite dimensional analysis. We also introduce a dual
space $\frak{J}$ of fields $J=\left\{ J^{x}:x\in \Lambda \right\} $ which we
call the \textit{source fields}. Our convention will be that the source
fields carry a `contravariant' index while the field, and its realizations,
carry a `covariant' index. The duality between fields and sources will be
written as 
\begin{equation*}
\left\langle \varphi ,J\right\rangle =\varphi _{x}J^{x}.
\end{equation*}
In the case where $\Lambda $ is a finite set, say of cardinality $N$, a
realization $\varphi \in \Phi $ can be viewed as a set of $N$ numbers and so
we can identify $\Phi $, and likewise $\frak{J}$, as $\mathbb{R}^{N}$. In
this case $\varphi _{x}J^{x}$ means $\sum_{x\in \Lambda }\varphi _{x}J^{x}$
and so the notation just implies an Einstein summation convention over $%
\Lambda $. If $d\varphi _{x}$ denotes standard Lebesgue measure, and let $S%
\left[ \cdot \right] $ be some functional, called the action, such that $\Xi
=\int_{\Phi }\exp \left\{ S\left[ \varphi \right] \right\} d\varphi <\infty $%
, then, we have the finite-dimensional probability measure, $\nu $, on $\Phi 
$ determined by 
\begin{equation*}
d\nu \left( \varphi \right) =\frac{1}{\Xi }e^{S\left[ \varphi \right]
}\,\prod_{x\in \Lambda }d\varphi _{x}.
\end{equation*}

The general situation which we are really interested is where $\Lambda $ is
continuous. If we want $\Lambda =\mathbb{R}^{d}$, then we should take $\frak{%
J}$ to be the space of Schwartz functions on $\mathbb{R}^{d}$ and $\Phi $ to
be the tempered distributions. (The field realizations being more singular
than the sources!) Here, the duality is denoted is $\left\langle \varphi
,J\right\rangle =\int_{\mathbb{R}^{d}}\varphi _{x}J^{x}dx$ and we again
shorten to just $\varphi _{x}J^{x}$. We shall therefore adopt an Einstein
summation/integration convention over all indices from now on. The
appropriate way to consider randomizing the field in the
infinite-dimensional case, will then be to consider probability measures on
the Borel sets of $\frak{J}$, that is, on the $\sigma $-algebra generated by
the weak topology of the Schwartz functions\footnote{%
At this stage we shall stop with the functional analysis details and proceed
as if every thing is well-defined. The mathematically inclined reader can
fill in the details, while everyone else could well live without them.}.

\section{Field Calculus}

In this section, I want to present a way of handling functional derivatives
exploiting the commutativity of multiple point derivatives. The notation
that results has similarities to one introduced by Guichardet \cite{Gui} for
investigating Bosonic Fock spaces. We shall denote by $\dfrac{\delta }{%
\delta J^{x}}$ the functional derivative wrt. $J^{x}$. For $X$ a finite
subset of $\Lambda $ we shall adopt the notations 
\begin{equation*}
J^{X}=\prod_{x\in X}J^{x},\qquad \dfrac{\delta ^{|X|}}{\delta J^{X}}%
=\prod_{x\in X}\frac{\delta }{\delta J^{x}}.
\end{equation*}
(In both cases, we need only the set $X$ - the ordering of the elements is
irrelevant!) We of course have $\dfrac{\delta }{\delta J^{x}}\left(
J^{y}\right) =\delta _{x}^{y}$. Similarly, we shall write $\varphi _{X}$ for 
$\prod_{x\in \Lambda }\varphi _{x}$, and $\dfrac{\delta ^{|X|}}{\delta
\varphi _{X}}=\prod_{x\in X}\dfrac{\delta }{\delta \varphi _{x}}$.

\subsection{Analytic Functionals}

A multi-linear map $T:\times ^{n}\frak{J}\mapsto \mathbb{C}$ is called a
tensor of covariant rank $n$ and it will be determined by the components $%
T_{x_{1}\cdots x_{n}}$ such that $T\left( J_{\left( 1\right) },\cdots
,J_{\left( n\right) }\right) =T_{x_{1}\cdots x_{n}}\,J_{\left( 1\right)
}^{x_{1}}\cdots J_{\left( n\right) }^{x_{n}}$. Likewise, we refer to a
multilinear map from $\times ^{n}\Phi $ to the complex numbers as a tensor
of contravariant rank $n$.

A functional $F=F\left[ J\right] $ is said to be \textit{analytic} in $J$ if
it admits a series expansion of the form $F\left[ J\right] =\sum_{n\geq 0}%
\frac{1}{n!}f_{x_{1}\cdots x_{n}}\,J^{x_{1}}\cdots J^{x_{n}}$ where $%
f_{x_{1}\cdots x_{n}}\ $are the components of a completely symmetric
covariant tensor and as usual the repeated dummy indices are
summed/integrated over. A more compact notation is to write the series
expansion as 
\begin{equation*}
F\left[ J\right] =\sum_{X}\frac{1}{|X|!}f_{X}J^{X}.
\end{equation*}
It is easy to see that $\dfrac{\delta }{\delta J^{y_{1}}}\cdots \dfrac{%
\delta }{\delta J^{y_{m}}}F\left[ J\right] =\sum_{n\geq 0}\frac{1}{n!}%
f_{y_{1}\cdots y_{m}x_{1}\cdots x_{n}}J^{x_{1}}\cdots J^{x_{n}}$ and this
now reads as 
\begin{equation*}
\dfrac{\delta ^{|Y|}}{\delta J^{Y}}\left( \sum_{X}\frac{1}{|X|!}%
f_{X}J^{X}\right) =\sum_{X}\frac{1}{|X|!}f_{Y\cup X}J^{X}.
\end{equation*}

\subsection{The Leibniz rule}

Functional derivatives obey all the standard rules of calculus, including
the Leibniz rule and we have the natural extension to multiple derivatives: 
\begin{equation}
\frac{\delta ^{|X|}}{\delta J^{X}}\left( FG\right) =\sum_{Y\subseteq X}\frac{%
\delta ^{|Y|}F}{\delta J^{Y}}\,\frac{\delta ^{|X/Y|}G}{\delta J^{X/Y}}.
\label{Leibniz rule}
\end{equation}
The generalization of this for several factors is 
\begin{equation*}
\frac{\delta ^{|X|}}{\delta J^{X}}\left( F_{1}\cdots F_{m}\right)
=\sum\nolimits_{Y_{1},\cdots ,Y_{m}\subseteq X}^{^{\prime }}\frac{\delta
^{|Y_{1}|}F_{1}}{\delta J^{Y_{1}}}\cdots \frac{\delta ^{|Y_{m}|}F_{m}}{%
\delta J^{Y_{m}}}
\end{equation*}
where the sum is over all partitions of $X$ into $m$ labelled parts (as
opposed to the unlabeled ones we have considered up to now).

\subsection{Differential Formulas}

We now derive a useful result.

\begin{lemma}
Let $W=W\left[ J\right] $ then 
\begin{equation}
\frac{\delta ^{|X|}}{\delta J^{X}}e^{W}=e^{W}\sum_{\mathcal{A}\in \frak{P}%
\left( X\right) }\prod_{A\in \mathcal{A}}\frac{\delta ^{\left| A\right| }W}{%
\delta J^{A}}.  \label{diff exp W}
\end{equation}
\end{lemma}

\begin{proof}
This is easily seen by induction on $n$. As $\dfrac{\delta }{\delta J^{x}}%
e^{W}=e^{W}\dfrac{\delta W}{\delta J^{x}}$, the identity is true for $n=1$.
Now assume that it is true for $n$, then 
\begin{equation*}
\dfrac{\delta }{\delta J^{x}}\frac{\delta ^{|X|}}{\delta J^{X}}e^{W}=e^{W}%
\dfrac{\delta W}{\delta J^{x}}\sum_{\mathcal{A}\in \frak{P}\left( X\right)
}\prod_{A\in \mathcal{A}}\frac{\delta ^{\left| A\right| }W}{\delta J^{A}}%
+e^{W}\dfrac{\delta W}{\delta J^{x}}\sum_{\mathcal{A}\in \frak{P}\left(
X\right) }\dfrac{\delta }{\delta J^{x}}\prod_{A\in \mathcal{A}}\frac{\delta
^{\left| A\right| }W}{\delta J^{A}}
\end{equation*}
however, the first term on the right hand side is a sum over all parts of $%
X\cup \left\{ x\right\} $ having $x$ occurring as a singleton, while the
second term, when differentiated wrt. $J^{x}$, will be a sum over all parts
of $X\cup \left\{ x\right\} $ having $x$ in some part containing at least
one element of $X$. Thus we may write the above as 
\begin{equation*}
\frac{\delta ^{|X|+1}}{\delta J^{X\cup \left\{ x\right\} }}e^{W}=e^{W}\sum_{%
\mathcal{A}\in \frak{P}\left( X\cup \left\{ x\right\} \right) }\prod_{A\in 
\mathcal{A}}\frac{\delta ^{\left| A\right| }W}{\delta J^{A}}.
\end{equation*}
The identity the follows by induction.
\end{proof}

\subsection{Legendre Transforms}

Suppose that $W=W\left[ J\right] $ is a convex analytic function. That is, $%
W $ is a real-valued analytic functional with the property that 
\begin{equation}
W\left[ tJ_{1}+\left( 1-t\right) J_{2}\right] \leq tW\left[ J_{1}\right]
+\left( 1-t\right) W\left[ J_{2}\right]  \label{convex}
\end{equation}
for all $0<t<1$ and $J_{1},J_{2}\in \frak{J}$. The Legendre-Frenchel
transform of $W$ is then defined by 
\begin{equation}
\Gamma \left[ \varphi \right] =\inf_{J\in \frak{J}}\left\{ W\left[ J\right]
-\left\langle \varphi ,J\right\rangle \right\} .  \label{Legendre}
\end{equation}
$\Gamma \left[ \varphi \right] $ will then be a concave (i.e. $-\Gamma $ is
convex) analytic functional in $\varphi $ and we may invert the formula as
follows: 
\begin{equation}
W\left[ J\right] =\sup_{\varphi \in \Phi }\left\{ \Gamma \left[ \varphi %
\right] +\left\langle \varphi ,J\right\rangle \right\} .
\label{Legendre inverted}
\end{equation}

If the functional $W$ is taken to be strictly convex, that is, if we have
strict inequality in $\left( \ref{convex}\right) $, then the infimum is
attained at a unique source $\bar{J}=\bar{J}\left[ \varphi \right] $ for
each fixed $\varphi $, and so $\Gamma \left[ \varphi \right] =W\left[ \bar{J}%
\left[ \varphi \right] \right] -\left\langle \varphi ,\bar{J}\left[ \varphi %
\right] \right\rangle $. Moreover, we may invert $\bar{J}:\Phi \mapsto \frak{%
J}$ to get a mapping $\bar{\phi}:\frak{J}\mapsto \Phi $, and for fixed $J$
the supremum is given by $\bar{\phi}\left[ J\right] $ and so $W\left[ J%
\right] =\Gamma \left[ \bar{\phi}\left[ J\right] \right] +\left\langle \bar{%
\phi}\left[ J\right] ,J\right\rangle $. The extremal conditions are then 
\begin{equation*}
\bar{J}^{x}\equiv -\frac{\delta \Gamma }{\delta \varphi _{x}},\qquad \bar{%
\phi}_{x}\equiv \frac{\delta W}{\delta J^{x}}.
\end{equation*}

Let $W^{\prime \prime }\left[ J\right] $ be the symmetric tensor with
entries $W_{xy}^{\prime \prime }\left[ J\right] =\dfrac{\delta ^{2}W}{\delta
J^{x}\delta J^{y}}$. This will be positive definite - it will be interpreted
below as the covariance of the field in the presence of the source $J$.
Likewise, if we let $\Gamma ^{\prime \prime }\left[ \varphi \right] $ be the
linear operator with entries $\dfrac{\delta ^{2}\Gamma }{\delta \varphi
_{x}\delta \varphi _{y}}$, then we have 
\begin{eqnarray*}
W_{xy}^{\prime \prime }\left[ J\right] &=&\dfrac{\delta \bar{\phi}_{x}}{%
\delta J^{y}}, \\
\Gamma ^{\prime \prime xy}\left[ \varphi \right] &=&-\dfrac{\delta \bar{J}%
^{y}}{\delta \varphi _{x}},
\end{eqnarray*}
and so we conclude that $W^{\prime \prime }\left[ J\right] $ and $-\bar{%
\Gamma}^{\prime \prime }\left[ J\right] =\Gamma ^{\prime \prime }\left[ \bar{%
\phi}\left[ J\right] \right] $ will be inverses for each other. In other
words, 
\begin{equation}
\dfrac{\delta ^{2}W}{\delta J^{x}\delta J^{y}}\left. \dfrac{\delta
^{2}\Gamma }{\delta \varphi _{y}\delta \varphi _{z}}\right| _{\varphi =\bar{%
\phi}\left[ J\right] }=-\delta _{x}^{z}.  \label{WGamma}
\end{equation}

\begin{lemma}
Let $F:\Phi \mapsto \mathbb{R}$ be a functional and suppose $\bar{F}:\frak{J}%
\mapsto \mathbb{R}$ is then given by $\bar{F}\left[ J\right] :=F\left[ \bar{%
\phi}\left( J\right) \right] $ then 
\begin{equation*}
\frac{\delta \bar{F}}{\delta J^{x}}=\frac{\delta ^{2}W}{\delta J^{x}\delta
J^{y}}\,\left. \frac{\delta F}{\delta \varphi }\right| _{\bar{\phi}\left[ J%
\right] }
\end{equation*}
\end{lemma}

\begin{proof}
This is just the chain rule, as $\dfrac{\delta \bar{\phi}^{y}}{\delta J^{x}}=%
\dfrac{\delta ^{2}W}{\delta J^{x}\delta J^{y}}$.
\end{proof}

\bigskip

\begin{lemma}
\textit{Let us introduce the tensor coefficients} 
\begin{equation}
\bar{\Gamma}^{X}\left[ J\right] :=\left. \frac{\delta ^{|X|}\Gamma }{\delta
\varphi _{X}}\right| _{\bar{\phi}\left[ J\right] },
\end{equation}
then \textit{they satisfy the differential equations} 
\begin{equation}
\frac{\delta ^{|Y|}}{\delta J^{Y}}\bar{\Gamma}^{X}=\sum_{\mathcal{A}\in 
\frak{P}\left( Y\right) }\bar{\Gamma}^{X\cup Z_{\mathcal{A}}}\left(
\prod_{A\in \mathcal{A}}\frac{\delta ^{|A|+1}W}{\delta J^{A\cup \left\{
z_{A}\right\} }}\right) ,  \label{diff Gamma}
\end{equation}
\textit{where each }$z_{A}$\textit{\ is a dummy variable associated with
each component part }$A$\textit{\ and }$Z_{\mathcal{A}}=\left\{ z_{A}:A\in 
\mathcal{A}\right\} $\textit{. There is, as usual, an implied contraction
over the repeated }$z_{A}$'s in $\bar{\Gamma}^{X\cup Z_{\mathcal{A}}}$\ and
the $\dfrac{\delta ^{|A|+1}W}{\delta J^{A\cup \left\{ z_{A}\right\} }}$.
\end{lemma}

\begin{proof}
We shall prove this inductively. First note that for $F=\Gamma ^{X}$, we
have $\dfrac{\delta }{\delta J^{y}}\bar{\Gamma}^{X}=\dfrac{\delta ^{2}W}{%
\delta J^{y}\delta J^{z}}\,\bar{\Gamma}^{X\cup \left\{ z\right\} }$ by the
previous lemma. Assuming that the relation $\left( \ref{diff Gamma}\right) $
is true for a set $Y$, let $y$ be a free index then 
\begin{gather*}
\frac{\delta }{\delta J^{y}}\frac{\delta ^{|Y|}}{\delta J^{Y}}\bar{\Gamma}%
^{X}=\frac{\delta }{\delta J^{y}}\sum_{\mathcal{A}\in \frak{P}\left(
Y\right) }\bar{\Gamma}^{X\cup Z_{\mathcal{A}}}\left( \prod_{A\in \mathcal{A}}%
\frac{\delta ^{|A|+1}W}{\delta J^{A\cup \left\{ z_{A}\right\} }}\right) \\
=\sum_{\mathcal{A}\in \frak{P}\left( Y\right) }\left\{ \bar{\Gamma}^{X\cup
Z_{\mathcal{A}}}\frac{\delta }{\delta J^{y}}\left( \prod_{A\in \mathcal{A}}%
\frac{\delta ^{|A|+1}W}{\delta J^{A\cup \left\{ z_{A}\right\} }}\right) +%
\bar{\Gamma}^{X\cup Z_{\mathcal{A}}\cup \left\{ z\right\} }\frac{\delta ^{2}W%
}{\delta J^{\left\{ y,z\right\} }}\left( \prod_{A\in \mathcal{A}}\frac{%
\delta ^{|A|+1}W}{\delta J^{A\cup \left\{ z_{A}\right\} }}\right) \right\} \\
\equiv \sum_{\mathcal{A}\in \frak{P}\left( Y\cup \left\{ y\right\} \right) }%
\bar{\Gamma}^{X\cup Z_{\mathcal{A}}}\left( \prod_{A\in \mathcal{A}}\frac{%
\delta ^{|A|+1}W}{\delta J^{A\cup \left\{ z_{A}\right\} }}\right)
\end{gather*}
and so the relation holds for $Y\cup \left\{ y\right\} $.
\end{proof}

\bigskip

\begin{theorem}
We have the followi\textit{ng recurrence relation} 
\begin{equation}
\frac{\delta ^{|X|+1}W}{\delta J^{X\cup \left\{ y\right\} }}=\sum\limits_{%
\mathcal{A}\in \frak{P}^{f}\left( X\right) }\frac{\delta ^{2}W}{\delta
J^{y}\delta J^{p}}\,\bar{\Gamma}^{\left\{ p\right\} \cup Z_{\mathcal{A}%
}}\,\left( \prod_{A\in \mathcal{A}}\frac{\delta ^{|A|+1}W}{\delta J^{A\cup
\left\{ z_{A}\right\} }}\right) ,  \label{sdag}
\end{equation}
\textit{where, again, each }$z_{A}$\textit{\ is a dummy variable associated
with each component part }$A$\textit{\ and }$Z_{\mathcal{A}}=\left\{
z_{A}:A\in \mathcal{A}\right\} $\textit{. }
\end{theorem}

\begin{proof}
Taking $\dfrac{\delta ^{|X|}}{\delta J^{X}}$ of $\left( \ref{WGamma}\right) $
and using the multi-derivative form of the Leibniz rule, we find 
\begin{eqnarray*}
0 &=&\dfrac{\delta ^{|X|}}{\delta J^{X}}\left( \frac{\delta ^{2}W}{\delta
J^{x}\delta J^{y}}\,\bar{\Gamma}^{\left\{ y,z\right\} }\right) \\
&=&\sum_{Y\subseteq X}\frac{\delta ^{|Y|+2}W}{\delta J^{Y\cup \left\{
x,y\right\} }}\,\frac{\delta ^{|X/Y|}}{\delta J^{X/Y}}\bar{\Gamma}^{\left\{
y,z\right\} } \\
&=&\sum_{Y\subseteq X}\frac{\delta ^{|Y|+2}W}{\delta J^{Y\cup \left\{
x,y\right\} }}\sum_{\mathcal{A}\in \frak{P}\left( X/Y\right) }\bar{\Gamma}%
^{\left\{ y,z\right\} \cup Z_{\mathcal{A}}}\left( \prod_{A\in \mathcal{A}}%
\frac{\delta ^{|A|+1}W}{\delta J^{A\cup \left\{ z_{A}\right\} }}\right) .
\end{eqnarray*}
Now the $Y=X$ term in the last summation yields $\dfrac{\delta ^{|X1+2}W}{%
\delta J^{X\cup \left\{ x,y\right\} }}\bar{\Gamma}^{\left\{ y,z\right\} }$,
which will be the highest order derivative appearing in the expression, and
we take it over to the left hand side, we then multiply both sides by $-%
\dfrac{\delta ^{2}W}{\delta J^{x}\delta J^{z}}$ which is the inverse of $%
\bar{\Gamma}^{\left\{ y,z\right\} }$, finally we get the expression 
\begin{equation*}
\frac{\delta ^{|X|+2}W}{\delta J^{X\cup \left\{ x,y\right\} }}%
=\sum_{Y\subset X}\sum_{\mathcal{A}\in \frak{P}\left( X/Y\right) }\frac{%
\delta ^{2}W}{\delta J^{y}\delta J^{p}}\frac{\delta ^{|Y|+2}W}{\delta
J^{Y\cup \left\{ x,q\right\} }}\left( \prod_{A\in \mathcal{A}}\frac{\delta
^{|A|+1}W}{\delta J^{A\cup \left\{ z_{A}\right\} }}\right) \,\bar{\Gamma}%
^{\left\{ p,q\right\} \cup Z_{\mathcal{A}}}.
\end{equation*}
We note that the sum over all $Y\cup \left\{ x\right\} $ (where $Y\subset X$%
, but not $Y=X$) and partitions of $X/Y$ can be reconsidered as a sum over
all partitions of $X\cup \left\{ x\right\} $, excepting the coarsest one.
Let us do this and set $X^{\prime }=X\cup \left\{ x\right\} $ and $Z_{%
\mathcal{A}}^{\prime }=Z_{\mathcal{A}}\cup \left\{ q\right\} $; dropping the
primes then yields $\left( \ref{sdag}\right) $.
\end{proof}

\section{Green Functions}

\subsection{Field Moments}

The Green functions are the field moments $\left\langle \phi _{x_{1}}\cdots
\phi _{x_{n}}\right\rangle $, where $X=\left\{ x_{1},\cdots ,x_{n}\right\} $
is a subset of $\Lambda $, and they form the components of a symmetric
tensor of covariant rank $n=|X|$ under the conventions above. We shall
assume that field moments to all orders exist. It is convenient to introduce
the shorthand notation 
\begin{equation}
\left\langle \phi _{X}\right\rangle =\left\langle \prod_{x\in X}\phi
_{x}\right\rangle \text{.}  \label{greens function}
\end{equation}
The moment generating function is given by the Laplace transform 
\begin{equation}
Z\left[ J\right] =\int_{\Phi }e^{\left\langle \varphi ,J\right\rangle
}\,d\nu \left( \varphi \right) \equiv \left\langle e^{\phi
_{x}J^{x}}\right\rangle .  \label{Z}
\end{equation}
In particular $Z\left[ 0\right] =1$ and we have the expansion 
\begin{equation*}
Z\left[ J\right] =\sum_{n\geq 0}\frac{1}{n!}\left\langle \phi _{x_{1}}\cdots
\phi _{x_{n}}\right\rangle \,J^{x_{1}}\cdots J^{x_{n}}\equiv \sum_{X}\frac{1%
}{|X|!}\left\langle \phi _{X}\right\rangle J^{X}.
\end{equation*}
The field moments are then recovered from $Z\left[ J\right] $ using
functional differentiation $\left. \dfrac{\delta ^{n}Z}{\delta
J^{x_{1}}\cdots \delta J^{x_{n}}}\right| _{J=0}=\left\langle \phi
_{x_{1}}\cdots \phi _{x_{n}}\right\rangle $, or, in shorthand, this reads as 
\begin{equation*}
\left\langle \phi _{X}\right\rangle =\left. \dfrac{\delta ^{|X|}Z\left[ J%
\right] }{\delta J^{X}}\right| _{J=0}.
\end{equation*}

\subsection{Cumulant Field Moments}

The cumulant field moments $\left\langle \!\left\langle \phi
_{X}\right\rangle \!\right\rangle $ (also known as Ursell functions in
statistical mechanics) are defined through 
\begin{equation}
\left\langle \phi _{X}\right\rangle =\sum_{\mathcal{A}\in \frak{P}\left(
X\right) }\prod_{A\in \mathcal{A}}^{c}\left\langle \!\left\langle \phi
_{A}\right\rangle \!\right\rangle  \label{greens-cumulants}
\end{equation}
and using M\"{o}bius inversion, we get 
\begin{equation}
\left\langle \!\left\langle \phi _{X}\right\rangle \!\right\rangle =\sum_{%
\mathcal{A}\in \frak{P}\left( X\right) }\mu \left( \mathcal{A}\right)
\prod_{A\in \mathcal{A}}\left\langle \phi _{A}\right\rangle
\label{cumulant-greens}
\end{equation}
where $\mu \left( \mathcal{A}\right) =-\prod_{j\geq 1}\left\{ -\left(
j-1\right) !\right\} ^{n_{j}\left( \mathcal{A}\right) }$. The first couple
of cumulant moments are 
\begin{eqnarray*}
\left\langle \!\left\langle \phi _{x}\right\rangle \!\right\rangle
&=&\left\langle \phi _{x}\right\rangle , \\
\left\langle \!\left\langle \phi _{x}\phi _{y}\right\rangle \!\right\rangle
&=&\left\langle \phi _{x}\phi _{y}\right\rangle -\left\langle \phi
_{x}\right\rangle \left\langle \phi _{y}\right\rangle , \\
\left\langle \!\left\langle \phi _{x}\phi _{y}\phi _{z}\right\rangle
\!\right\rangle &=&\left\langle \phi _{x}\phi _{y}\phi _{z}\right\rangle
-\left\langle \phi _{x}\phi _{y}\right\rangle \left\langle \phi
_{z}\right\rangle -\left\langle \phi _{y}\phi _{z}\right\rangle \left\langle
\phi _{x}\right\rangle -\left\langle \phi _{z}\phi _{x}\right\rangle
\left\langle \phi _{y}\right\rangle \\
&&+2\left\langle \phi _{x}\right\rangle \left\langle \phi _{y}\right\rangle
\left\langle \phi _{z}\right\rangle .
\end{eqnarray*}
Note that $\left\langle \!\left\langle \phi _{x}\right\rangle
\!\right\rangle $ is just the mean field while $\left\langle \!\left\langle
\phi _{x}\phi _{y}\right\rangle \!\right\rangle $ is the covariance of $\phi
_{x}$ and $\phi _{y}$ in the state.

\begin{theorem}
\textit{The cumulant Green functions are generated through the} 
\begin{equation}
W\left[ J\right] =\sum_{X}\frac{1}{|X|!}\left\langle \!\left\langle \phi
_{X}\right\rangle \!\right\rangle \,J^{X}\equiv \ln Z\left[ J\right]
\label{W}
\end{equation}
\end{theorem}

\begin{proof}
For convenience, we write $\left( CJ\right) _{j}$ for $\left\langle
\!\left\langle \phi _{x_{1}}\cdots \phi _{x_{j}}\right\rangle
\!\right\rangle \,J^{x_{1}}\cdots J^{x_{j}}$, then 
\begin{equation*}
Z\left[ J\right] \equiv \sum_{n\geq 0}\frac{1}{n!}\sum_{\mathcal{A}\in \frak{%
P}_{n}}\prod_{A\in \mathcal{A}}\left( CJ\right) _{\left| A\right| }.
\end{equation*}
Now suppose that we have a partition consisting of $n_{j}$ parts of size $j$
then the occupation sequence is $\mathbf{n}=\left( n_{1},n_{2},n_{3},\cdots
\right) $. The number of parts in the partition is therefore $\sum_{j}n_{j}$
while the number of indices being partition is $n=\sum_{j}j\,n_{j}$. Recall
that the number of different partitions leading to the same occupation
sequence $\mathbf{n}$ is given by $\rho \left( \mathbf{n}\right) $ in
equation $\left( \ref{no. partitions}\right) $\ and so 
\begin{eqnarray*}
Z\left[ J\right] &=&\sum_{\mathbf{n}}\frac{1}{n_{1}!n_{2}!n_{3}!\cdots }%
\dfrac{1}{\left( 1!\right) ^{n_{1}}\left( 2!\right) ^{n_{2}}\left( 3!\right)
^{n_{3}}\cdots }\left( cJ\right) _{1}^{n_{1}}\left( cJ\right)
_{2}^{n_{2}}\cdots \\
&=&\sum_{\mathbf{n}}\prod_{j\geq 1}\frac{1}{n_{j}!}\left[ \frac{\left(
CJ\right) _{j}}{j!}\right] ^{n_{j}} \\
&=&\prod_{j\geq 1}\sum_{n}\frac{1}{n!}\left[ \frac{\left( CJ\right) _{j}}{j!}%
\right] ^{n} \\
&=&\prod_{j\geq 1}\exp \left\{ \frac{\left( CJ\right) _{j}}{j!}\right\} \\
&=&\exp \sum_{j\geq 1}\frac{\left( CJ\right) _{j}}{j!}
\end{eqnarray*}
and so $Z=\exp W$.
\end{proof}

\bigskip

Note that we have used the $\sum \prod \leftrightarrow \prod \sum $ trick
from our section on prime decompositions. We could have alternatively used
the formula $\left( \ref{diff exp W}\right) $ to derive the same result,
however, the above proof suggests that we should think of cumulant moments
as somehow being the `primes' from which the ordinary moments are calculated.

\subsection{Presence of Sources}

Let $J$ be a source field. Given a probability measure $\nu $, we may
introduce a modified probability measure $\nu ^{J}$, absolutely continuous
wrt. $\nu $, and having Radon-Nikodym derivative 
\begin{equation*}
\frac{d\nu ^{J}}{d\nu }\left( \varphi \right) =\frac{1}{Z_{\nu }\left[ J%
\right] }\exp \left\{ \left\langle \varphi ,J\right\rangle _{\nu }\right\} .
\end{equation*}
The corresponding state is referred to as the state modified by the presence
of a source field $J$.

Evidently, we just recover the reference measure $\nu $ when we put $J=0$.
The Laplace transform of the modified state will be $Z_{\nu ^{J}}\left[ K%
\right] =\left\langle e^{\phi _{x}K^{x}}\right\rangle _{\nu ^{J}}$ and it is
readily seen that this reduces to 
\begin{equation*}
Z_{\nu ^{J}}\left[ K\right] =\frac{Z\left[ J+K\right] }{Z\left[ J\right] }.
\end{equation*}
In particular, the cumulants are obtained through $W_{\nu ^{J}}\left[ K%
\right] =W_{\nu }\left[ J+K\right] -W_{\nu }\left[ J\right] $ and we find $%
\left\langle \!\left\langle \phi _{X}\right\rangle \!\right\rangle _{\nu
^{J}}=\left. \dfrac{\delta ^{|X|}W^{J}\left[ K\right] }{\delta K^{X}}\right|
_{K=0}$.

It is, however, considerably simpler to treat $J$ as a free parameter and
just consider $\nu $ as being the family $\left\{ \nu ^{J}:J\in \frak{J}%
\right\} $. In these terms, we have

\begin{equation}
\left\langle \!\left\langle \phi _{X}\right\rangle \!\right\rangle _{\nu
^{J}}=\dfrac{\delta ^{|X|}W_{\nu }\left[ J\right] }{\delta J^{X}}=\sum_{Y}%
\frac{1}{|Y|!}\left\langle \!\left\langle \phi _{X\cup Y}\right\rangle
\!\right\rangle _{\nu }J^{Y},  \label{cumulants from W}
\end{equation}

We point out that it is likewise more convenient to write 
\begin{equation*}
\left\langle \,\phi _{X}\,\right\rangle _{\nu ^{J}}=\frac{1}{Z_{\nu }\left[ J%
\right] }\left\langle \,\phi _{X}\,e^{\left\langle \phi ,J\right\rangle
}\right\rangle _{\nu }=\frac{1}{Z_{\nu }\left[ J\right] }\dfrac{\delta
^{|X|}Z_{\nu }\left[ J\right] }{\delta J^{X}}
\end{equation*}
and again we can drop the superscripts $J$.

\bigskip

The \textit{mean field in the presence of the source}, $\bar{\phi}\left[ J%
\right] \in \Phi $, is defined to be $\bar{\phi}_{x}\left[ J\right]
=\left\langle \,\phi _{x}\,\right\rangle _{\nu ^{J}}$ and is given by the
expression 
\begin{equation}
\bar{\phi}_{x}\left[ J\right] =\sum_{n\geq 0}\frac{1}{n!}\left\langle
\!\left\langle \phi _{x}\phi _{x_{1}}\cdots \phi _{x_{n}}\right\rangle
\!\right\rangle _{\nu }\,J^{x_{1}}\cdots J^{x_{n}}  \label{mean field}
\end{equation}
and, of course, reduces to $\left\langle \phi _{x}\right\rangle _{\nu }$
when $J=0$.

\subsection{States}

Our basic example of a state is a Gaussian state. We also show how we might
perturb one state to get another.

\bigskip

\subsubsection{Gaussian States}

Let $L$ be a linear, symmetric operator on $\Phi $ with well-defined inverse 
$G$. We shall write $g^{xy}$ for the components of $L$ and $g_{xy}$ for the
components of $G$. That is, the equation $L\varphi =J$, or $g^{xy}\varphi
_{y}=J^{x}$ will have unique solution $\varphi =GJ$, or $\varphi
_{x}=g_{xy}J^{y}$. As $G$ is positive definite, symmetric it can be used as
a metric. It can also be used to construct a Gaussian state.

We construct an Gaussian state explicitly in the finite dimensional case
where $|\Lambda |=N<\infty $ by setting 
\begin{equation*}
d\gamma \left( \varphi \right) =\frac{1}{\sqrt{\left( 2\pi \right) ^{N}\det G%
}}\exp \left\{ -\frac{1}{2}g^{xy}\varphi _{x}\varphi _{y}\right\}
\,\prod_{x\in \Lambda }d\varphi _{x}
\end{equation*}
which we may say is determined from the a quadratic action given by $%
S_{\gamma }\left[ \varphi \right] =-\frac{1}{2}g^{xy}\varphi _{x}\varphi
_{y} $. The moment generating function is then given by 
\begin{equation}
Z_{\gamma }\left[ J\right] =\exp \left\{ \frac{1}{2}g_{xy}J^{x}J^{y}\right\}
.  \label{free gaussian}
\end{equation}
In the infinite dimensional case, we may use $\left( \ref{free gaussian}%
\right) $ as the definition of the measure.

The measure is completely characterized by the fact that the only
non-vanishing cumulant is $\left\langle \phi _{x}\phi _{y}\right\rangle
_{\gamma }=g_{xy}$ and if we now use $\left( \ref{greens-cumulants}\right) $
to construct the Green's functions we see that all odd moments vanish while 
\begin{equation}
\left\langle \phi _{x\left( 1\right) }\cdots \phi _{x\left( 2k\right)
}\right\rangle _{\text{$\gamma $}}=\sum_{\mathcal{P}_{2k}}g_{x\left(
p_{1}\right) x\left( q_{1}\right) }\cdots g_{x\left( p_{k}\right) x\left(
q_{k}\right) }  \label{Gaussian field monents}
\end{equation}
where the sum is over all pair partitions of $\left\{ 1,\cdots ,2k\right\} $%
. The right-hand side will of course consist of $\dfrac{\left( 2k\right) !}{%
2^{k}k!}$ terms.

\bigskip

\subsubsection{Perturbations of a State}

Suppose we are given a probability measure $\mu _{0}$. and suppose that $%
S_{I}\left[ \cdot \right] $ is some analytic functional on $\Phi $, say $%
S_{I}\left[ \varphi \right] =\sum_{n\geq 0}\frac{1}{n!}v^{y_{1}\cdots y_{n}}$
$\varphi _{y_{1}}\cdots \varphi _{y_{n}}$, or more compactly 
\begin{equation}
S_{I}\left[ \varphi \right] =\sum_{X}\frac{1}{|X|!}v^{X}\varphi _{X}.
\label{Interaction}
\end{equation}
A probability measure $\mu $, absolutely continuous wrt. $\mu _{0}$, is then
prescribed by taking its Radon-Nikodym to be 
\begin{equation*}
\frac{d\nu }{d\nu _{0}}\left( \varphi \right) =\frac{1}{\Xi }\exp \left\{
S_{I}\left[ \varphi \right] \right\}
\end{equation*}
provided, of course, that the normalization $\Xi \equiv \left\langle \exp
\left\{ S_{I}\left[ \varphi \right] \right\} \right\rangle _{0}<\infty $.

\bigskip

The generating functional for $\mu $ will then be 
\begin{equation*}
Z_{\mu }\left[ J\right] =\frac{1}{\Xi }\left\langle e^{S_{I}\left[ \phi %
\right] }e^{\left\langle \phi ,J\right\rangle }\right\rangle _{\mu _{0}}=%
\frac{1}{\Xi }\exp \left\{ S_{I}\left[ \dfrac{\delta }{\delta J}\right]
\right\} \left\langle e^{\left\langle \phi ,J\right\rangle }\right\rangle
_{\mu _{0}}
\end{equation*}
or, more directly, 
\begin{equation}
Z_{\mu }\left[ J\right] =\frac{1}{\Xi }\exp \left\{ S_{I}\left[ \dfrac{%
\delta }{\delta J}\right] \right\} \,Z_{\mu _{0}}\left[ J\right] .
\label{perturbed mgf}
\end{equation}

We remark that, for finite dimensional fields, we may think of $\mu $ being
determined by the action $S\left[ \varphi \right] =S_{0}\left[ \varphi %
\right] +S_{I}\left[ \varphi \right] $ where $S_{0}$ is the action of $\mu
_{0}$.

\subsection{The Dyson-Schwinger Equation}

We now derive a functional differential equation for the generating function.

\bigskip

\begin{lemma}
The Gaussian \textit{generating functional }$Z_{\gamma }$\textit{\ satisfies
the differential equations} 
\begin{equation*}
\left\{ F_{\gamma }^{x}\left[ \frac{\delta }{\delta J}\right] +J^{x}\right\}
Z_{\gamma }\left[ J\right] =0
\end{equation*}
where $F_{\gamma }^{x}\left[ \varphi \right] =\dfrac{\delta }{\delta \varphi
_{x}}S_{\gamma }\left[ \varphi \right] $.
\end{lemma}

\begin{proof}
Explicitly, we have $Z_{\gamma }=\exp \left\{ \frac{1}{2}J^{x}g_{xy}J^{y}%
\right\} $ so that $\dfrac{\delta }{\delta J^{x}}Z_{\gamma
}=g_{xy}J^{y}Z_{\gamma }$ which can be rearranged as 
\begin{equation*}
\left( -g^{xy}\frac{\delta }{\delta J^{y}}+J^{x}\right) Z_{\gamma }=0.
\end{equation*}
However, $S_{\gamma }\left[ \varphi \right] =-\frac{1}{2}g^{xy}\varphi
_{x}\varphi _{y}$ and so $F_{\gamma }^{x}\left[ \varphi \right]
=-g^{xy}\varphi _{x}$.
\end{proof}

\begin{lemma}
Suppose that $Z_{\mu _{0}}$ satisfies a differential equation of the form 
\begin{equation*}
\left\{ F_{\mu _{0}}^{x}\left[ \frac{\delta }{\delta J}\right]
+J^{x}\right\} Z_{\mu _{0}}\left[ J\right] =0
\end{equation*}
and that $\mu $, is absolutely continuous wrt. $\mu _{0}$, is given by $%
\frac{d\nu }{d\nu _{0}}\left( \varphi \right) =\frac{1}{\Xi }\exp \left\{
S_{I}\left[ \varphi \right] \right\} $. Then $Z_{\mu }$ satisfies 
\begin{equation*}
\left\{ F_{I}^{x}\left[ \frac{\delta }{\delta J}\right] +F_{\mu _{0}}^{x}%
\left[ \frac{\delta }{\delta J}\right] +J^{x}\right\} Z_{\mu }\left[ J\right]
=0
\end{equation*}
where $F_{I}^{x}\left[ \varphi \right] =\dfrac{\delta }{\delta \varphi _{x}}%
S_{I}\left[ \varphi \right] $.
\end{lemma}

\begin{proof}
We observe that from $\left( \ref{perturbed mgf}\right) $ we have 
\begin{equation*}
J^{x}Z_{\mu }\left[ J\right] =\frac{1}{\Xi }J^{x}\exp \left\{ S_{I}\left[ 
\dfrac{\delta }{\delta J}\right] \right\} \,Z_{\mu _{0}}\left[ J\right]
\end{equation*}
and using the commutation identity 
\begin{equation*}
\left[ J^{x},\exp \left\{ S_{I}\left[ \dfrac{\delta }{\delta J}\right]
\right\} \right] =-F_{I}^{x}\left[ \frac{\delta }{\delta J}\right] \exp
\left\{ S_{I}\left[ \dfrac{\delta }{\delta J}\right] \right\}
\end{equation*}
we find 
\begin{eqnarray*}
J^{x}Z_{\mu }\left[ J\right] &=&\frac{1}{\Xi }\exp \left\{ S_{I}\left[ 
\dfrac{\delta }{\delta J}\right] \right\} \,J^{x}Z_{\mu _{0}}\left[ J\right]
-\frac{1}{\Xi }F_{I}^{x}\left[ \frac{\delta }{\delta J}\right] \exp \left\{
S_{I}\left[ \dfrac{\delta }{\delta J}\right] \right\} \,Z_{\mu _{0}}\left[ J%
\right] \\
&=&-F_{\mu _{0}}^{x}\left[ \frac{\delta }{\delta J}\right] Z_{\mu }\left[ J%
\right] -F_{I}^{x}\left[ \frac{\delta }{\delta J}\right] Z_{\mu }\left[ J%
\right]
\end{eqnarray*}
which gives the result.
\end{proof}

Putting these two lemmas together we obtain the following result

\bigskip

\begin{theorem}[Dyson-Schwinger]
\textit{The generating functional }$Z_{\mu }$\textit{\ for a probability
measure absolutely continuous wrt. }$\gamma $ \textit{satisfies the
differential equation} 
\begin{equation}
\left\{ F^{x}\left[ \frac{\delta }{\delta J}\right] +J^{x}\right\} Z\left[ J%
\right] =0  \label{DS}
\end{equation}
\textit{where }$F^{x}\left[ \varphi \right] =\dfrac{\delta S\left[ \varphi %
\right] }{\delta \varphi _{x}}=-\frac{1}{2}g^{xy}\varphi _{y}+F_{I}^{x}\left[
\varphi \right] $.
\end{theorem}

\bigskip

\begin{corollary}
\textit{Under the conditions of the above theorem, if }$S_{I}\left[ \varphi %
\right] =\sum_{X}\frac{1}{|X|!}v^{X}\varphi _{X}$\textit{, then the ordinary
moments of }$\mu $\textit{\ satisfy the algebraic equations} 
\begin{equation}
\left\langle \phi _{\left\{ x\right\} \cup X}\right\rangle =\sum_{x^{\prime
}\in X}g_{xx^{\prime }}\left\langle \phi _{X/\left\{ x^{\prime }\right\}
}\right\rangle +g_{xy}\sum_{Y}\frac{1}{|Y|!}v^{\left\{ y\right\} \cup
Y}\,\left\langle \phi _{X\cup Y}\right\rangle .  \label{DS Green's function}
\end{equation}
\end{corollary}

\begin{proof}
We now have $F^{x}=-g^{xy}\varphi _{y}+\sum_{X}\dfrac{1}{|X|!}v^{\left\{
x\right\} \cup X}\varphi _{X}$. The Dyson-Schwinger equation then becomes 
\begin{equation*}
-g^{xy}\frac{\delta Z}{\delta J^{y}}+\sum_{Y}\frac{1}{|Y|!}v^{\left\{
x\right\} \cup Y}\,\frac{\delta ^{|Y|}Z}{\delta J^{Y}}+J^{x}Z=0
\end{equation*}
and we consider applying the further differentiation $\dfrac{\delta ^{|X|}}{%
\delta J^{X}}$ to obtain 
\begin{equation*}
-g^{xy}\left\langle \phi _{\left\{ y\right\} \cup X}\right\rangle
_{J}+\sum_{Y}\frac{1}{|Y|!}v^{\left\{ x\right\} \cup Y}\,\left\langle \phi
_{X\cup Y}\right\rangle _{J}+\dfrac{\delta ^{|X|}}{\delta J^{X}}\left(
J^{x}Z\right) =0.
\end{equation*}
The result follows from setting $J=0$.
\end{proof}

\bigskip

We may write $\left( \ref{DS Green's function}\right) $ in index notation as 
\begin{eqnarray*}
\left\langle \phi _{x}\phi _{x_{1}}\cdots \phi _{x_{m}}\right\rangle
&=&\sum_{i=1}^{m}g_{xx_{i}}\left\langle \phi _{x_{1}}\cdots \widehat{\phi
_{x_{i}}}\cdots \phi _{x_{m}}\right\rangle \\
&&+g_{xy}\sum_{n\geq 0}\frac{1}{n!}v^{yy_{1}\cdots y_{n}}\,\left\langle \phi
_{y_{1}}\cdots \phi _{y_{n}}\phi _{x_{1}}\cdots \phi _{x_{m}}\right\rangle
\end{eqnarray*}
where the hat indicates an omission. This hierarchy of equations for the
Green's functions is equivalent to the Dyson-Schwinger equation.

We remark that the first term on the right hand side of $\left( \ref{DS
Green's function}\right) $ contains the moments $\left\langle \phi
_{X/\left\{ x^{\prime }\right\} }\right\rangle $ which are of order two
smaller than the left hand side $\left\langle \phi _{\left\{ x\right\} \cup
X}\right\rangle $. The second term on the right hand side of $\left( \ref{DS
Green's function}\right) $ contains the moments of higher order and so we
generally cannot use this equation recursively. In the Gaussian case,
however, we have $\left\langle \phi _{\left\{ x\right\} \cup X}\right\rangle
_{\gamma }=\sum_{x^{\prime }\in X}g_{xx^{\prime }}\left\langle \phi
_{X/\left\{ x^{\prime }\right\} }\right\rangle _{\gamma }$ from which we can
deduce $\left( \ref{Gaussian field monents}\right) $ by just knowing that $%
\left\langle \phi _{x}\right\rangle =0$ and $\left\langle \phi _{x}\phi
_{y}\right\rangle =g_{xy}$.

\bigskip

The Dyson-Schwinger equations may alternatively stated for $W$, they are 
\begin{equation*}
-g^{xy}\frac{\delta W}{\delta J^{y}}+\sum_{X}\frac{1}{|X|!}v^{\left\{
x\right\} \cup X}\,\sum_{\mathcal{A}\in \frak{P}\left( X\right) }\prod_{A\in 
\mathcal{A}}\frac{\delta ^{\left| A\right| }W}{\delta J^{A}}+J^{x}=0.
\end{equation*}

\subsection{Field-Source Relations}

Recall that $\bar{\phi}_{x}\left[ J\right] $\ is the mean field value in the
presence of a source $J\in \frak{J}$ and is given by 
\begin{equation}
\bar{\phi}_{x}=\frac{\delta W}{\delta J^{x}}.
\end{equation}
Let us assume again that the map $J\mapsto \bar{\phi}$ from $\frak{J}$ to $%
\Phi $ is invertible and denote the inverse by $\bar{J}$. That is, $\phi _{x}%
\left[ \bar{J}\left[ \varphi \right] \right] =\varphi _{x}$. Moreover, we
suppose that $\bar{J}$ admits an analytic expansion of the type $\bar{J}^{x}%
\left[ \varphi \right] =-\sum_{n\geq 0}\frac{1}{n!}\Gamma ^{xy_{1}\cdots
y_{n}}\varphi _{y_{1}}\cdots \varphi _{y_{n}}$, or 
\begin{equation*}
\bar{J}^{x}\left[ \varphi \right] =-\sum_{X}\dfrac{1}{|X|!}\Gamma ^{\left\{
x\right\} \cup X}\varphi _{X}.
\end{equation*}

We introduce the functional 
\begin{equation*}
\Gamma \left[ \varphi \right] =\sum_{n\geq 0}\frac{1}{n!}\Gamma
^{y_{1}\cdots y_{n}}\varphi _{y_{1}}\cdots \varphi _{y_{n}}=\sum_{X}\dfrac{1%
}{|X|!}\Gamma ^{X}\varphi _{X}
\end{equation*}
so that 
\begin{equation}
\bar{J}^{x}\left[ \varphi \right] =-\frac{\delta \Gamma }{\delta \varphi _{x}%
}.  \label{source-Gamma}
\end{equation}
We then recognize $\Gamma \left[ \varphi \right] $ as the Legendre-Frenchel
transform of $W\left[ J\right] $, that is, $W\left[ J\right] =\sup_{\varphi
\in \Phi }\left\{ \Gamma \left[ \varphi \right] +\left\langle \varphi
,J\right\rangle \right\} $ with the supremum attained at the mean-field $%
\varphi =\bar{\phi}\left[ J\right] $.

\bigskip

For the Gaussian state we have $W_{\gamma }\left[ J\right] =\frac{1}{2}%
g_{xy}J^{x}J^{y}$ and we see that $\bar{\phi}_{x}=g_{xy}J^{y}$. The inverse
map is therefore $\bar{J}^{x}\left[ \varphi \right] =g^{xy}\varphi _{y}$ and
so we obtain 
\begin{equation*}
\Gamma _{\text{$\gamma $}}\left[ \varphi \right] =-\frac{1}{2}g^{xy}\varphi
_{x}\varphi _{y}.
\end{equation*}

More generally, we have $\Gamma =\Gamma _{\gamma }+\Gamma _{I}$ where 
\begin{equation*}
\Gamma _{I}\left[ \varphi \right] =\Gamma ^{x}\varphi _{x}+\frac{1}{2}\pi
^{xy}\varphi _{x}\varphi _{y}+\sum_{n\geq 3}\frac{1}{n!}\Gamma ^{x_{1}\cdots
x_{n}}\varphi _{x_{1}}\cdots \varphi _{x_{n}}
\end{equation*}
where it is customary to set $\Gamma ^{xy}=-g^{xy}+\pi ^{xy}$. Without loss
of generality we shall assume that the lead term $\Gamma ^{x}$ is equal to
zero: this means that the state is centred so that $\bar{\phi}\left[ 0\right]
=0$. It follows that 
\begin{equation}
\bar{J}^{x}\left[ \varphi \right] =\left( g^{xy}-\pi ^{xy}\right) \varphi
_{y}-\frac{1}{2}\Gamma ^{xyz}\varphi _{y}\varphi _{z}-\cdots  \label{tree}
\end{equation}
and (substituting $\varphi =\bar{\phi}$) we may rearrange this to get 
\begin{equation}
\bar{\phi}_{x}=g_{xy}\left( J^{y}+\pi ^{yz}\bar{\phi}_{z}+\frac{1}{2}\Gamma
^{yzw}\bar{\phi}_{z}\bar{\phi}_{w}+\cdots \right)  \label{tree1}
\end{equation}
We note that $\bar{\phi}$ appears on the right-hand side of $\left( \ref
{tree1}\right) $\ in a generally nonlinear manner: let us rewrite this as $%
\bar{\phi}=GJ+f\left( \bar{\phi}\right) $ where $f$ satisfies $f\left(
0\right) =0$. We may re-iterate $\left( \ref{tree1}\right) $ to get a
so-called tree expansion 
\begin{equation*}
\bar{\phi}=GJ+f\left( GJ+f\left( GJ+f\left( GJ+\cdots \right) \right) \right)
\end{equation*}
and we know that this expansion should be re-summed to give the series
expansion in terms of $J$ as in $\left( \ref{mean field}\right) $.

\bigskip

Now $\Gamma ^{\prime \prime xy}\left[ \varphi \right] =-g^{xy}+\Gamma
_{I}^{\prime \prime xy}\left[ \varphi \right] $ so we conclude that 
\begin{equation}
W^{\prime \prime }\left[ J\right] =\frac{1}{L-\bar{\Gamma}_{I}^{\prime
\prime }\left[ J\right] }  \label{self}
\end{equation}
where $\left( \bar{\Gamma}_{I}^{\prime \prime }\left[ J\right] \right)
^{xy}\equiv \pi ^{xy}+\sum_{|X|\geq 1}\frac{1}{|X|!}\Gamma ^{\left\{
x,y\right\} \cup X}\bar{\phi}_{X}\left[ J\right] $. This relation may be
alternatively written as the series 
\begin{equation}
W^{\prime \prime }\left[ J\right] =G\frac{1}{1-\bar{\Gamma}_{I}^{\prime
\prime }\left[ J\right] G}=G+G\bar{\Gamma}_{I}^{\prime \prime }\left[ J%
\right] G+G\bar{\Gamma}_{I}^{\prime \prime }\left[ J\right] G\bar{\Gamma}%
_{I}^{\prime \prime }\left[ J\right] G+\cdots  \label{self1}
\end{equation}
In particular, when we set $J=0$, $W^{\prime \prime }\left[ 0\right] $ is
the covariance matrix while $\bar{\Gamma}_{I}^{\prime \prime }\left[ 0\right]
\equiv \pi $, provided we assume that the state is centred ($\bar{\phi}=0$
when $J=0$), and we obtain the series expansion 
\begin{equation}
W^{\prime \prime }\left[ 0\right] =G+G\pi G+G\pi G\pi G+\cdots  \label{self2}
\end{equation}

\bigskip

We now wish to determine a formula relating the cumulant moments in the
presence of the source $J$ to the tensor coefficients $\bar{\Gamma}^{X}\left[
J\right] $.

\begin{theorem}
We have the followi\textit{ng recurrence relation} 
\begin{equation}
\left\langle \!\left\langle \phi _{X\cup \left\{ y\right\} }\right\rangle
\!\right\rangle _{\nu ^{J}}=\sum\limits_{\mathcal{A}\in \frak{P}^{f}\left(
X\right) }\left\langle \!\left\langle \phi _{y}\phi _{p}\right\rangle
\!\right\rangle _{\nu ^{J}}\,\bar{\Gamma}^{\left\{ p\right\} \cup Z_{%
\mathcal{A}}}\left[ J\right] \,\left( \prod_{A\in \mathcal{A}}\left\langle
\!\left\langle \phi _{A\cup \left\{ z_{A}\right\} }\right\rangle
\!\right\rangle _{\nu ^{J}}\right) ,  \label{sdhag}
\end{equation}
\textit{where, again, each }$z_{A}$\textit{\ is a dummy variable associated
with each component part }$A$\textit{\ and }$Z_{\mathcal{A}}=\left\{
z_{A}:A\in \mathcal{A}\right\} $\textit{. }
\end{theorem}

This, of course, just follows straight from $\left( \ref{sdag}\right) $. The
crucial thing about $\left( \ref{sdhag}\right) $ is that the right hand side
contains lower order cumulants and so can be used recursively. Let us
iterate once : 
\begin{gather*}
\left\langle \!\left\langle \phi _{\left\{ y\right\} \cup X}\right\rangle
\!\right\rangle _{\nu ^{J}}=\left\langle \!\left\langle \phi _{\left\{
y,p\right\} }\right\rangle \!\right\rangle _{\nu ^{J}}\sum\limits_{\mathcal{A%
}\in \frak{P}^{f}\left( X\right) }\bar{\Gamma}^{\left\{ p\right\} \cup Z_{%
\mathcal{A}}}\left[ J\right] \left( \prod_{A\in \mathcal{A}}\left\langle
\!\left\langle \phi _{A\cup \left\{ z_{A}\right\} }\right\rangle
\!\right\rangle _{\nu ^{J}}\right) \\
=\left\langle \!\left\langle \phi _{\left\{ y,p\right\} }\right\rangle
\!\right\rangle _{\nu ^{J}}\sum\limits_{\mathcal{A}\in \frak{P}^{f}\left(
X\right) }\bar{\Gamma}_{J}^{\left\{ p\right\} \cup Z_{\mathcal{A}}}\left[ J%
\right] \prod_{A\in \mathcal{A}}\left\langle \!\left\langle \phi _{\left\{
z_{A},q\right\} }\right\rangle \!\right\rangle _{\nu ^{J}} \\
\times \sum\limits_{\mathcal{B}\in \frak{P}^{f}\left( A\right) }\bar{\Gamma}%
^{\left\{ q\right\} \cup Z_{\mathcal{B}}}\left[ J\right] \left( \prod_{B\in 
\mathcal{B}}\left\langle \!\left\langle \phi _{B\cup \left\{ z_{B}\right\}
}\right\rangle \!\right\rangle _{\nu ^{J}}\right) .
\end{gather*}
What happens is that each part $A$\ of the first partition gets properly
divided up into sub-parts and we continue until eventually break down $X$
into its singleton parts. However, this is just a top-down description of a
hierarchy on $X$. Proceeding in this way we should obtain a sum over all
hierarchies of $X$. At the root of the tree, we have the factor $%
\left\langle \!\left\langle \phi _{y}\phi _{p}\right\rangle \!\right\rangle
_{\nu ^{J}}$ and for each node/part $A$, appearing anywhere in the sequence,
labelled by dummy index $z_{A}$ say, we will break it into a proper
partition $\mathcal{B}\in \frak{P}\left( A\right) $ and obtain a
multiplicative factor $\left\langle \!\left\langle \phi _{\left\{
z_{A},q\right\} }\right\rangle \!\right\rangle _{\nu ^{J}}\,\bar{\Gamma}%
^{\left\{ q\right\} \cup Z_{\mathcal{B}}}\left[ J\right] $ where $Z_{%
\mathcal{B}}$ will be the set of labels for each part $B\in \mathcal{B}$.

If we set $X=\left\{ x_{1},x_{2}\right\} $ then we have only one hierarchy
to sum over and we find $\left\langle \!\left\langle \phi _{y}\phi
_{x_{1}}\phi _{x_{2}}\right\rangle \!\right\rangle _{\nu ^{J}}=\left\langle
\!\left\langle \phi _{y}\phi _{q}\right\rangle \!\right\rangle _{\nu
^{J}}\left\langle \!\left\langle \phi _{x_{1}}\phi _{z_{1}}\right\rangle
\!\right\rangle _{\nu ^{J}}\left\langle \!\left\langle \phi _{x_{2}}\phi
_{z_{2}}\right\rangle \!\right\rangle _{\nu ^{J}}\,\bar{\Gamma}^{qz_{1}z_{2}}%
\left[ J\right] $.

It is useful to use the bottom-up description to give the general result.
First we introduce some new coefficients defined by 
\begin{equation*}
\Upsilon _{x_{1}\cdots x_{n}}^{Y}\left[ J\right] :=\left\langle
\!\left\langle \phi _{x_{1}}\phi _{z_{1}}\right\rangle \!\right\rangle _{\nu
^{J}}\cdots \left\langle \!\left\langle \phi _{x_{n}}\phi
_{z_{n}}\right\rangle \!\right\rangle _{\nu ^{J}}\,\bar{\Gamma}^{\left\{
z_{1},\cdots ,z_{n}\right\} \cup Y}.
\end{equation*}
with the exceptional case $\Upsilon _{xy}:=\left\langle \!\left\langle \phi
_{x}\phi _{y}\right\rangle \!\right\rangle _{\nu ^{J}}$. Then we find that 
\begin{eqnarray}
\left\langle \!\left\langle \phi _{\left\{ y\right\} \cup X}\right\rangle
\!\right\rangle _{\nu ^{J}} &=&\sum\limits_{\mathcal{H}=\left\{ \mathcal{A}%
^{\left( 1\right) },\cdots ,\mathcal{A}^{\left( m\right) }\right\} \in \frak{%
H}\left( X\right) }\Upsilon _{\left\{ y\right\} \cup Z_{\mathcal{A}^{\left(
m\right) }}}  \notag \\
&&\times \prod_{A^{\left( m\right) }\in \mathcal{A}^{\left( m\right)
}}\Upsilon _{Z_{\mathcal{A}^{\left( m-1\right) }}}^{z_{A^{\left( m\right)
}}}\prod_{A^{\left( m-1\right) }\in \mathcal{A}^{\left( m-1\right)
}}\Upsilon _{Z_{\mathcal{A}^{\left( m-2\right) }}}^{z_{A^{\left( m-1\right)
}}}\cdots \prod_{A^{\left( 1\right) }\in \mathcal{A}^{\left( 1\right)
}}\Upsilon _{Z_{\mathcal{A}^{\left( 2\right) }}}.  \notag \\
&&  \label{sum over hierarchies}
\end{eqnarray}
For instance, we have the following expansions for the lowest cumulants: 
\begin{eqnarray*}
\left\langle \!\left\langle \phi _{y}\phi _{x_{1}}\phi _{x_{2}}\right\rangle
\!\right\rangle _{\nu ^{J}} &=&\Upsilon _{yx_{1}x_{2}}, \\
\left\langle \!\left\langle \phi _{y}\phi _{x_{1}}\phi _{x_{2}}\phi
_{x_{3}}\right\rangle \!\right\rangle _{\nu ^{J}} &=&\Upsilon
_{yx_{1}x_{2}x_{3}}+\left( \Upsilon _{yx_{1}}^{\quad r}\Upsilon
_{rx_{2}x_{3}}+\cdots \right) , \\
\left\langle \!\left\langle \phi _{y}\phi _{x_{1}}\phi _{x_{2}}\phi
_{x_{3}}\phi _{x_{4}}\right\rangle \!\right\rangle _{\nu ^{J}} &=&\Upsilon
_{yx_{1}x_{2}x_{3}x_{4}}+\left( \Upsilon _{yx_{1}}^{\quad \;r}\Upsilon
_{rx_{2}x_{3}x_{4}}+\cdots \right) +\left( \Upsilon _{yx_{1}x_{2}}^{\quad
\;r}\Upsilon _{rx_{3}x_{4}}+\cdots \right) \\
&&+\left( \Upsilon _{yx_{1}}^{\quad r}\Upsilon _{rx_{2}}^{\quad q}\Upsilon
_{qx_{3}x_{4}}+\cdots \right) +\left( \Upsilon _{y}^{\quad rq}\Upsilon
_{rx_{1}x_{2}}\Upsilon _{qx_{3}x_{4}}+\cdots \right) , \\
&&\text{etc.}
\end{eqnarray*}
The terms in round brackets involve permutations of the $x_{j}$ indices
leading to distinct terms. Thus, there are $1+\frac{1}{2}\binom{4}{2}=4$
terms making up the right-hand side for the fourth order cumulant: The first
term in round brackets corresponds to the hierarchy $\left\{ \left\{ \left\{
x_{1}\right\} \right\} ,\left\{ \left\{ x_{2}\right\} ,\left\{ x_{3}\right\}
\right\} \right\} $ and there are 3 such second order hierarchies. There are 
$1+\binom{5}{3}+\frac{1}{2}\binom{5}{1}\binom{4}{2}=26$ terms making up the
right-hand side for the fifth order cumulant.

\section{Wick Ordering}

\subsection{Wick Cumulants}

Up to now, we have been considering cumulant moments of the field. Let $%
F_{1},F_{2},\cdots ,F_{n}$ be some functionals of our field, then we may
define the object $\left\langle \!\left\langle \left[ F_{1}\right] \cdots %
\left[ F_{n}\right] \right\rangle \!\right\rangle $ through 
\begin{equation*}
\left\langle \!\left\langle \left[ F_{1}\right] \cdots \left[ F_{n}\right]
\right\rangle \!\right\rangle =\sum_{\mathcal{A}\in \frak{P}\left( X\right)
}\prod_{A\in \mathcal{A}}\left\langle \!\left\langle \prod_{i\in A}\left[
F_{i}\right] \right\rangle \!\right\rangle
\end{equation*}
where $\left\langle \left[ F_{1}\right] \cdots \left[ F_{n}\right]
\right\rangle \equiv \left\langle F_{1}\cdots F_{n}\right\rangle $. Thus $%
\left\langle \!\left\langle \left[ F\right] \left[ G\right] \right\rangle
\!\right\rangle =\left\langle FG\right\rangle -\left\langle F\right\rangle
\left\langle G\right\rangle $, etc.

As the notation hopefully suggests, we treat each component in a square
brackets as an indivisible whole object. For instance, we have that $%
\left\langle \!\left\langle \left[ \phi _{x}\right] \left[ \phi _{y}\phi _{z}%
\right] \right\rangle \!\right\rangle $ is a second order cumulant (even
though it has three field factors!) and it equates to $\left\langle \phi
_{x}\phi _{y}\phi _{z}\right\rangle -\left\langle \phi _{x}\right\rangle
\left\langle \phi _{y}\phi _{z}\right\rangle $.

\bigskip

\subsection{Wick Monomials}

The following definition of Wick monomial is due to Djah, Gottschalk and
Ouerdiane \cite{DjahGottschalkOuerdiane}. It is more general than the
traditional definition, which uses the vacuum state, and we review it in
terms of our field calculus notation: the theorems are otherwise after \cite
{DjahGottschalkOuerdiane} directly.

\begin{definition}
For each finite subset $X$, the Wick ordered monomial $:\phi _{X}:$ \ is the
operator defined through the property that 
\begin{equation}
\left\langle :\phi _{X}:\;F\right\rangle =\left\langle \!\left\langle
\prod_{x\in X}\left[ \phi _{x}\right] \left[ F\right] \right\rangle
\!\right\rangle
\end{equation}
for all appropriate $F$.
\end{definition}

\bigskip

Let us first remark that the appropriate functionals $F=F\left[ \phi \right] 
$ are those such that $F=F\left[ \varphi \right] $ is square-integrable wrt.
the measure $\nu $. Secondly, it should be emphasized that the definition of 
$:\phi _{X}:$ depends on the choice of state or, equivalently, on $\nu $.

\bigskip

\begin{lemma}
The ordinary field operator products $\phi _{X}$ can be put together from
the Wick monomials according to the formula 
\begin{equation*}
\phi _{X}=\left\langle \phi _{X}\right\rangle +\sum_{\mathcal{A}\in \frak{P}%
\left( X\right) }\sum_{A\in \mathcal{A}}\left( \prod_{B\in \mathcal{A}%
}^{B\neq A}\left\langle \!\left\langle \phi _{B}\right\rangle
\!\right\rangle \right) \;:\phi _{A}:.
\end{equation*}
\end{lemma}

\begin{proof}
First of all, we observe that we can use the short-hand form $\left\langle
\!\left\langle \phi _{X\cup \left\{ \alpha \right\} }\right\rangle
\!\right\rangle $ for $\left\langle \!\left\langle \prod_{x\in X}\left[ \phi
_{x}\right] \left[ F\right] \right\rangle \!\right\rangle $ by setting $\phi
_{\alpha }=F$ where $\alpha $ is an exceptional label, and we augment our
configuration space to $\Lambda \cup \left\{ \alpha \right\} $. Now the
ordinary moment $\left\langle \phi _{X\cup \left\{ \alpha \right\}
}\right\rangle $ can be expanded as $\sum_{\mathcal{A}\in \frak{P}\left(
X\cup \left\{ \alpha \right\} \right) }\prod_{A\in \mathcal{A}}\left\langle
\!\left\langle \phi _{A}\right\rangle \!\right\rangle $ and the partitions
of $X\cup \left\{ \alpha \right\} $ can be set out into two types: those
that contain $\left\{ \alpha \right\} $ as a singleton part, and those that
don't. This leads to 
\begin{eqnarray*}
\left\langle \phi _{X\cup \left\{ \alpha \right\} }\right\rangle &=&\left(
\sum_{\mathcal{A}\in \frak{P}\left( X\right) }\prod_{A\in \mathcal{A}%
}\left\langle \!\left\langle \phi _{A}\right\rangle \!\right\rangle \right)
\left\langle \!\left\langle \phi _{\alpha }\right\rangle \!\right\rangle
+\sum_{\mathcal{A}\in \frak{P}\left( X\right) }\sum_{A\in \mathcal{A}}\left(
\prod_{B\in \mathcal{A}}^{B\neq A}\left\langle \!\left\langle \phi
_{B}\right\rangle \!\right\rangle \right) \left\langle \!\left\langle \phi
_{A\cup \alpha }\right\rangle \!\right\rangle \\
&\equiv &\left\langle \phi _{X}\right\rangle \left\langle \phi _{\alpha
}\right\rangle +\sum_{\mathcal{A}\in \frak{P}\left( X\right) }\sum_{A\in 
\mathcal{A}}\left( \prod_{B\in \mathcal{A}}^{B\neq A}\left\langle
\!\left\langle \phi _{B}\right\rangle \!\right\rangle \right) \left\langle
:\phi _{A}:\;\phi _{\alpha }\right\rangle .
\end{eqnarray*}
However, as $\phi _{\alpha }$ was arbitrary, we obtain the required identity.
\end{proof}

\bigskip

\begin{corollary}
The Wick monomials satisfy the following recursion relation 
\begin{equation*}
:\phi _{X}:=\phi _{X}-\left\langle \phi _{X}\right\rangle -\sum_{\mathcal{A}%
\in \frak{P}^{f}\left( X\right) }\sum_{A\in \mathcal{A}}\left( \prod_{B\in 
\mathcal{A}}^{B\neq A}\left\langle \!\left\langle \phi _{B}\right\rangle
\!\right\rangle \right) \;:\phi _{A}:.
\end{equation*}
\end{corollary}

\bigskip

\begin{theorem}
Let $X,Y_{1},\cdots ,Y_{m}$ be disjoint finite subsets, then 
\begin{equation*}
\left\langle :\phi _{Y_{1}}:\cdots :\phi _{Y_{m}}:\;\phi _{X}\right\rangle
=\sum_{\mathcal{A}\in \frak{P}^{^{\prime }}\left( Y_{1},\cdots
,Y_{m};X\right) }\prod_{A\in \mathcal{A}}\left\langle \!\left\langle \phi
_{A}\right\rangle \!\right\rangle
\end{equation*}
where $\frak{P}^{^{\prime }}\left( Y_{1},\cdots ,Y_{m};X\right) $ is the set
of all partitions of $Y_{1}\cup \cdots \cup Y_{m}\cup X$ having no subset of
any of the $Y_{j}$'s as a part.
\end{theorem}

\begin{proof}
Let $n=\sum_{j=1}^{m}|Y_{j}|$, the result will be established by strong
induction. The case $n=0$, to begin with, is just the expansion of $%
\left\langle \phi _{X}\right\rangle $ in terms of its connected Green's
function. If the results is assumed to hold up to value $n$ then 
\begin{gather*}
\left\langle :\phi _{Y_{1}}:\cdots :\phi _{Y_{m+1}}:\;\phi _{X}\right\rangle
=\left\langle :\phi _{Y_{1}}:\cdots :\phi _{Y_{m}}:\;\phi _{Y_{m}\cup
X}\right\rangle \\
-\left\langle :\phi _{Y_{1}}:\cdots :\phi _{Y_{m}}:\;\phi _{X}\right\rangle
\left\langle \phi _{Y_{m+1}}\right\rangle \\
-\sum_{\mathcal{A}\in \frak{P}^{f}\left( Y_{m+1}\right) }\sum_{A\in \mathcal{%
A}}\left( \prod_{B\in \mathcal{A}}^{B\neq A}\left\langle \!\left\langle \phi
_{B}\right\rangle \!\right\rangle \right) \left\langle :\phi _{Y_{1}}:\cdots
:\phi _{Y_{m}}::\phi _{A}:\;\phi _{X}\right\rangle
\end{gather*}
and by induction up to order $n$ we may rearrange this as 
\begin{eqnarray}
&&\sum_{\mathcal{A}\in \frak{P}^{^{\prime }}\left( Y_{1},\cdots
,Y_{m};Y_{m+1}\cup X\right) }\prod_{A\in \mathcal{A}}\left\langle
\!\left\langle \phi _{A}\right\rangle \!\right\rangle -\sum_{\mathcal{A}\in 
\frak{P}^{^{\prime }}\left( Y_{1},\cdots ,Y_{m};X\right) }\sum_{\mathcal{B}%
\in \frak{P}\left( Y_{m+1}\right) }\prod_{A\in \mathcal{A}}\left\langle
\!\left\langle \phi _{A}\right\rangle \!\right\rangle \prod_{B\in \mathcal{B}%
}\left\langle \!\left\langle \phi _{B}\right\rangle \!\right\rangle  \notag
\\
&&-\sum_{\mathcal{A}\in \frak{P}^{f}\left( Y_{m+1}\right) }\sum_{A\in 
\mathcal{A}}\sum_{\mathcal{C}\in \frak{P}^{^{\prime }}\left( Y_{1},\cdots
,Y_{m},A;X\right) }\prod_{B\in \mathcal{A}}^{B\neq A}\left\langle
\!\left\langle \phi _{B}\right\rangle \!\right\rangle \prod_{C\in \mathcal{C}%
}\left\langle \!\left\langle \phi _{C}\right\rangle \!\right\rangle .
\label{radish}
\end{eqnarray}

The claim is that this equals $\sum_{\mathcal{A}\in \frak{P}^{^{\prime
}}\left( Y_{1},\cdots ,Y_{m},Y_{m+1};X\right) }\prod_{A\in \mathcal{A}%
}\left\langle \!\left\langle \phi _{A}\right\rangle \!\right\rangle $. Now
the first term appearing in $\left( \ref{radish}\right) $ is the sum over $%
\frak{P}^{^{\prime }}\left( Y_{1},\cdots ,Y_{m};Y_{m+1}\cup X\right) $ which
is strictly larger than $\frak{P}^{^{\prime }}\left( Y_{1},\cdots
,Y_{m},Y_{m+1};X\right) $ by means of including unwanted partitions of $%
Y_{1}\cup \cdots \cup Y_{m+1}\cup X$ in which subsets of $Y_{m+1}$ appear as
parts. The second term of $\left( \ref{radish}\right) $\ subtracts all such
contributions from partitions for which every part $B$ that contains an
element of $Y_{m+1}$ only contains elements of $Y_{m+1}$. The third term
subtracts the contribution from the remaining unwanted partitions, namely
those where the offending parts $B$ make up only some subset $Y_{m+1}/A$ of $%
Y_{m+1}$.

This establishes the result by induction.
\end{proof}

\chapter{Stochastic Integrals}

Stochastic processes are intermediatory between random variables and quantum
fields. That's certainly not the view that most probabilists would have of
their subject but for our purposes it is true enough. Indeed, there exists a
natural theory of quantum stochastic calculus extending the classical theory
of It\^{o} to the quantum domain.

We can treat stochastic processes as a special case of quantum fields over
one time dimension. More exactly,we may view stochastic processes as
regularized forms of white noise processes - these are stochastic processes
only in some distributional sense, but can be realized as quantum fields
theory . Instead, we shall work with the more traditional viewpoint used by
probabilists. We first recall some basic ideas.

We say that a random variable is second order if it has finite first and
second moment and that a second order variable $X$ is the mean-square limit
of a sequence of second order variables $\left( X_{n}\right) _{n}$ if $%
\mathbb{E}\left[ \left( X-X_{n}\right) ^{2}\right] \rightarrow 0$. A
stochastic process is a family of random variables $\left\{ X_{t}:t\geq
0\right\} $ labeled by time parameter $t$. The process is second order if $%
\mathbb{E}\left[ X_{t}\right] ,\,\mathbb{E}\left[ X_{t}X_{s}\right] <\infty $%
, for all times $t,s>0$.

A process $\left\{ X_{t}:t\geq 0\right\} $ is said to have independent
increments if $X_{t}-X_{s}$ and $X_{t^{\prime }}-X_{s^{\prime }}$ are
independent whenever the time intervals $\left( s,t\right) $ and $\left(
s^{\prime },t^{\prime }\right) $ do not overlap. The increments are
stationary if the probability distribution of each $X_{t}-X_{s}$ depends
only on $t-s$, where we assume $t>s$. The most important examples of second
order processes are the Wiener process $\left\{ W_{t}:t\geq 0\right\} $ and
the Poisson process $\left\{ N_{t}:t\geq 0\right\} $. The Wiener process is
characterized by having increments $W_{t}-W_{s}$, $\left( t>s\right) $,
distributed according to a Gaussian law of mean zero and variance $t-s$. The
Poisson process is characterized by having increments $N_{t}-N_{s}$, $\left(
t>s\right) $, distributed according to a Poisson law of intensity $t-s$.

In general, given a pair of stochastic processes $\left\{ X_{t}\right\} $
and $\left\{ Y_{t}\right\} $, we can try and give meaning to their
stochastic integral $\int_{a}^{b}X_{t}dY_{t}$. What we do is to divide the
interval $\left[ a,b\right] $ up as $a=t_{0}<t_{1}<\cdots <t_{n}=b$ and
consider the finite sum $\sum_{j=0}^{n}X_{t_{j}}\left(
Y_{t_{j+1}}-Y_{t_{j}}\right) $ and consider the limit in which $\max \left\{
t_{j+1}-t_{j}\right\} \rightarrow 0$. If the sequence of finite sums has a
mean-square limit then we call it the It\^{o} integral $%
\int_{a}^{b}X_{t}dY_{t}$.

There is a problem, however. The Wiener process has increments that do not
behave in the way we expect infinitesimals to behave. Let $\Delta t>0$ and
set $\Delta W_{t}=W_{t+\Delta t}-W_{t}$, then, far from being negligible, $%
\left( \Delta W_{t}\right) ^{2}$ is a random variable of mean value $\Delta t
$. The consequence is that the usual rules of calculus will not apply to
stochastic integrals wrt. the Wiener process. Remarkably, if we work with
suitable process there is a self-consistent theory, the It\^{o} calculus,
which extends the usual notions. In particular, the integration by parts
formula is replaced by the It\^{o} formula,\ which states that 
\begin{equation*}
d\left( X_{t}Y_{t}\right) =\left( dX_{t}\right) Y_{t}+X_{t}\left(
dY_{t}\right) +\left( dX_{t}\right) \left( dY_{t}\right) 
\end{equation*}
and is to be understood under the integral sign. We now present some
combinatorial results on multiple It\^{o} integrals: more details can be
found in the paper of Rota and Wallstrom \cite{RotaWallstrom}.

\section{Multiple Stochastic Integrals}

Let $X_{t}^{\left( j\right) }$ be stochastic processes for $j=1,\dots ,n$.
We use the following natural (notational) conventions: 
\begin{equation*}
X_{t}^{\left( n\right) }\cdots X_{t}^{\left( 1\right) }=\int_{\left[ 0,t%
\right] }dX_{t_{n}}^{\left( n\right) }\cdots \int_{\left[ 0,t\right]
}dX_{t_{1}}^{\left( 1\right) }\equiv \int_{\left[ 0,t\right]
^{n}}dX_{t_{n}}^{\left( n\right) }\cdots dX_{t_{1}}^{\left( 1\right) }.
\end{equation*}

In the following we denote by $\Delta _{\sigma }^{n}\left( t\right) $ the $%
n- $simplex in $\left[ 0,t\right] ^{n}$ determined by a permutation $\sigma
\in \frak{S}_{n}$: that is, 
\begin{equation*}
\Delta _{\sigma }^{n}\left( t\right) =\left\{ \left( t_{n},\cdots
,t_{1}\right) \in \left( 0,t\right) ^{n}:t_{\sigma \left( n\right)
}>t_{\sigma \left( n-1\right) }>\cdots >t_{\sigma \left( 1\right) }\right\} .
\end{equation*}
We denote by $\Delta ^{n}\left( t\right) $ the simplex determined by the
identity permutation: that is $t>t_{n}>t_{n-1}>\cdots >t_{1}>0$. Clearly $%
\cup _{\sigma \in \frak{S}_{n}}\Delta _{\sigma }^{n}\left( t\right) $ is $%
\left[ 0,t\right] ^{n}$ with the absence of the hypersurfaces (diagonals) of
dimension less than $n$ corresponding to the $t_{j}$'s being equal.
Moreover, the $\Delta _{\sigma }^{n}\left( t\right) $ are distinct for
different $\sigma $.

We also define the off-diagonal integral ``$-\hspace{-0.12in}\int $'' to be
the expression with all the diagonal terms subtracted out. Explicitly 
\begin{equation*}
-\hspace{-0.15in}\int_{\left[ 0,t\right] ^{n}}dX_{t_{n}}^{\left( n\right)
}\cdots dX_{t_{1}}^{\left( 1\right) }:=\sum_{\sigma \in \frak{S}%
_{n}}\int_{\Delta _{\sigma }^{n}\left( t\right) }dX_{t_{n}}^{\left( n\right)
}\cdots dX_{t_{1}}^{\left( 1\right) }.
\end{equation*}

Take $s_{1},s_{2},\dots $ to be real variables and let $\mathcal{A}=\left\{
A_{1},\dots ,A_{m}\right\} $ be a partition of $\left\{ 1,\dots ,n\right\} $%
\ then, for each $i\in \left\{ 1,\dots ,n\right\} $, define $s_{\mathcal{A}%
}\left( i\right) $ to be the variable $s_{j}$ where $i$ lies in the part $%
A_{j}$.

\bigskip

\begin{lemma}
The multiple stochastic integral can be decomposed as 
\begin{equation*}
\int_{\left[ 0,t\right] ^{n}}dX_{t_{n}}^{\left( n\right) }\cdots
dX_{t_{1}}^{\left( 1\right) }=\sum_{\mathcal{A}\in \frak{P}_{n}}-\hspace{%
-0.15in}\int_{\left[ 0,t\right] ^{N\left( \Gamma \right) }}dX_{s_{\mathcal{A}%
}\left( n\right) }^{\left( n\right) }\cdots dX_{s_{\mathcal{A}}\left(
1\right) }^{\left( 1\right) }
\end{equation*}
\end{lemma}

\bigskip

Note that there $n=1$ case is immediate as the $\int $ and $-\hspace{-0.12in}%
\int $ integrals coincide. In the situation $n=2$ we have by the It\^{o}
formula 
\begin{eqnarray*}
X_{t}^{\left( 2\right) }X_{t}^{\left( 1\right) }
&=&\int_{0}^{t}dX_{t_{2}}^{\left( 2\right) }\,X_{t_{2}}^{\left( 1\right)
}+\int_{0}^{t}X_{t_{1}}^{\left( 2\right) }\,dX_{t_{1}}^{\left( 1\right)
}+\int_{0}^{t}dX_{s}^{\left( 2\right) }dX_{s}^{\left( 1\right) } \\
&=&\int_{t>t_{2}>t_{1}>0}dX_{t_{2}}^{\left( 2\right) }dX_{t_{1}}^{\left(
1\right) }+\int_{t>t_{1}>t_{2}>0}dX_{t_{2}}^{\left( 2\right)
}dX_{t_{1}}^{\left( 1\right) }+\int_{0}^{t}dX_{s}^{\left( 2\right)
}dX_{s}^{\left( 1\right) } \\
&\equiv &-\hspace{-0.15in}\int_{\left[ 0,t\right] ^{2}}dX_{t_{2}}^{\left(
2\right) }dX_{t_{1}}^{\left( 1\right) }+-\hspace{-0.15in}\int_{\left[ 0,t%
\right] }dX_{s}^{\left( 2\right) }dX_{s}^{\left( 1\right) }
\end{eqnarray*}
and this is the required relation.

The higher order terms are computed through repeated applications of the
It\^{o} formula. An inductive proof is arrived at along the following lines.
Let $X_{t}^{\left( n+1\right) }$ be another quantum stochastic integral,
then the It\^{o} formula is 
\begin{equation*}
X_{t}^{\left( n+1\right) }Y_{t}=-\hspace{-0.15in}\int_{\left[ 0,t\right]
^{2}}dX_{t_{n+1}}^{\left( n+1\right) }dY_{t_{n}}+-\hspace{-0.15in}\int_{%
\left[ 0,t\right] }dX_{s}^{\left( n+1\right) }dY_{s}
\end{equation*}
and we take $Y_{t}=X_{t}^{\left( n\right) }\cdots X_{t}^{\left( 1\right) }$.
Assume the formula is true form $n$. The first term will be the sum over all
partitions of $\left\{ n+1,n,\dots ,1\right\} $ in which $\left\{
n+1\right\} $ appears as a singleton, the second term will be the sum over
all partitions of $\left\{ n+1,n,\dots ,1\right\} $ in which $n+1$ appears
as an extra in some part of a partition of $\left\{ n,\dots ,1\right\} $. In
this way we arrive at the appropriate sum over $\frak{P}_{n+1}$.

\bigskip

\begin{corollary}
The inversion formula for off-diagonal integrals is 
\begin{equation*}
-\hspace{-0.15in}\int_{\left[ 0,t\right] ^{n}}dX_{t_{n}}^{\left( n\right)
}\cdots dX_{t_{1}}^{\left( 1\right) }\equiv \sum_{\mathcal{A}\in \frak{P}%
_{n}}\mu \left( \mathcal{A}\right) \,\int_{\left[ 0,t\right] ^{N\left(
\Gamma \right) }}dX_{s_{\Gamma }\left( n\right) }^{\left( n\right) }\cdots
dX_{s_{\Gamma }\left( 1\right) }^{\left( 1\right) }
\end{equation*}
where $\mu \left( \mathcal{A}\right) $\ is the M\"{o}bius function
introduced earlier.
\end{corollary}

\subsection{Multiple Martingale Integrals}

We suppose that $X_{t}$ is a martingale so that, in the particular, $\mathbb{%
E}\left[ \int_{[0,t]}\phi \left( s\right) dX_{s}\right] =0$ where $\phi $ is
any adapted integrable function. In general, multiple integrals with respect
to $X$ will not have zero expectation, however,this will be the case for the
off-diagonal integrals: 
\begin{equation*}
\mathbb{E}\left[ -\hspace{-0.15in}\int_{\left[ 0,t\right] ^{n}}dX_{t_{n}}^{%
\left( n\right) }\cdots dX_{t_{1}}^{\left( 1\right) }\right] \equiv 0.
\end{equation*}
This property is in fact the main reason for introducing off-diagonal
integrals in the first place.

In the special case of classical processes we can employ the commutativity
to write 
\begin{equation*}
\int_{\Delta ^{n}\left( t\right) }dX_{t_{n}}^{\left( n\right) }\cdots
dX_{t_{1}}^{\left( 1\right) }\equiv \frac{1}{n!}-\hspace{-0.15in}\int_{\left[
0,t\right] ^{n}}dX_{t_{n}}^{\left( n\right) }\cdots dX_{t_{1}}^{\left(
1\right) }.
\end{equation*}
We define a family of random variables $E^{X_{t}}$ by 
\begin{equation*}
E^{X_{t}}:=\sum_{n\geq 0}\frac{1}{n!}-\hspace{-0.15in}\int_{\left[ 0,t\right]
^{n}}dX_{t_{n}}^{\left( n\right) }\cdots dX_{t_{1}}^{\left( 1\right) }
\end{equation*}
for $t\geq 0$. We refer to process $t\mapsto E^{X_{t}}$ as an exponentiated
martingale.

\subsection{Wiener Integrals}

Let us take $X_{t}$ to be the Wiener process $W_{s}$. We will have 
\begin{equation*}
-\hspace{-0.15in}\int_{\left[ 0,t\right] ^{n}}dW_{t_{n}}^{\left( n\right)
}\cdots dW_{t_{1}}^{\left( 1\right) }\equiv \sum_{\mathcal{A}\in \frak{P}%
_{n}}\mu \left( \mathcal{A}\right) \,\int_{\left[ 0,t\right] ^{N\left( 
\mathcal{A}\right) }}dW_{s_{\mathcal{A}}\left( n\right) }^{\left( n\right)
}\cdots dW_{s_{\mathcal{A}}\left( 1\right) }^{\left( 1\right) }
\end{equation*}
Here we have $dW_{s}dW_{s}=ds$: this means that if we sum over partitions $%
\mathcal{A}$ in the inversion formula then we need only consider those
consisting of singletons and pairs only. We will then have $E\left( \mathcal{%
A}\right) =n_{1}+2n_{2}=n$ and $N\left( \mathcal{A}\right)
=n_{1}+n_{2}=n-n_{2}$. The combinatorial factor is $F\left( \mathcal{A}%
\right) =1$.

Now there are $\binom{n}{n_{1}}=\binom{n}{2n_{2}}$ ways to choose the
singletons and $\frac{\left( 2n_{2}\right) !}{2^{n_{2}}n_{2}!}$ ways to
choose the pairs. This yields 
\begin{eqnarray*}
-\hspace{-0.15in}\int_{\left[ 0,t\right] ^{n}}dW_{t_{n}}^{\left( n\right)
}\cdots dW_{t_{1}}^{\left( 1\right) } &=&\sum_{n_{2}=0}^{\left[ n/2\right]
}\left( -1\right) ^{n-n_{2}}\frac{n!}{2^{n_{2}}n_{2}!\left( n-2n_{2}\right) !%
}\left( W_{t}\right) ^{n-2n_{2}}t^{n_{2}} \\
&\equiv &t^{n/2}H_{n}\left( \frac{W_{t}}{\sqrt{t}}\right) 
\end{eqnarray*}
where $H_{n}\left( x\right) =\sum_{k=0}^{\left[ n/2\right] }\left( -1\right)
^{n-k}\frac{n!}{2^{k}k!\left( n-2k\right) !}x^{n-k}$ are the well-known
Hermite polynomials. The implication that $\int_{\Delta ^{n}\left( t\right)
}dW_{t_{n}}^{\left( n\right) }\cdots dW_{t_{1}}^{\left( 1\right) }=\frac{1}{%
n!}t^{n/2}H_{n}\left( \frac{W_{t}}{\sqrt{t}}\right) $ is a result due
originally to It\^{o} .

Using the relation $\sum_{n=0}^{\infty \ }\frac{t^{n}}{n!}H_{n}\left(
x\right) =\exp \left( xt-t^{2}/2\right) $ for generating the Hermite
polynomials, we see that the exponentiated random variable corresponding to
the choice of the Wiener process is $E^{zW_{t}}=e^{zW_{t}-z^{2}t/2}.$

More generally, if we take $W_{t}\left( f\right) =\int_{\left[ 0,t\right]
}f\left( s\right) dW_{s}$ for $f$ square-integrable then its exponentiated
random variable is 
\begin{equation*}
E^{W_{t}\left( f\right) }=\exp \left\{ W_{t}\left( f\right) -\frac{1}{2}%
\int_{\left[ 0,t\right] }f\left( s\right) ^{2}ds\right\} .
\end{equation*}

\subsection{Compensated Poisson Process Integrals}

Let $N_{t}$ be the Poisson process and $Y_{t}=N_{t}-t$ be the compensated
process. We have the differential rule $\left( dN_{t}\right) ^{p}=dN_{t}$
for all positive integer powers $p$. The process $Y_{t}$ is a martingale and
we have $\left( dY_{t}\right) ^{p}=dN_{t}=dY_{t}+dt$ for all positive
integer $p$ with the obvious exception of $p=1$.

It is convenient to replace sums over partitions with sums over occupation
numbers. This time we find 
\begin{equation*}
-\hspace{-0.15in}\int_{\left[ 0,t\right] ^{n}}dY_{t_{n}}^{\left( n\right)
}\cdots dY_{t_{1}}^{\left( 1\right) }\equiv \sum_{\mathbf{n}}^{E\left( 
\mathbf{n}\right) =n}\frac{\left( -1\right) ^{n_{2}+2n_{3}+3n_{4}+\cdots }}{%
\left( 1\right) ^{n_{1}}\left( 2\right) ^{n_{2}}\left( 3\right)
^{n_{3}}\cdots }\;\frac{n!}{n_{1}!n_{2}!n_{3}!\cdots }\left( Y_{t}\right)
^{n_{1}}\left( Y_{t}+t\right) ^{n_{2}}\left( Y_{t}+t\right) ^{n_{3}}\cdots
\end{equation*}
Again we use a $\sum \prod \longleftrightarrow \prod \sum $ trick! 
\begin{eqnarray*}
E^{zY_{t}} &=&\sum_{\mathbf{n}}\frac{\left( -1\right)
^{n_{2}+2n_{3}+3n_{4}+\cdots }}{\left( 1\right) ^{n_{1}}\left( 2\right)
^{n_{2}}\left( 3\right) ^{n_{3}}\cdots }\;\frac{z^{n_{1}+2n_{2}+3n_{3}+%
\cdots }}{n_{1}!n_{2}!n_{3}!\cdots }\left( Y_{t}\right) ^{n_{1}}\left(
Y_{t}+t\right) ^{n_{2}}\left( Y_{t}+t\right) ^{n_{3}}\cdots \\
&=&e^{-zt}\sum_{\mathbf{n}}\frac{\left( -1\right)
^{n_{2}+2n_{3}+3n_{4}+\cdots }}{\left( 1\right) ^{n_{1}}\left( 2\right)
^{n_{2}}\left( 3\right) ^{n_{3}}\cdots }\;\frac{z^{n_{1}+2n_{2}+3n_{3}+%
\cdots }}{n_{1}!n_{2}!n_{3}!\cdots }\left( Y_{t}+t\right) ^{n_{1}}\left(
Y_{t}+t\right) ^{n_{2}}\left( Y_{t}+t\right) ^{n_{3}}\cdots \\
&=&e^{-zt}\prod_{k=1}^{\infty }\left\{ \sum_{n_{k}=0}^{\infty }\left(
-1\right) ^{\left( k-1\right) n_{k}}\frac{1}{n_{k}!}\left( \frac{z^{k}\left(
Y_{t}+1\right) }{k}\right) ^{n_{k}}\right\} \\
&=&e^{-zt}\prod_{k=1}^{\infty }\exp \left\{ \left( -1\right) ^{k-1}\frac{%
z^{k}}{k}\left( Y_{t}+t\right) \right\}
\end{eqnarray*}
and using $\sum_{k=1}^{\infty }\left( -1\right) ^{k-1}\frac{z^{k}}{k}=\ln
\left( 1+z\right) $ we end up with 
\begin{equation*}
E^{zY_{t}}=e^{-zt}\left( 1+z\right) ^{Y_{t}+t}.
\end{equation*}
This means that, in terms of the Poisson process $N_{t}$, the exponentiated
random variable associated with the compensated Poisson process $Y_{t}$ is $%
E^{Y_{t}}\left( z\right) =e^{-zt}\left( 1+z\right) ^{N_{t}}$ and we have 
\begin{equation*}
-\hspace{-0.15in}\int_{\left[ 0,t\right] ^{n}}dY_{t_{n}}^{\left( n\right)
}\cdots dY_{t_{1}}^{\left( 1\right) }=C_{n}\left( N_{t},t\right)
\end{equation*}
where $C_{n}\left( x,t\right) $ are the Charlier polynomials determined by
the generating relation $\sum_{n\geq 0}\frac{z^{n}}{n!}C_{n}\left(
x,t\right) =e^{-zt}\left( 1+z\right) ^{x}$. Explicitly we have 
\begin{equation*}
C_{n}\left( x,t\right) =\sum_{k}\left( -t\right) ^{n-k}\binom{n}{k}%
x^{\downarrow k}=\sum_{k,r}\left( -t\right) ^{n-k}\binom{n}{k}s\left(
k,r\right) x^{k}.
\end{equation*}

\section{It\^{o}-Fock Isomorphism}

Here we examine stochastic processes and exploit the fact that they can be
viewed as commuting one-dimensional quantum fields (with base space
typically taken as time). It should be remarked that the Wiener and Poisson
process of classical probability can be naturally viewed as combinations of
creation, annihilation and conservation processes introduced by Hudson and
Parthasarathy \cite{HP} as the basis of a quantum stochastic calculus which
extends the It\^{o} calculus to operator-valued processes. In fact the It%
\^{o} correction can be alternatively understood as the additional term that
arises from Wick ordering of stochastic integrals with respect to white
noise creation/annihilation operators, see \cite{Gough}, for instance, and
also \cite{GOUGHCMP}.

\subsection{It\^{o}-Fock Isomorphism}

Let $X_{t}$ be a classical martingale with canonical probability space $%
\left( \Omega _{X},\mathcal{F}_{X},\mathbb{P}\right) $ and let $\frak{h}%
_{X}=L^{2}\left( \Omega _{X},\mathcal{F}_{X},\mathbb{P}\right) $. We
consider the function $F\left( t\right) :=\mathbb{E}\left[ X_{t}^{2}\right] $
and this defines a monotone increasing function. We shall understand $%
dF\left( t\right) $ to be the Stieltjes integrator in the following.

It turns out that our considerations so far allow us to construct a natural
isomorphism between $\frak{h}_{X}$ and the Fock space $\Gamma _{+}\left(
L^{2}\left( \mathbb{R}^{+},dF\right) \right) $, see e.g. \cite{Partha}.

For f$\in L^{2}\left( \mathbb{R}^{+},dF\right) $, we define the random
variable $\tilde{X}\left( f\right) :=\int_{[0,\infty )}$f$\left( s\right)
dX_{s}$ and $\tilde{X}_{t}\left( f\right) :=\tilde{X}\left( 1_{\left[ 0,t%
\right] }f\right) $.

\bigskip

\begin{lemma}
Let $\tilde{Z}_{t}^{\left( n\right) }\left( f\right) =\int_{\Delta
^{n}\left( t\right) }d\tilde{X}_{t_{n}}\left( f\right) \cdots \tilde{X}%
_{t_{1}}\left( f\right) $\ then 
\begin{equation*}
\mathbb{E}\left[ \tilde{Z}_{t}^{\left( n\right) }\left( f\right) \tilde{Z}%
_{s}^{\left( m\right) }\left( g\right) \right] =\frac{1}{n!}\left[
\int_{0}^{t\wedge s}f\left( u\right) g\left( u\right) \,dF\left( u\right) %
\right] ^{n}\,\delta _{n,m}.
\end{equation*}
\end{lemma}

\begin{proof}
For simplicity, we ignore the intensities. Let $Z_{t}^{\left( n\right)
}=\int_{\Delta ^{n}\left( t\right) }dX_{t_{n}}\cdots dX_{t_{1}}$ then we
have $\mathbb{E}\left[ Z_{t}^{\left( n\right) }Z_{s}^{\left( 0\right) }%
\right] =\mathbb{E}\left[ Z_{t}^{\left( n\right) }\right] =0$ whenever $n>0$%
. Next suppose that $n$ and $m$ are positive integers, then 
\begin{eqnarray*}
\mathbb{E}\left[ Z_{t}^{\left( n\right) }Z_{s}^{\left( m\right) }\right] &=&%
\mathbb{E}\left[ \int_{0}^{t}dX_{u}\,Z_{u}^{\left( n-1\right)
}\;\int_{0}^{s}dX_{v}\,Z_{v}^{\left( m-1\right) }\right] \\
&=&\int_{0}^{t\wedge s}dF\left( u\right) \;\mathbb{E}\left[ Z_{u}^{\left(
n-1\right) }\,Z_{u}^{\left( m-1\right) }\right]
\end{eqnarray*}
and we may re-iterate until we reduce at least one of the orders to zero. We
then have 
\begin{eqnarray*}
\mathbb{E}\left[ Z_{t}^{\left( n\right) }Z_{s}^{\left( m\right) }\right]
&=&\delta _{n,m}\;\int_{0}^{t\wedge s}dF\left( u_{n}\right)
\int_{0}^{u_{n}}dF\left( u_{n-1}\right) \cdots \int_{0}^{u_{2}}dF\left(
u_{1}\right) \\
&=&\frac{1}{n!}F\left( t\wedge s\right) ^{n}\,\delta _{n,m}.
\end{eqnarray*}
The proof with the intensities from $L^{2}\left( \mathbb{R}^{+},dF\right) $
included is then a straightforward generalization.
\end{proof}

\bigskip

\begin{theorem}
The Hilbert spaces $\frak{h}_{X}=L^{2}\left( \Omega _{X},\mathcal{F}_{X},%
\mathbb{P}\right) $\ and $\Gamma _{+}\left( L^{2}\left( \mathbb{R}%
^{+},dF\right) \right) $\ are naturally isomorphic.
\end{theorem}

\begin{proof}
Consider the map into the exponential vectors (\ref{Expvectors}) given by 
\begin{equation*}
E^{\tilde{X}\left( f\right) }\mapsto \varepsilon \left( f\right)
\end{equation*}
for each f$\in L^{2}\left( \mathbb{R}^{+},dF\right) $. We know that the
exponential vectors are dense in Fock space and in a similar way the
exponential martingales $E^{\tilde{X}\left( f\right) }$ are dense in $\frak{h%
}_{X}$. The map may then be extended to one between the two Hilbert spaces.

Unitarity follows from the observation that 
\begin{equation*}
\mathbb{E}\left[ E^{\tilde{X}\left( f\right) }E^{\tilde{X}\left( g\right) }%
\right] =e^{\int_{[0,\infty )}fg\,\,dF}
\end{equation*}
which is an immediate consequence of the previous lemma.
\end{proof}

\bigskip

The choice of the Wiener process is especially widely used. Here we have the
identification 
\begin{equation*}
L^{2}\left( \Omega _{W},\mathcal{F}_{W},\mathbb{P}\right) \cong \Gamma
_{+}\left( L^{2}\left( \mathbb{R}^{+},dt\right) \right)
\end{equation*}
which goes under the name of the Wiener-It\^{o}-Segal isomorphism. This
result is one of the corner stones of Hida's theory \cite{Hida} of white
noise analysis. The same Fock space occurs when we consider the compensated
Poisson process also \cite{Partha}.

\end{document}